\documentclass[prd,aps,nofootinbib,twocolumn,10pt]{revtex4}
\usepackage{graphicx}

\usepackage{bbold}
\usepackage{epsfig,epsf}
\usepackage{amsmath}
\usepackage{amsthm}
\usepackage{amsfonts}
\usepackage{amssymb}
\usepackage{dsfont}
\usepackage{multirow}
\usepackage{appendix}
\usepackage{slashed}
\usepackage[active]{srcltx}
\usepackage{psfrag}
\usepackage{dcolumn}

\def  \nn     {\nonumber}

\begin{document}

\title{ Exploratory study of X(4140) in QCD sum rules }
\author{Arzu T{\"u}rkan}
\affiliation{{\"O}zye\u{g}in University, Department of Natural and Mathematical Sciences,
\c{C}ekmek\"{o}y, \.{I}stanbul, Turkey}
\author{H{\"u}seyin Da\u{g}}
\affiliation{{\"O}zye\u{g}in University, Department of Natural and Mathematical Sciences,
\c{C}ekmek\"{o}y, \.{I}stanbul, Turkey}

\begin{abstract}
In this work, we chose three molecular and three diquark-antidiquark currents with the quark content $c\bar{c}s\bar{s}$ and $J^{PC}=0^{++},1^{++},2^{++}$, and estimated the masses and the meson coupling constants of the ground states coupling to these currents in the framework of QCD sum rules. In operator product expansion, we considered the terms including dimension eight, and we performed pole contribution tests carefully. According to our results, all of these currents couple to the ground states with degenerate masses which are in 10 MeV vicinity of X(4140). Therefore, with a QCD sum rules analysis, it is not possible to conclude that X(4140) has a dominant molecular or diquark-antidiquark content. However, there may be three states degenerate in mass, with positive charge conjugation and different isospins. 

\end{abstract}

\maketitle

\section{Introduction}

%
%
In the past fifteen  years, charmonium like resonances which do not seem to have a simple $c\bar{c}$ structure were observed by several experiments. All of these new states, decay into final states containing charmonium, however their final decay rates into open charm pairs are unexpectedly small. Thus, they are considered as good candidates of exotic hadrons, which are assumed to have other structure than the ordinary mesons and baryons. A list of these states, and their current experimental status can be found in Refs. \cite{Ozet1,Ozet2,Nielsen_Ozet}.

Among these newly observed resonances, X(4140) was experimentally observed by several collaborations in the invariant mass spectrum of $J/\psi \phi$ final states, and very recently its quantum numbers were announced to be $J^{PC}=1^{++}$ by LHCb Collaboration\cite{CDF1,Belle,CDF2,LHCb1,CMS,D0,Babar,D0_2,LHCb}. However, the decay width of the state observed by LHCb is unexpectedly wider than the previous observations, and the origin of this difference still remains unsolved. In Table \ref{Tab::Int_EXP} we summarized the current experimental results which confirms the existence  of X(4140) . Despite these observations, its structure has not been totally understood yet, as well as other exotic mesons.

\begin{table}[h]
\begin{center}
\caption{Experimental results on the mass and decay width of X(4140), given in chronological order. The average is taken from Ref. \cite{LHCb}, and it gives the status of X(4140) measurements before the recent LHCb results.}
\label{Tab::Int_EXP}
\begin{tabular}{lclclclcl}
\hline\hline
Year & ~Experiment~ & Mass (MeV) &~~~ Width (MeV) ~~~ \\ \hline\hline
2008&CDF \cite{CDF1}&$4143.0\pm 2.9\pm 1.2$&$11.7^{+8.3}_{-5.0}\pm 3.7$\\
2009&Belle \cite{Belle}& $4143.0$ &$11.7$\\
2011& CDF \cite{CDF2}& $4143.4^{+ 2.9}_{-3.0}\pm 0.6$ &$15.3^{+10.4}_{-6.1}\pm 2.5$\\
2011& LHCb \cite{LHCb1}& $4143.4$ & $15.3$\\
2013& CMS \cite{CMS}&$4148.0\pm 2.4\pm 6.3$&$28^{+15}_{-11}\pm 19$\\
2013& D0 \cite{D0}&$4159.0\pm 4.3\pm 6.6$&$19.9\pm 12.6^{+1.0}_{-8.0}$\\
2014& Babar \cite{Babar}& $4143.4$ & $15.3$\\
2015& D0 \cite{D0_2}& $4152.5\pm 1.7^{+ 6.2}_{-5.4}$ &$16.3\pm 5.6\pm 11.4$\\
& Average \cite{LHCb}&$4147.1\pm2.4$&$15.7\pm6.3$\\\hline
2017 &LHCb \cite{LHCb}&$4146.5\pm 4.5^{+ 4.6}_{-2.8}$ &$83\pm 21^{21}_{-14}$\\
\hline\hline
\end{tabular}
\end{center}%
\end{table}

In the literature, there are several studies exploring the structure of X(4140)\cite{Stancu,Mahajan,Branz,Prelovsek,Nielsen,Zhang,Huang,Wang,Wang2,Wang3,Wang4,Wang5,CZ2011,Agaev}. Among these studies, the tetraquark model proposed in \cite{Stancu} predicted the quantum numbers and the mass of the X(4140) consistent with LHCb. However, lattice QCD calculation with diquark operators found no evidence for the existence of an axial vector X(4140)\cite{Prelovsek}. Within these theoretical efforts, studies with QCD sum rules (QCDSR) are puzzling since most of them predicts the quantum numbers of X(4140) inconsistent with the  LHCb. In Refs. \cite{Nielsen,Zhang,Huang,Wang3}, the authors predicted X(4140) to be a scalar $D_s^*\bar{D}_s^*$ molecule, however Refs. \cite{Wang,Wang2} claim the opposite with similar currents. In Refs. \cite{Wang,Wang2}, using scalar and axial vector tetraquark currents, the authors calculated masses of the ground states but the results were incompatible with X(4140). However with a similar axial vector diquark-antidiquark current, a mass close to recent results of LHCb was predicted in \cite{CZ2011}. These inconsistencies in between the predictions of QCD sum rules studies, and with the LHCb results, motivated us to perform a complete QCD sum rules investigation with the aim of contributing to this unconcluded topic.
 
In the present work, we calculated the mass and the meson coupling constants of the ground states coupling to $D^*_s\bar{D}^*_s$ or diquark-antidiquark currents with the content $sc\bar{c}\bar{s}$. In order to make a comparison with the aforementioned QCD sum rules studies, we chose these currents to be scalar, axialvector and tensor. To this end, we calculated the two point correlation functions of these where the quark, gluon and mixed condansates were considered up to dimension eight. We performed a very detailed numerical analysis to achieve the convergence of the series in operator product expansion(OPE) and the dominance of the ground state resonance in the continuum. Performed investigations should allow us to find an explanation to this unconcluded puzzle.    

This work is organised as follows. In section \ref{S2}, we present the sum rules calculations of masses and meson couplings of the ground states coupling to currents under investigation. In section \ref{S3}, we performed a numerical analysis for the success of the evaluated sum rules and extracted the numerical values of masses and meson couplings. Here we also compared our results with other theoretical predictions. Section IV contains our final conclusions. The explicit expressions of the spectral densities of the two point correlators are given in Appendix.
 
%
\section{QCD Sum Rules for Masses and Meson Coupling Constants}\label{S2}

In order to study the physical properties of X(4140), we started with six interpolating currents, interpreting X(4140) as a scalar, a vector or a tensor $D_s^*\bar{D}_s^*$ molecule, or as a scalar, a vector or a tensor diquark-antidiquark. The following molecular currents

\begin{equation}\label{curr_S_m}
J^{(1)}(x)=\overline{s}^{i}(x)\gamma_{\mu}c^{i}(x)~\overline{c}^{j}(x)\gamma^{\mu}s^{j}(x),
\end{equation}
\begin{eqnarray}\label{curr_AV_m}
\nn J^{(1)}_{\mu}(x)=\frac{1}{\sqrt{2}}\left(\overline{s}^{i}(x)\gamma_{5}c^{i}(x)~\overline{c}^{j}(x)\gamma_{\mu}s^{j}(x)\right. \\
-\left. \overline{s}^{i}(x)\gamma_{\mu}c^{i}(x)~\overline{c}^{j}(x)\gamma_{5}s^{j}(x)\right),
\end{eqnarray}
\begin{eqnarray}\label{curr_T_m}
\nn J^{(1)}_{\mu\nu}(x)=\frac{1}{\sqrt{2}}\left(\overline{s}^{i}(x)\gamma_{\mu}c^{i}(x)~\overline{c}^{j}(x)\gamma_{\nu}s^{j}(x)\right.\\
+\left.\overline{s}^{i}(x)\gamma_{\nu}c^{i}(x)~\overline{c}^{j}(x)\gamma_{\mu}s^{j}(x)\right) ,
\end{eqnarray}
and the following diquark-antidiquark currents
\begin{eqnarray}\label{curr_S_t}
J^{(2)}(x)=\varepsilon^{ijk}\varepsilon^{imn}(s^{j}(x)C\gamma_{\mu}c^{k}(x)~\overline{s}^{m}(x)\gamma^{\mu}C\overline{c}^{n}(x)),
\end{eqnarray}
%
\begin{eqnarray}\label{curr_AV_t}
\nn J^{(2)}_{\mu}(x)=\frac{\varepsilon^{ijk}\varepsilon^{imn}}{\sqrt{2}}\left({s}^{j}(x)C\gamma_{5}c^{k}(x)~\overline{s}^{m}(x)\gamma_{\mu}C \overline c^{n}(x)\right.\\
\left.-{s}^{j}(x)C\gamma_{\mu}c^{k}(x)~\overline{s}^{m}(x)\gamma_{5}C \overline c^{n}(x)\right),
\end{eqnarray}
\begin{eqnarray}\label{curr_T_t}
\nn J^{(2)}_{\mu\nu}(x)=\frac{\varepsilon^{ijk}\varepsilon^{imn}}{\sqrt{2}}\left(s^{j}(x)C\gamma_{\mu}c^{k}(x)~\overline{s}^{m}(x)\gamma_{\nu}C\overline{c}^{n}(x)\right.\\
\left.+s^{j}(x)C\gamma_{\nu}c^{k}(x)~\overline{s}^{m}(x)\gamma_{\mu}C\overline{c}^{n}(x)\right),
\end{eqnarray}
are chosen to examine the QCDSR analyses that were done in literature\cite{Nielsen,Zhang,Huang,Wang,Wang2,Wang3,Wang4,Wang5,CZ2011}, where $C$ is the charge conjugation matrix and $i,\ j \ , ...$ are color indices. In Eqs. (\ref{curr_S_m} - \ref{curr_T_t}), the superscripts (1) and (2) denote that the current has either molecular or diquark-antidiquark structure, and they will be denoted by superscript (a) whenever a compact formalism is required. The sum rules to obtain masses and meson coupling constants of the ground state mesons coupling to these currents are constructed from the following two point correlation functions 
\begin{equation}\label{corr0}
\Pi^{(a)} (q)= i \int d^4x e^{i q\cdot x} \langle 0 \vert  {\cal T} (J^{(a)}(x) {J^{(a)}}^{\dagger}(0)) \vert 0 \rangle,
\end{equation}
\begin{equation}\label{corr0_1}
\Pi_{\mu\nu}^{(a)} (q)= i \int d^4x e^{i q\cdot x} \langle 0 \vert  {\cal T} (J_\mu^{(a)}(x) {J_\nu^{(a)}}^{\dagger}(0)) \vert 0 \rangle,
\end{equation}
\begin{equation}\label{corr0_2}
\Pi_{\mu\nu\alpha\beta}^{(a)} (q)= i \int d^4x e^{i q\cdot x} \langle 0 \vert  {\cal T} (J_{\mu\nu}^{(a)}(x) {J_{\alpha\beta}^{(a)}}^{\dagger}(0)) \vert 0 \rangle.
\end{equation}
In QCDSR approach, the correlation function's dependence on momentum $q$ enables us to extract the physical properties of a hadron, by evaluating the correlator twice in different momentum regions and relating these two expressions to obtain sum rules. For $q^2>>0$, i.e., large distances, the interpolating current and its conjugate are interpreted as annihilation and creation operators of the mesons which has the same quark content and quantum numbers as the chosen current. In this case, the correlation functions in Eqs. (\ref{corr0}-\ref{corr0_2}) are saturated with a complete set of hadrons and integrals over $x$ are performed. These interpretations of the correlators are called ``the physical (or phenomenological) side". For $q^2<<0$, the correlation functions can be calculated by using OPE. By using OPE, contributions from quark, gluon and mixed condansates are included in the evaluation of the correlators in Eqs. (\ref{corr0}-\ref{corr0_2}). The evaluations of the correlators with the help of OPE are oftenly called ``the OPE (or QCD) side". 

The physical sides of the correlation functions in Eqs. (\ref{corr0}-\ref{corr0_2}) can be expressed as 

\begin{equation}\label{corr_phen_1}
\Pi ^{(a)\mathrm{Phys}}(q)=\frac{\langle
0|J^{(a)}|X(q)\rangle \langle
X(q)|{J^{(a)}}^{\dag }|0\rangle }{m_{X}^{2}-q^{2}}+...\ ,
\end{equation}
\begin{equation}\label{corr_phen_1x1}
\Pi_{\mu\nu}^{(a)\mathrm{Phys}}(q)=\frac{\langle
0|J_{\mu}^{(a)}|X(q)\rangle \langle
X(q)|{J_{\nu}^{(a)}}^{\dag }|0\rangle }{m_{X}^{2}-q^{2}}+...\ ,
\end{equation}
\begin{equation}\label{corr_phen_1x2}
\Pi_{\mu\nu\alpha\beta}^{(a)\mathrm{Phys}}(q)=\frac{\langle
0|J_{\mu\nu}^{(a)}|X(q)\rangle \langle
X(q)|{J_{\alpha\beta}^{(a)}}^{\dag }|0\rangle }{m_{X}^{2}-q^{2}}+...\ ,
\end{equation}
where $m_{X}$ is the mass of the ground state meson coupling to the chosen current and dots denote the higher resonance contributions which are parameterized via continuum threshold parameter $s_0$. The scalar, axial vector and tensor matrix elements are defined as

\begin{equation}\label{corr2a}
 \langle 0|J^{(a)}|X(q)\rangle = \lambda^{(a)}_{0},
\end{equation}
\begin{equation}\label{corr2b}
\langle 0|J^{(a)}_\mu |X(q)\rangle = \lambda^{(a)}_{1} \varepsilon_{\mu},
\end{equation}
\begin{equation}\label{corr2c}
 \langle 0|J^{(a)}_{\mu\nu}|X(q)\rangle = \lambda^{(a)}_{2} \varepsilon_{\mu\nu},
\end{equation}
where  subscript $0,1,2$ denote the spin of the ground state coupling to chosen current and $ \varepsilon_{\mu} $ and $ \varepsilon_{\mu\nu} $ are vector and tensor polarizations satisfying the following relations
\begin{eqnarray}\label{polarizationt1and2}
\varepsilon_{\mu}\varepsilon_{\nu}^{*}&=&T_{\mu\nu},\\
\varepsilon_{\mu\nu}\varepsilon_{\alpha\beta}^{*}&=&\frac{1}{2}T_{\mu\alpha}T_{\nu\beta}+
\frac{1}{2} T_{\mu\beta}T_{\nu\alpha}-\frac{1}{3}T_{\mu\nu}T_{\alpha\beta},
\end{eqnarray}\vskip0.3cm
where $T_{\mu\nu}=-g_{\mu\nu}+q_\mu q_\nu/m_{X}^2$.

In terms of hadronic states, the correlators of the scalar, axial vector and tensor currents are obtained as

\begin{equation}\label{phenS}
\Pi^{(a)\rm Phys}(q)=\frac{{\lambda^{(a)}_{0}}^2}
{m_{X}^2-q^2}~,
\end{equation}

\begin{equation}\label{phenAV}
\Pi^{(a)\rm Phys}
_{\mu\nu}(q)=\frac{{\lambda^{(a)}_{1}}^2}
{m_{X}^2-q^2}
g_{\mu\nu}+
\mbox{other structures},
\end{equation}
\begin{eqnarray}\label{phenT}
\nonumber\Pi^{(a)\rm Phys}
_{\mu\nu\alpha\beta}(q)=\frac{{\lambda^{(a)}_{2}}^2}
{m_{X}^2-q^2}
\left\{\frac{1}{2}(g_{\mu\alpha}~g_{\nu\beta}+g_{\mu\beta}~g_{\nu\alpha})\right\}\\
+
\mbox{other structures}\ \ , \ \ \
\end{eqnarray}
where, only the Lorentz structures that are used for QCDSR analysis in this work are shown.

On the OPE side of the QCDSR calculations, light and heavy quark fields are contracted and the correlation functions for the currents given in Eqs. (\ref{curr_S_m} - \ref{curr_T_t}) are obtained in terms of full s and c quark propagators as

\onecolumngrid

\begin{eqnarray}
\Pi ^{\mathrm{(1)OPE}}(q)=i\int d^{4}xe^{iq\cdot x} &&\mathrm{Tr}\left[ S_{s}^{i^{\prime
}i}(-x)\gamma _{\mu }S_{c}^{i i^{\prime }}(x)\gamma _{\nu }\right] \mathrm{Tr}\left[ S_{c}^{j^{\prime
}j}(-x)\gamma _{\mu } S_{s}^{jj^{\prime }}(x)\gamma _{\nu }\right] ,  \label{eq:CorrM_1}
\end{eqnarray}
\begin{eqnarray}
\Pi_{\mu\nu} ^{\mathrm{(1)OPE}}(q)=i\int d^{4}xe^{iq\cdot x}  &&  \Big \lbrace \mathrm{Tr}\left[ S_{s}^{i^{\prime
}i}(-x)\gamma _{5}S_{c}^{ii^{\prime }}(x)\gamma _{5}\right]  \mathrm{Tr}\left[S_{c}^{j^{\prime
}j}(-x) \gamma_{\mu }S_{s}^{jj^{\prime }}(x)\gamma_{\nu }\right]\notag \\ &&-\mathrm{Tr}\left[ S_{s}^{i^{\prime
}i}(-x)\gamma _{5}S_{c}^{ii^{\prime }}(x)\gamma _{\nu}\right]  \mathrm{Tr}\left[S_{c}^{j^{\prime
}j}(-x) \gamma_{\mu }S_{s}^{jj^{\prime }}(x)\gamma_{5 }\right]\notag \\ &&-\mathrm{Tr}\left[ S_{s}^{i^{\prime
}i}(-x)\gamma _{\mu}S_{c}^{ii^{\prime }}(x)\gamma _{5}\right]  \mathrm{Tr}\left[S_{c}^{j^{\prime
}j}(-x) \gamma_{5 }S_{s}^{jj^{\prime }}(x)\gamma_{\nu }\right] \notag \\ &&+\mathrm{Tr}\left[ S_{s}^{i^{\prime
}i}(-x)\gamma _{\mu}S_{c}^{ii^{\prime }}(x)\gamma _{\nu}\right]  \mathrm{Tr}\left[S_{c}^{j^{\prime
}j}(-x) \gamma_{5 }S_{s}^{jj^{\prime }}(x)\gamma_{5 }\right]\Big \rbrace,  \label{eq:CorrM_2}
\end{eqnarray}
\begin{eqnarray}
\Pi_{\mu\nu\alpha\beta} ^{\mathrm{(1)OPE}}(q)=i\int d^{4}xe^{iq\cdot x} \notag &&  \frac{1}{2} \Big \lbrace \mathrm{Tr}\left[ S_{s}^{i^{\prime
}i}(-x)\gamma _{\mu}S_{c}^{ii^{\prime }}(x)\gamma _{\alpha}\right]  \mathrm{Tr}\left[S_{c}^{j^{\prime
}j}(-x) \gamma _{\nu }S_{s}^{jj^{\prime }}(x)\gamma _{\beta }\right]\notag \\ &&+\mathrm{Tr}\left[ S_{s}^{i^{\prime
}i}(-x)\gamma _{\mu}S_{c}^{ii^{\prime }}(x)\gamma _{\beta}\right]  \mathrm{Tr}\left[S_{c}^{j^{\prime
}j}(-x) \gamma _{\nu }S_{s}^{jj^{\prime }}(x)\gamma _{\alpha }\right]\notag \\ &&+\mathrm{Tr}\left[ S_{s}^{i^{\prime
}i}(-x)\gamma _{\nu}S_{c}^{ii^{\prime }}(x)\gamma _{\alpha}\right]  \mathrm{Tr}\left[S_{c}^{j^{\prime
}j}(-x) \gamma _{\mu }S_{s}^{jj^{\prime }}(x)\gamma _{\beta }\right] \notag \\ &&+\mathrm{Tr}\left[ S_{s}^{i^{\prime
}i}(-x)\gamma _{\nu}S_{c}^{ii^{\prime }}(x)\gamma _{\beta}\right]  \mathrm{Tr}\left[S_{c}^{j^{\prime
}j}(-x) \gamma _{\mu }S_{s}^{jj^{\prime }}(x)\gamma _{\alpha }\right] \Big \rbrace ,   \label{eq:CorrM_3}
\end{eqnarray}
\begin{eqnarray}
\Pi ^{\mathrm{(2)OPE}}(q)&=&i\int d^{4}xe^{iq\cdot x} \varepsilon ^{ijk}\varepsilon
^{imn}\varepsilon ^{i^{\prime }j^{\prime }k^{\prime }}\varepsilon
^{i^{\prime }m^{\prime }n^{\prime }} \notag \times\\   && \mathrm{Tr}\left[ \gamma _{\nu }\widetilde{S}_{s}^{jj^{\prime
}}(x)\gamma _{\mu }S_{c}^{kk^{\prime }}(x)\right] \mathrm{Tr}\left[ \gamma _{\mu }\widetilde{S}_{c}^{n^{\prime
}n}(-x) \gamma _{\nu }S_{s}^{m^{\prime }m}(-x)\right] ,  \label{eq:CorrT_1}
\end{eqnarray}
\begin{eqnarray}
\Pi_{\mu\nu} ^{\mathrm{(2)OPE}}(q)&=&i\int d^{4}xe^{iq\cdot x}\frac{\varepsilon ^{ijk}\varepsilon
^{imn}\varepsilon ^{i^{\prime }j^{\prime }k^{\prime }}\varepsilon
^{i^{\prime }m^{\prime }n^{\prime }}}{2} \notag \times\\ && \Big \lbrace \mathrm{Tr}\left[ \gamma _{5}\widetilde{S}_{s}^{jj^{\prime
}}(x)\gamma _{5}S_{c}^{kk^{\prime }}(x)\right]  \mathrm{Tr}\left[ \gamma _{\mu}\widetilde{S}_{c}^{n^{\prime
}n}(-x) \gamma _{\nu }S_{s}^{m^{\prime }m}(-x)\right]\notag \\ &-&\mathrm{Tr}\left[\gamma _{\nu} \widetilde{S}_{s}^{jj^{\prime
}}(x)\gamma _{5}S_{c}^{kk^{\prime }}(x)\right]  \mathrm{Tr}\left[ \gamma _{\mu}\widetilde{S}_{c}^{n^{\prime
}n}(-x) \gamma _{5}S_{s}^{m^{\prime }m}(-x)\right]\notag \\ &-&\mathrm{Tr}\left[ \gamma _{5 }\widetilde{S}_{s}^{jj^{\prime
}}(x)\gamma _{\mu}S_{c}^{kk^{\prime }}(x)\right] \mathrm{Tr}\left[\gamma _{5}\widetilde{S}_{c}^{n^{\prime
}n}(-x) \gamma _{\nu }S_{s}^{m^{\prime }m}(-x)\right] \notag \\ &+&\mathrm{Tr}\left[ \gamma _{\nu}\widetilde{S}_{s}^{jj^{\prime
}}(x)\gamma _{\mu}S_{c}^{kk^{\prime }}(x)\right]  \mathrm{Tr}\left[\gamma _{5}\widetilde{S}_{c}^{n^{\prime
}n}(-x) \gamma _{5 }S_{s}^{m^{\prime }m}(-x)\right] \Big \rbrace ,  \label{eq:CorrT_2}
\end{eqnarray}
%
%
\begin{eqnarray}
\Pi_{\mu\nu\alpha\beta} ^{\mathrm{(2)OPE}}(q)&=&i\int d^{4}xe^{iq\cdot x}\frac{\varepsilon ^{ijk}\varepsilon
^{imn}\varepsilon ^{i^{\prime }j^{\prime }k^{\prime }}\varepsilon
^{i^{\prime }m^{\prime }n^{\prime }}}{2} \notag \times\\ & & \Big \lbrace \mathrm{Tr}\left[ \gamma _{\beta}\widetilde{S}_{s}^{jj^{\prime
}}(x)\gamma _{\mu}S_{c}^{kk^{\prime }}(x)\right]  \mathrm{Tr}\left[ \gamma _{\nu}\widetilde{S}_{c}^{n^{\prime
}n}(-x) \gamma _{\alpha }S_{s}^{m^{\prime }m}(-x)\right]\notag \\ &+&\mathrm{Tr}\left[\gamma _{\alpha} \widetilde{S}_{s}^{jj^{\prime
}}(x)\gamma _{\mu}S_{c}^{kk^{\prime }}(x)\right]  \mathrm{Tr}\left[ \gamma _{\nu}\widetilde{S}_{c}^{n^{\prime
}n}(-x) \gamma _{\beta }S_{s}^{m^{\prime }m}(-x)\right]\notag \\ &+&\mathrm{Tr}\left[ \gamma _{\beta }\widetilde{S}_{s}^{jj^{\prime
}}(x)\gamma _{\nu}S_{c}^{kk^{\prime }}(x)\right]  \mathrm{Tr}\left[\gamma _{\mu}\widetilde{S}_{c}^{n^{\prime
}n}(-x) \gamma _{\alpha }S_{s}^{m^{\prime }m}(-x)\right] \notag \\ &+&\mathrm{Tr}\left[ \gamma _{\alpha}\widetilde{S}_{s}^{jj^{\prime
}}(x)\gamma _{\nu}S_{c}^{kk^{\prime }}(x)\right]  \mathrm{Tr}\left[\gamma _{\mu}\widetilde{S}_{c}^{n^{\prime
}n}(-x) \gamma _{\beta }S_{s}^{m^{\prime }m}(-x)\right] \Big \rbrace , \label{eq:CorrT_3}
\end{eqnarray}

\twocolumngrid
where $ \widetilde{S}_{q}^{ij}(x)=CS_{q}^{ijT}(x)C $, and $S_{s}^{ij}(x)$,  $S_{c}^{ij}(x)$ are the full propagators of s and c quarks respectively.  For the s quark, we chose the light quark propagator in the coordinate space which is in the form 
 \begin{eqnarray}\label{lightquarkpropagator}
S_{s}^{ij}(x)&=& i\frac{\!\not\!{x}}{
2\pi^2 x^4}\delta_{ij}
-\frac{m_s}{4\pi^2 x^2}\delta_{ij}-\frac{\langle
\bar{s}s\rangle}{12}\Big[1-i\frac{m_s}{4} \!\not\!{x} \Big]\delta_{ij} \nn\\
&&-\frac{x^{2}}{192} m_{0}^{2}
\langle
\bar{s}s\rangle\Big[1-i\frac{m_s}{6} \!\not\!{x} \Big]\delta_{ij}
\nonumber\\
&&-\frac{ig_s G^{\alpha\beta}_{i j}}{32\pi^{2} x^{2}}
\Big(\!\not\!{x} \sigma^{\alpha\beta}+\sigma^{\alpha\beta} \!\not\!{x} \Big)\delta_{ij}-i\frac{x^2 \!\not\!{x} g_{s}^{2} \langle
\bar{s}s\rangle^{2}}{7776}\delta_{ij}\nn\\
&&-\frac{x^4 \langle
\bar{s}s\rangle \langle
g_{s}^{2} G^2\rangle}{27648}+....
\end{eqnarray}
For the c quark, we used the following heavy quark propagator
\begin{eqnarray}\label{heavypropagator}
S_{c}^{ij}(x)&=&i \int\frac{d^4k e^{-ik \cdot x}}{(2\pi)^4} 
\left( \frac{\!\not\!{k}+m_c}{k^2-m_c^2}\delta_{ij}\right.\nn\\
&&-\frac{g_{s}G^{\alpha\beta}_{i j}}{4} \frac{\sigma^{\alpha\beta}(\!\not\!{k}+m_c)+(\!\not\!{k}+m_c)\sigma^{\alpha\beta} }{(k^2-m_c^2)^{2}} 
\nonumber\\
&&+\frac{g_{s}^{2} m_{c}}{12}\frac{k^{2}+m_{c}\!\not\!{k}}{(k^{2}-m_{c}^{2})^{4}}G^2\delta_{ij}\nn \\&&+\left.\cdots\ \right)\ \ , 
\end{eqnarray}  
given in Ref. \cite{Prop_C}, where $G_{\alpha\beta}=G^{A}_{\alpha\beta} t^{A}$, $G^2=G^{A}_{\alpha\beta} G^{A\alpha\beta}$ and $t^{A}=\lambda^{A}/2$ are the Gell-Mann matrices with $A=1,..,8$.  Similar to physical side, the correlation functions given in Eqs. (\ref{eq:CorrM_1}-\ref{eq:CorrT_3}) on the OPE side are also expanded in terms of Lorentz structures as
\begin{equation}\label{OPE_S}
\Pi^{(a)\rm OPE}(q)=\Gamma_0^{(a)} \mathbb{1},
\end{equation}
\begin{equation}\label{OPE_AV}
\Pi^{(a)\rm OPE}
_{\mu\nu}(q)=\Gamma_1^{(a)}
g_{\mu\nu}+
\mbox{other structures},
\end{equation}
\begin{eqnarray}\label{OPE_T}
\nn \Pi^{(a)\rm OPE}
_{\mu\nu\alpha\beta}(q)=\Gamma_2^{(a)}
\left\{\frac{1}{2}(g_{\mu\alpha}~g_{\nu\beta}+g_{\mu\beta}~g_{\nu\alpha})\right\}\\
+\ \ \mbox{other structures}\ \ ,\ \ \
\end{eqnarray}
where $\Gamma_J^{(a)}$ are the coefficients of the Lorentz structures that are selected in this work. A dispersion integral of the form
\begin{equation}
\label{disp_0} \Gamma^{(a)}_J (q^2)=\int_{s_{min}}^\infty ds \dfrac{\rho^{(a){\rm OPE}}_J (s)}{s-q^2},
\end{equation}
can be written for the selected coefficients, where $\rho^{(a)\rm OPE}_J=\mathrm{Im}[\Gamma^{(a)}_J/\pi]$ are the spectral densities, and $J$ is the total angular momentum of the state. According to quark hadron duality ansatz in QCDSR, it is assumed that the spectral density obtained from the continuum of the states given in Eq. (\ref{corr_phen_1}) in the physical side are related to the spectral density obtained from the OPE side via relation
 \begin{equation}
 \label{ansazt_1}\rho^{\text{cont}}(s)=\rho^{\text{OPE}}\Theta(s-s_0),
 \end{equation}
which enables us to isolate the ground state hadron from the infinite sum\cite{QCDSR1,QCDSR2,QCDSR3}. To improve the duality of the correlators, Borel transformation with respect to $q^2$ is applied. After applying these steps of traditional QCDSR analysis, the sum rules for the currents given in Eqs. (\ref{curr_S_m} - \ref{curr_T_t}) are obtained as
\begin{equation}
\label{SR_res}{\lambda^{(a)}_J}^2 e^{-m_X^2/M^2}=\int_{(2m_s+2m_c)^2}^{s_0} ds \rho^{(a)\rm OPE}_J (s) e^{-s/M^2}\ .
\end{equation}
In order to estimate the mass of the ground state hadron, one takes the derivative of Eq. (\ref{SR_res}) with respect to $1/M^2$ and divides it to Eq. (\ref{SR_res}) and obtains the mass of the ground state as
\begin{equation}\label{SR_mass}
m_{X}^{2}=\frac{\int_{(2m_{s}+2m_{c})^{2}}^{s_{0}}ds~s~
\rho ^{(a)\rm OPE}_J(s)e^{-s/M^{2}}}{\int_{(2m_{s}+2m_{c})^{2}}^{s_{0}}ds\rho
^{(a)\rm OPE}_J(s)e^{-s/M^{2}}}.  
\end{equation}
The expressions of the spectral densities obtained in this work are given in Appendix A. 

\section{Numerical Analysis}\label{S3}
The sum rules obtained in previous section depend on the values of the parameters such as quark, gluon and mixed condansates, and on the masses of c and s quarks. Values of these parameters are given in Table \ref{Tab::Param}. The c and s quark masses are chosen in the $\overline{\text{MS}}$ scheme at the scale $\mu=m_c$ and $\mu=2$GeV respectively, and their values are taken from the Particle Data Group \cite{PDG}, and the values of the condansates are taken from Ref. \cite{Nielsen}
\begin{table}[tbh]
\begin{center}
\caption{Input parameters}
\label{Tab::Param}
\begin{tabular}{lclcl}
\hline\hline
Parameters & Values \\ \hline\hline
$m_{c}$ & $(1.275\pm0.025)~\mathrm{GeV}$ \\
$m_{s} $ & $(95 \pm 5)~\mathrm{MeV} $ \\
$\langle \bar{q}q \rangle $ & $(-0.24\pm 0.01)^3 $ $\mathrm{GeV}^3$ \\
$\langle \bar{s}s \rangle $ & $0.8\ \langle \bar{q}q \rangle$ \\
$\langle\frac{\alpha_sG^2}{\pi}\rangle $ & $(0.012\pm0.004)$ $~\mathrm{GeV}%
^4 $ \\
$ m_{0}^2 $ & $(0.8\pm0.1)$ $\mathrm{GeV}^2$ \\ 
\hline\hline
\end{tabular}
\end{center}%
\end{table}
The expressions of the masses and meson coupling constants given in Eqs. (\ref{SR_res}) and (\ref{SR_mass}) depend also on the values of the continuum threshold ($s_0$) and the Borel Mass ($M^2$), which are parameters of the theory. In general, $s_0$ is related to the mass of the state under investigation as $(m_X+0.3)^2 GeV^2\leq s_0 \leq (m_X+0.5)^2 GeV^2$. In the present case, this restricts $s_0$ as $19.7 GeV^2\leq s_0 \leq 21.5 GeV^2$. In order to have reliable sum rules, $s_0$ and $M^2$ should jointly satisfy the pole dominance and OPE convergence requirements. In addition to these criteria, one has to choose working regions for these parameters in which the dependence of the obtained results for the masses and meson coupling constants should be minimal.  

\begin{figure}[ht]
\begin{center}
\includegraphics[width=8cm]{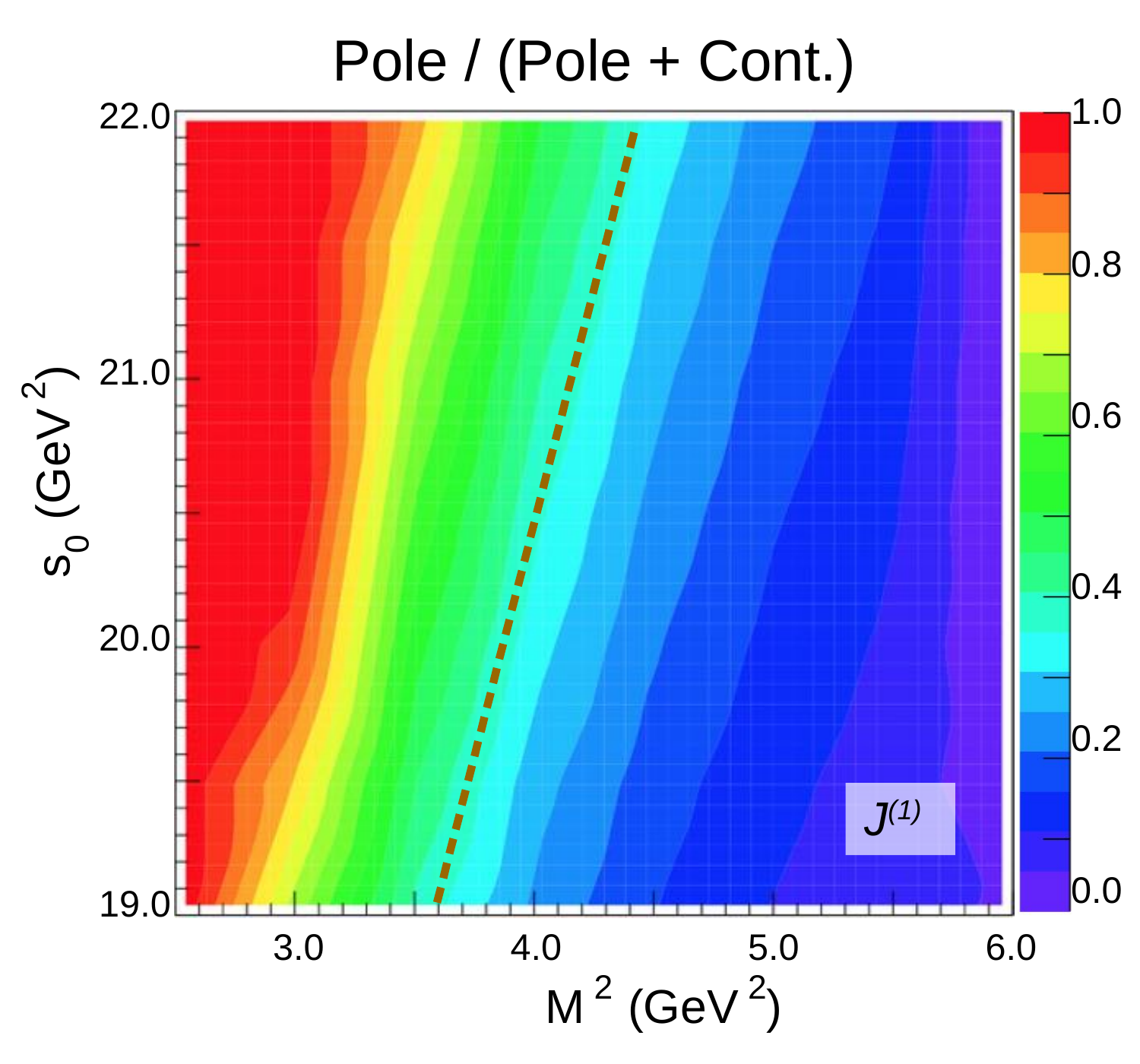}~~~~~~~~\includegraphics[width=8cm]{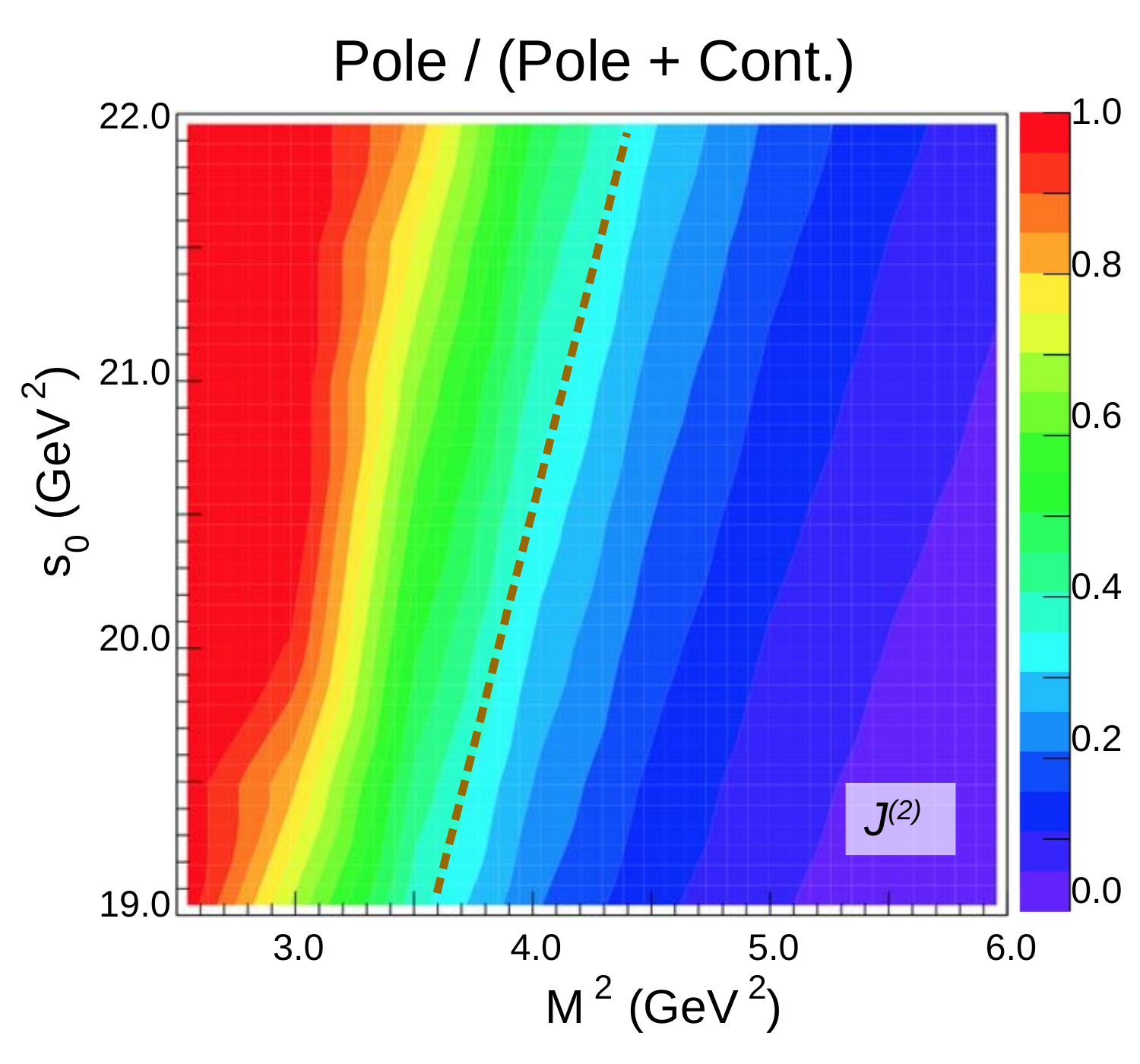}\\
\includegraphics[width=8cm]{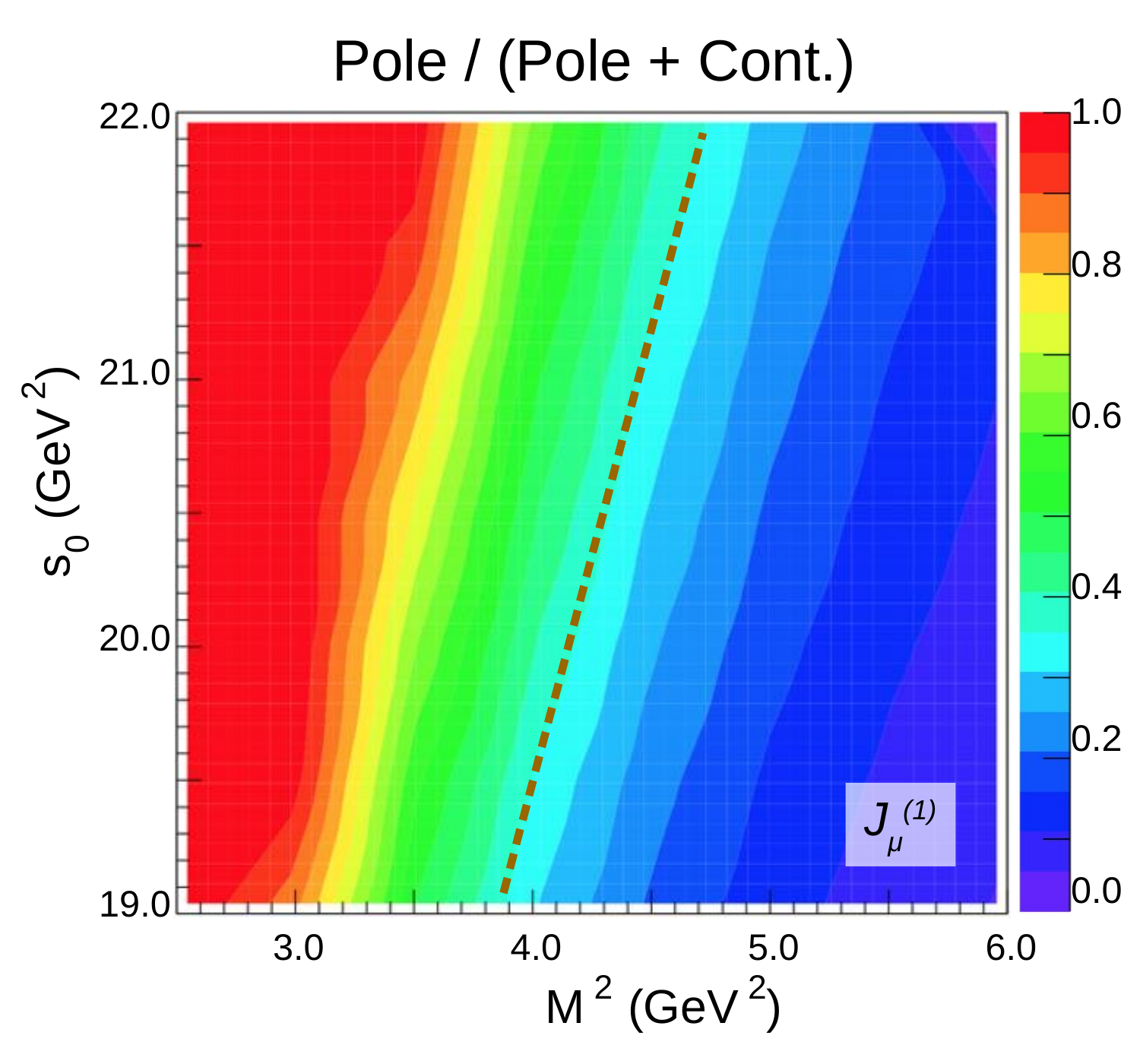}~~~~~~~~\includegraphics[width=8cm]{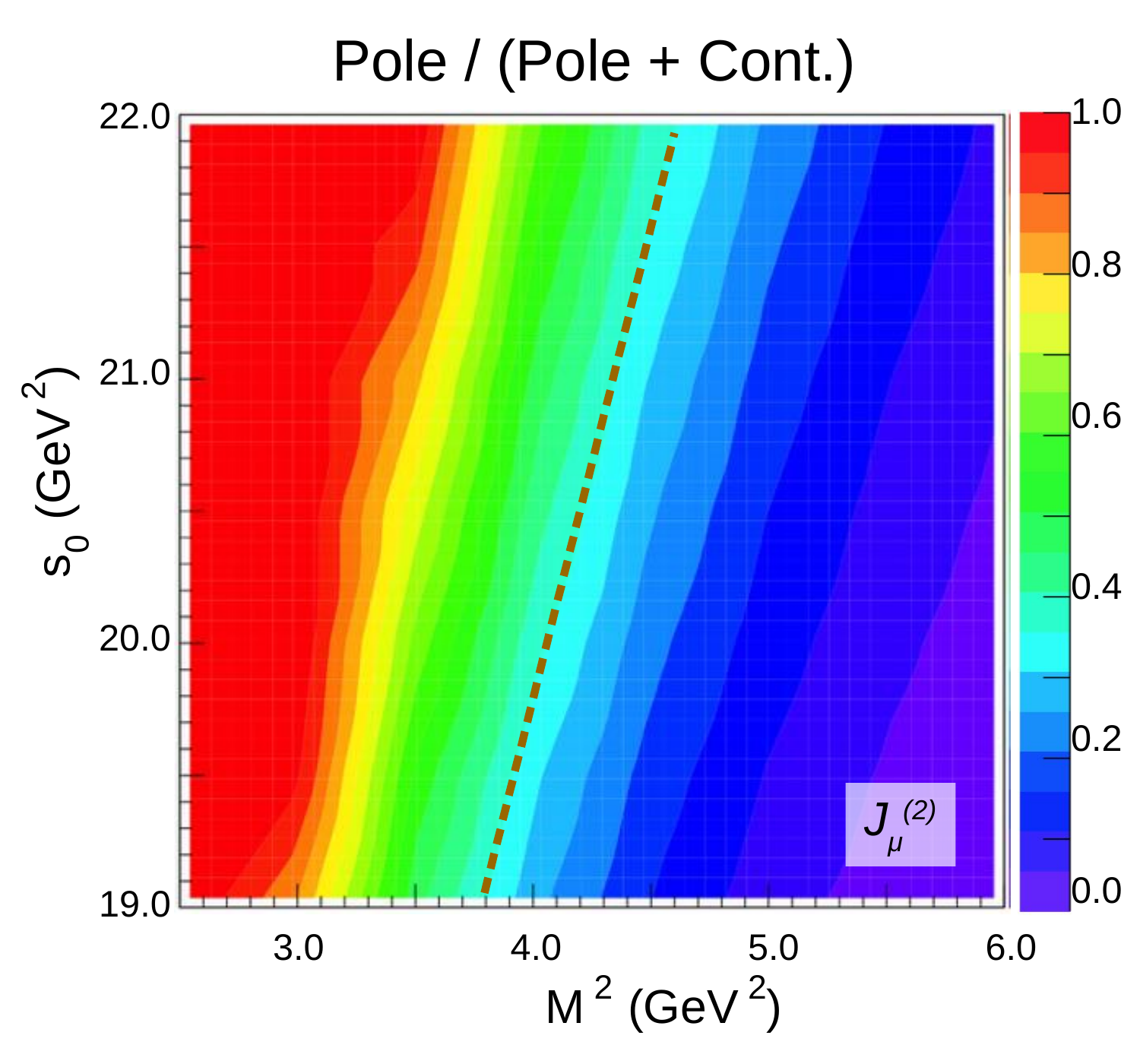}\\
\includegraphics[width=8cm]{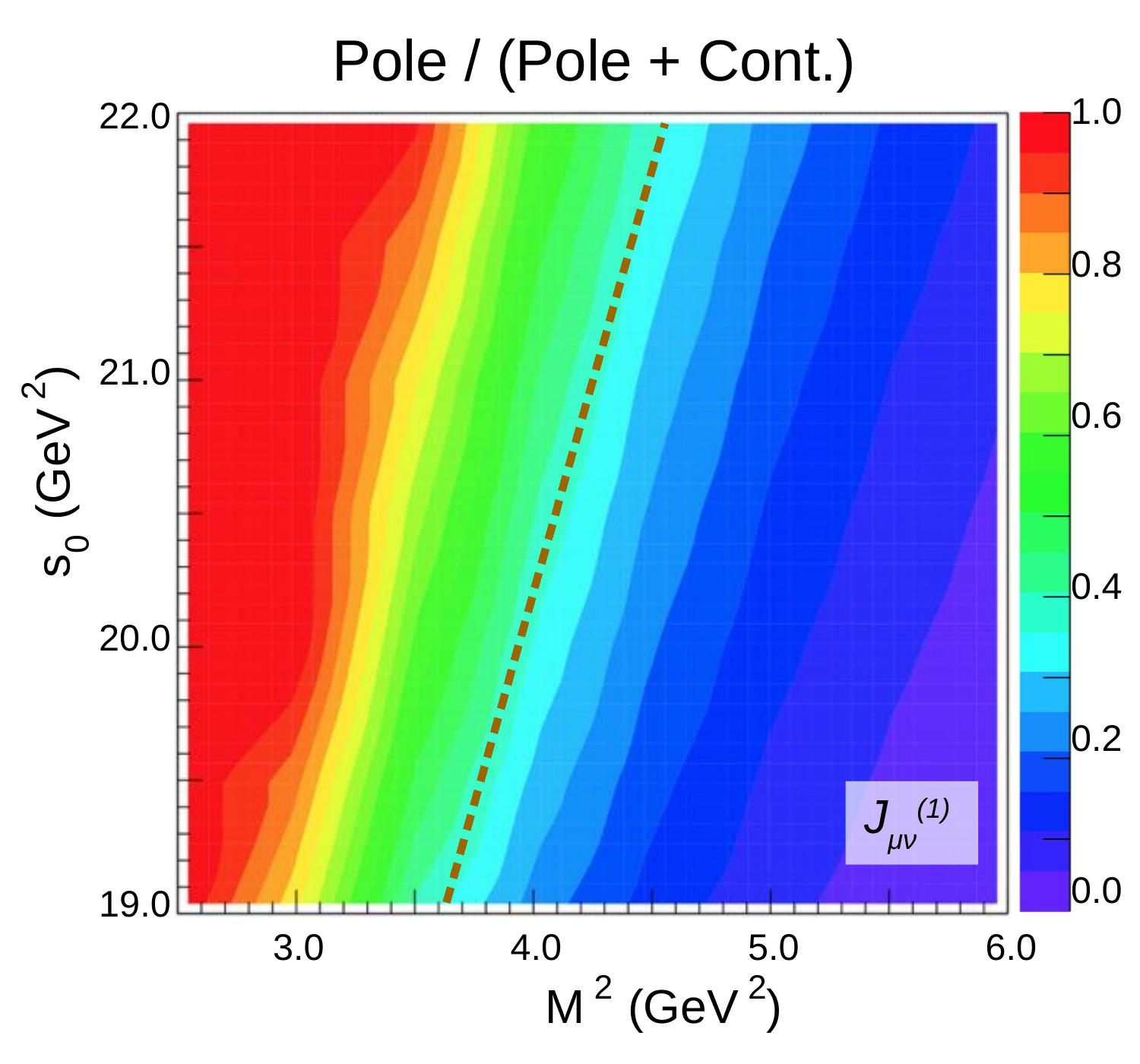}~~~~~~~~\includegraphics[width=8cm]{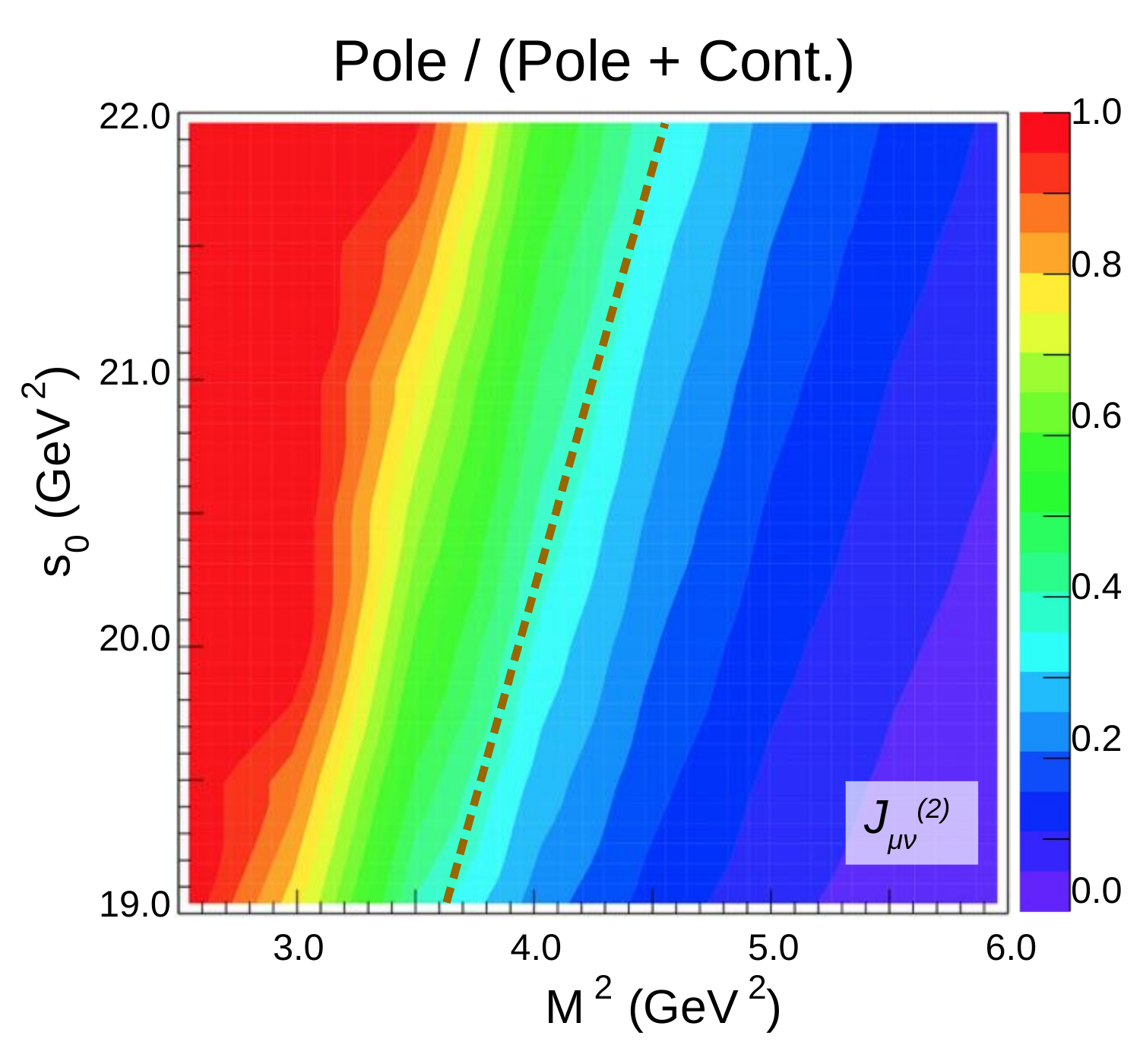}
 \vskip2mm \caption{\label{fig::PC}Pole dominance of the sum rules obtained in this work. The variation of the ratio of the pole to pole plus continuum with respect to $s_0$ and $M^2$, for  the scalar, axial vector and tensor molecular currents (left panel) and for the scalar, axial vector and tensor diquark-antidiquark currents (right panel). For each plot, on the left of the dashed line, pole dominance is achieved.}
\end{center}
\end{figure}

In QCDSR, the contribution coming from the pole of the ground state required to be greater than the contribution of the continuum. To analyze the pole dominance of the obtained sum rules, we plot the two parameter heat graphs of the ratio $\Pi(s_0, M^2) / \Pi(\infty, M^2)$ with respect to  $s_0$ and $M^2$ which are given in Fig. \ref{fig::PC}. Even though the sum rules obtained for exotic hadrons often requires more relaxed constraints \cite{Agaev,Nielsen2}, in this work we chose the regions in which the aforementioned ratio is greater than $40\%$, which are on the left of the dashed lines plotted in Fig. \ref{fig::PC}.

\begin{figure}[h]
\begin{center}
\includegraphics[width=8.5cm]{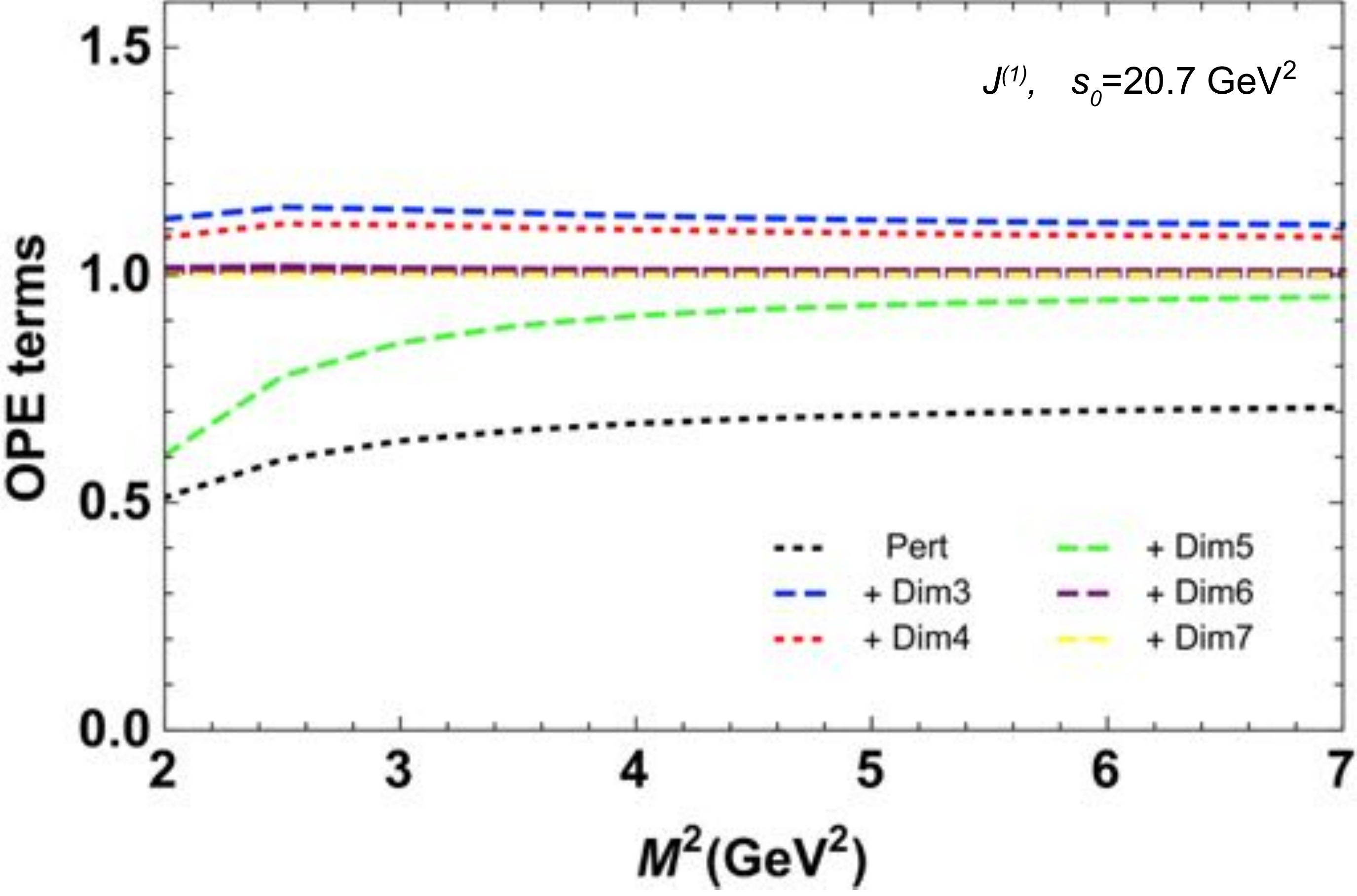}~~~~~~~~\includegraphics[width=8.5cm]{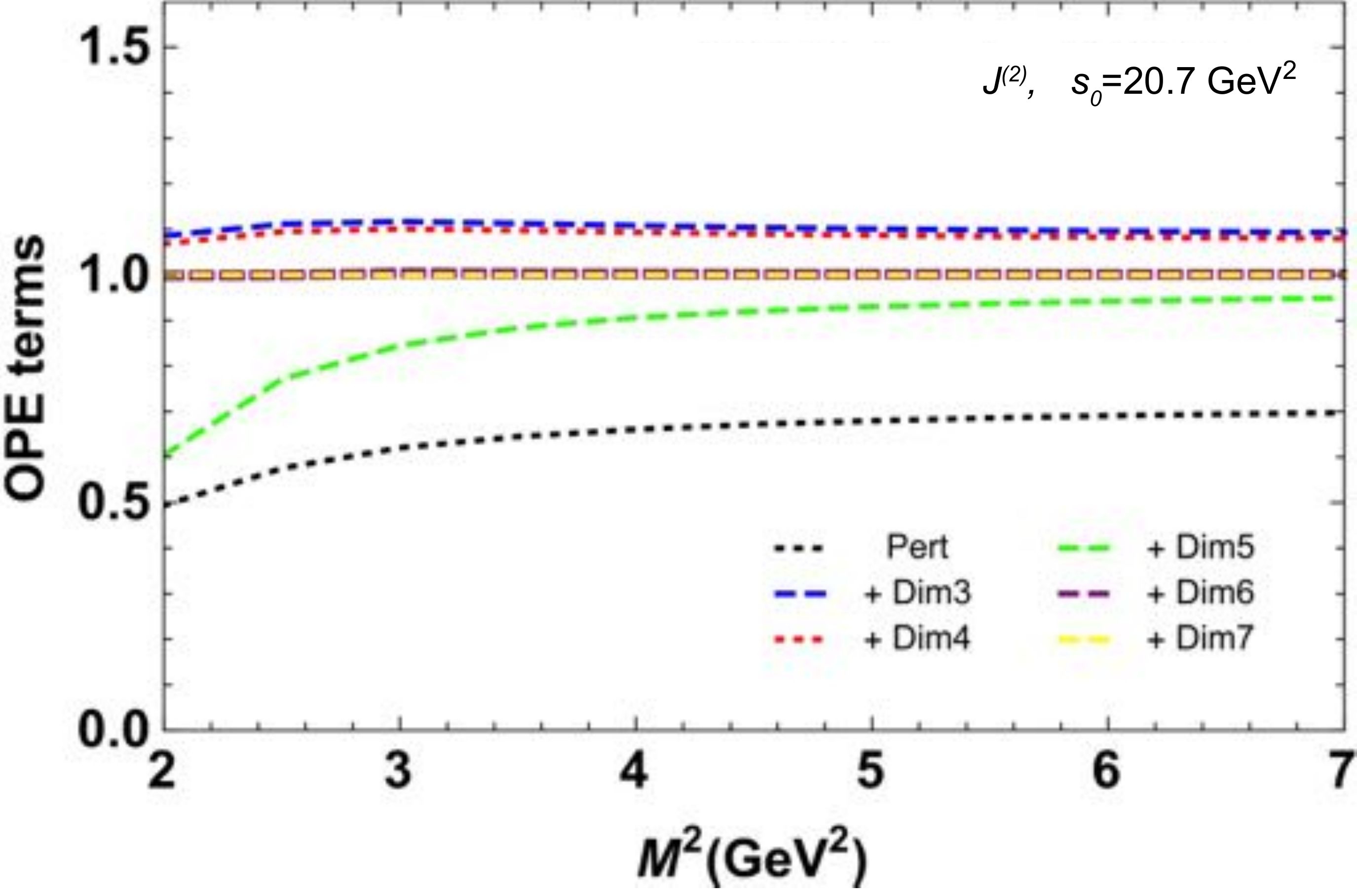}\\
\vskip1.2cm
\includegraphics[width=8.5cm]{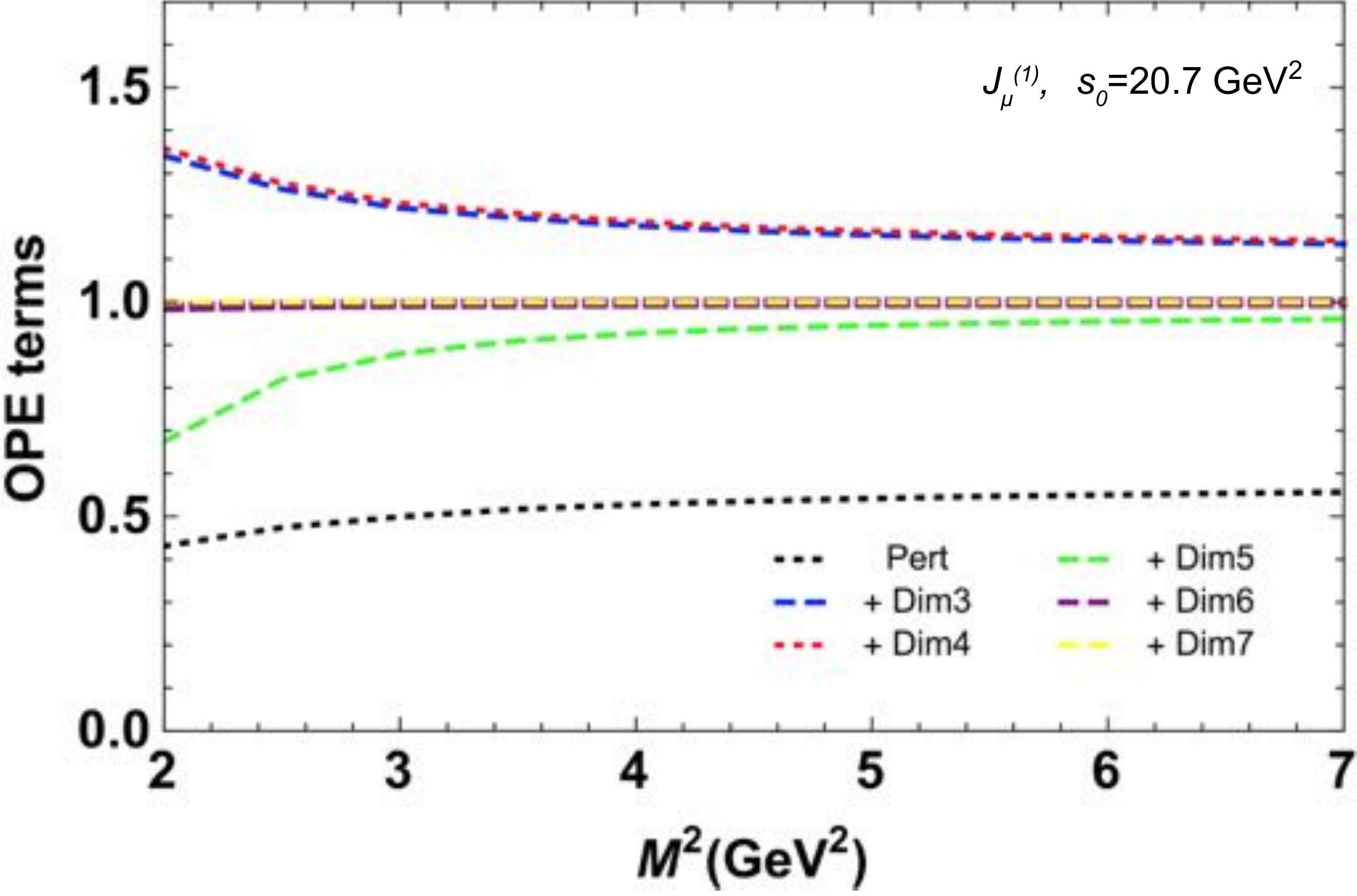}~~~~~~~~\includegraphics[width=8.5cm]{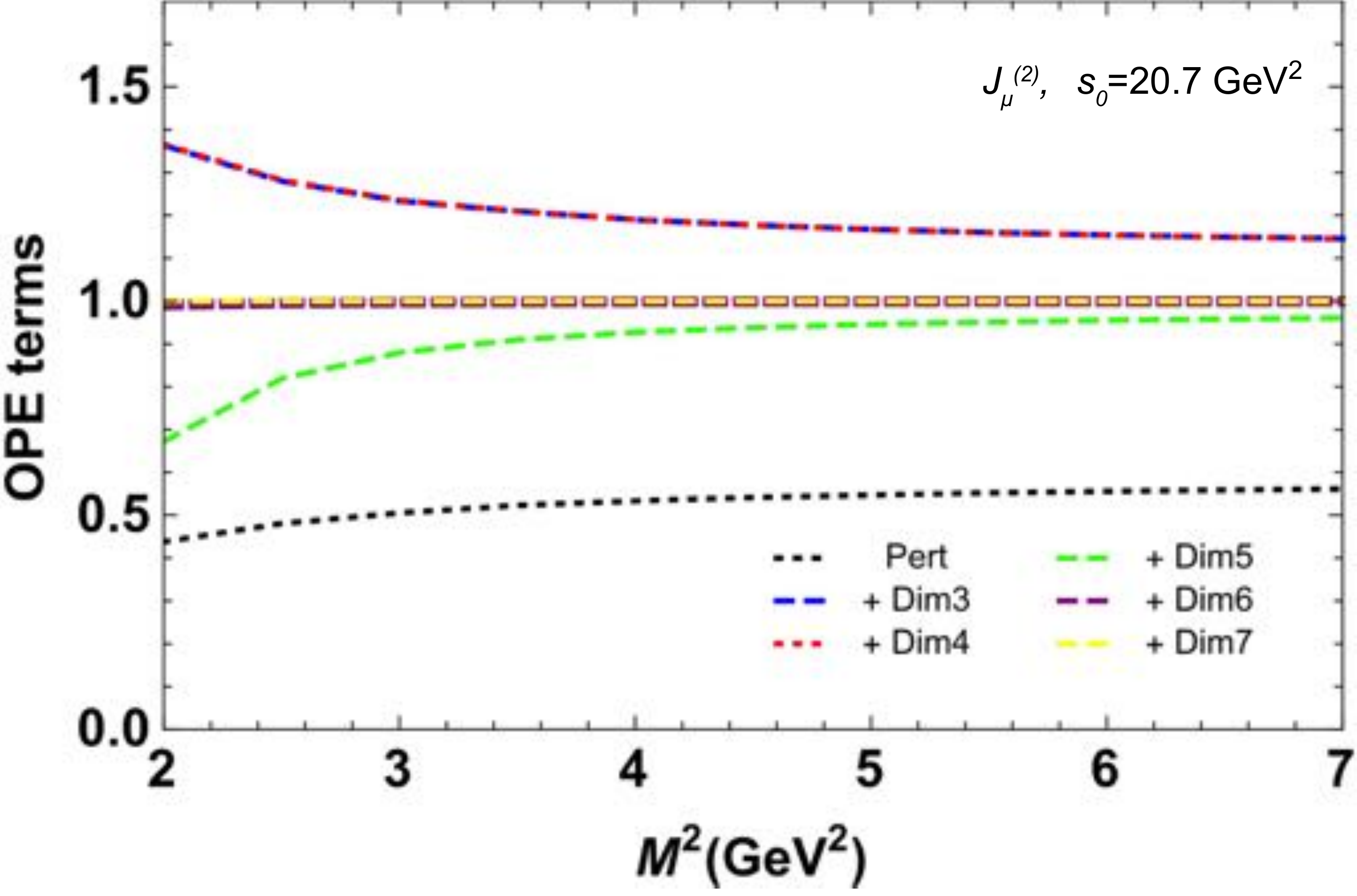}\\
\vskip1.2cm
\includegraphics[width=8.5cm]{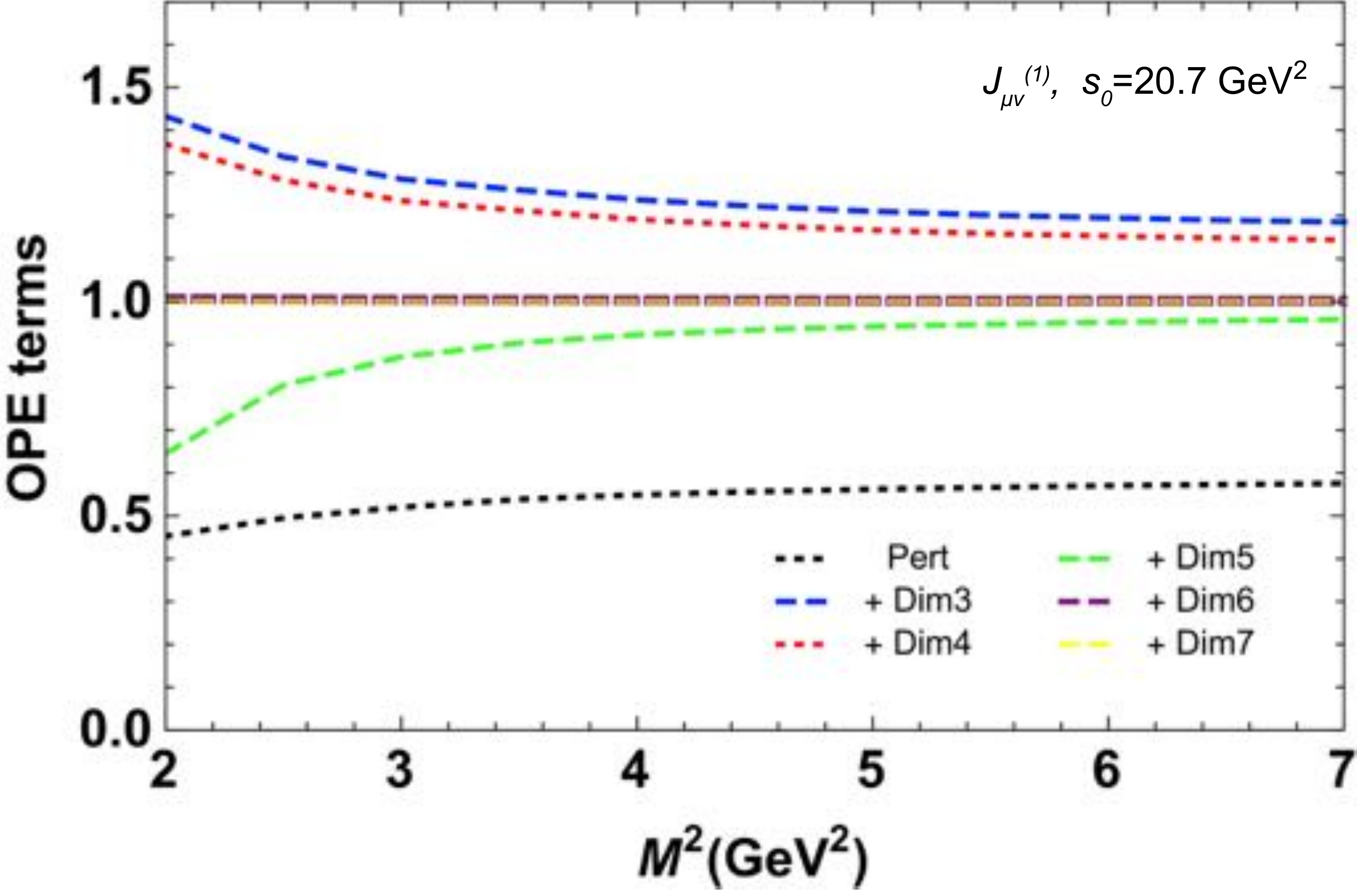}~~~~~~~~\includegraphics[width=8.5cm]{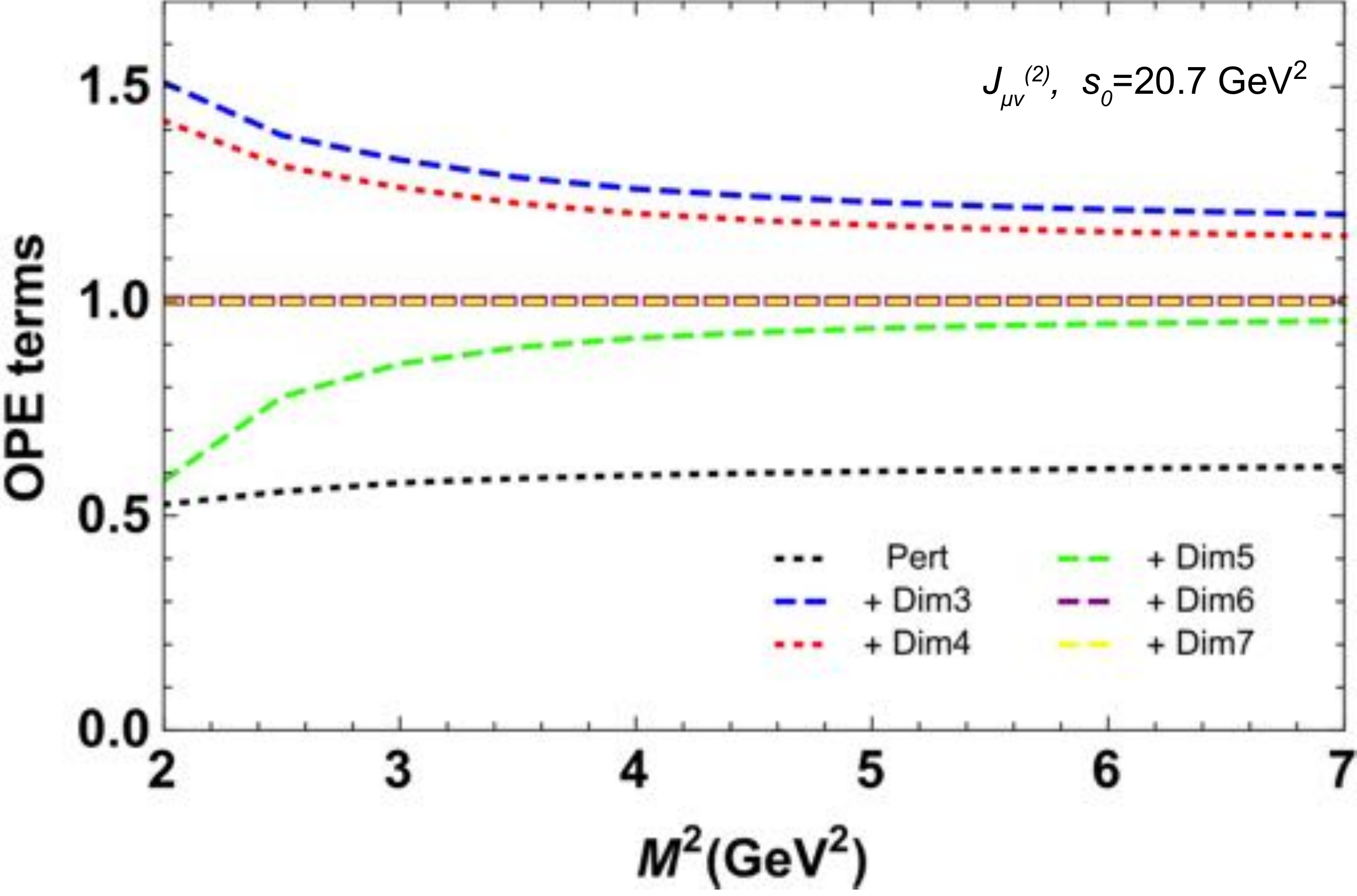}
 \vskip1.2cm \caption{\label{fig::OPE}The OPE convergence of the sum rules obtained in this work. The ratio of the sum of the terms up to specified dimension, to correlation function is plotted with respect to $M^2$ for the scalar, axial vector and tensor molecular currents (left panel) and for the scalar, axial vector and tensor diquark-antidiquark currents (right panel).}
\end{center}
\end{figure}

The OPE convergence of the obtained sum rules is analyzed as follows. In Fig. \ref{fig::OPE}, we plot the ratio of the sum of the terms with dimensions up to the specified term, to the correlator. It is seen from Fig. \ref{fig::OPE} that the correlators of the currents given in Eqs. (\ref{curr_S_m} - \ref{curr_T_t}) satisfy the following conditions.
\begin{itemize}
\item The contribution of the perturbative terms are greater than $50\%$.
\item The contribution of the terms with dimension five are smaller than $25\%$ for $M^2\geq 2.5 GeV^2$, and reduces for further values of $M^2$.
\item The contribution of the terms with dimension six to eight, either converge to zero or obtained as zero.
\end{itemize}
Thus for $M^2 \geq 2.5 GeV^2$, a good OPE convergence is ensured. The working regions of the sum rules obtained in this work are determined by combining the analysis up to this step, and they are given in Table \ref{Tab::WR}.

\begin{table}[hb]
\begin{center}
\caption{Working regions of the sum rules. }
\label{Tab::WR}       
\begin{tabular}{lclclclc|}
\hline\hline  Current & $ M_{\rm min}^2$  &$ M_{\rm max}^2$  &$21.5\geq s_0\geq$ \\
 & (GeV$^2$) &(GeV$^2$)&(GeV$^2$)  \\
\hline\hline $ J^{(1)} $  & 3.69 & 4.43  &$ 3.75 M^2 + 5.88 $ \\
\hline$ J^{(1)}_{\mu}  $ & 3.99&4.74&$ 3.66 M^2 + 5.10$ \\
\hline$ J^{(1)}_{\mu\nu}  $ & 3.95&4.74 &$3.53 M^2 + 5.76$  \\
\hline\hline$J^{(2)} $  & 3.65& 4.38 &$ 3.85 M^2 + 5.65$  \\
\hline$ J^{(2)}_{\mu} $  &3.89&4.63 &$ 3.75 M^2 + 5.13$  \\
\hline$ J^{(2)}_{\mu\nu}$& 3.73& 4.57 &$3.33 M^2 + 7.27$ \\
\hline\hline
\end{tabular}
\end{center}
\end{table}

\begin{figure}[h]
\begin{center}
\includegraphics[width=8.5cm]{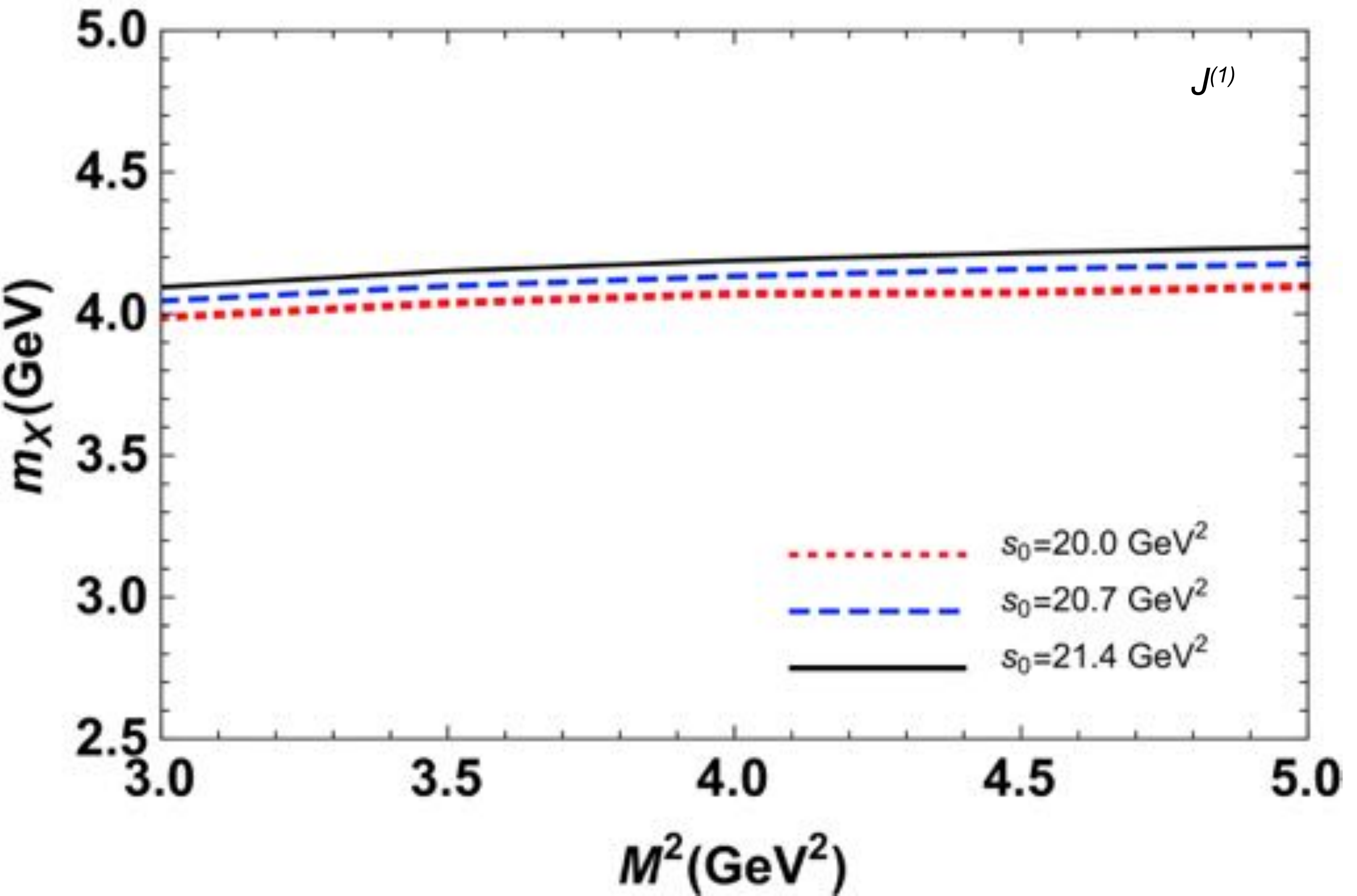}~~~~~~~~\includegraphics[width=8.5cm]{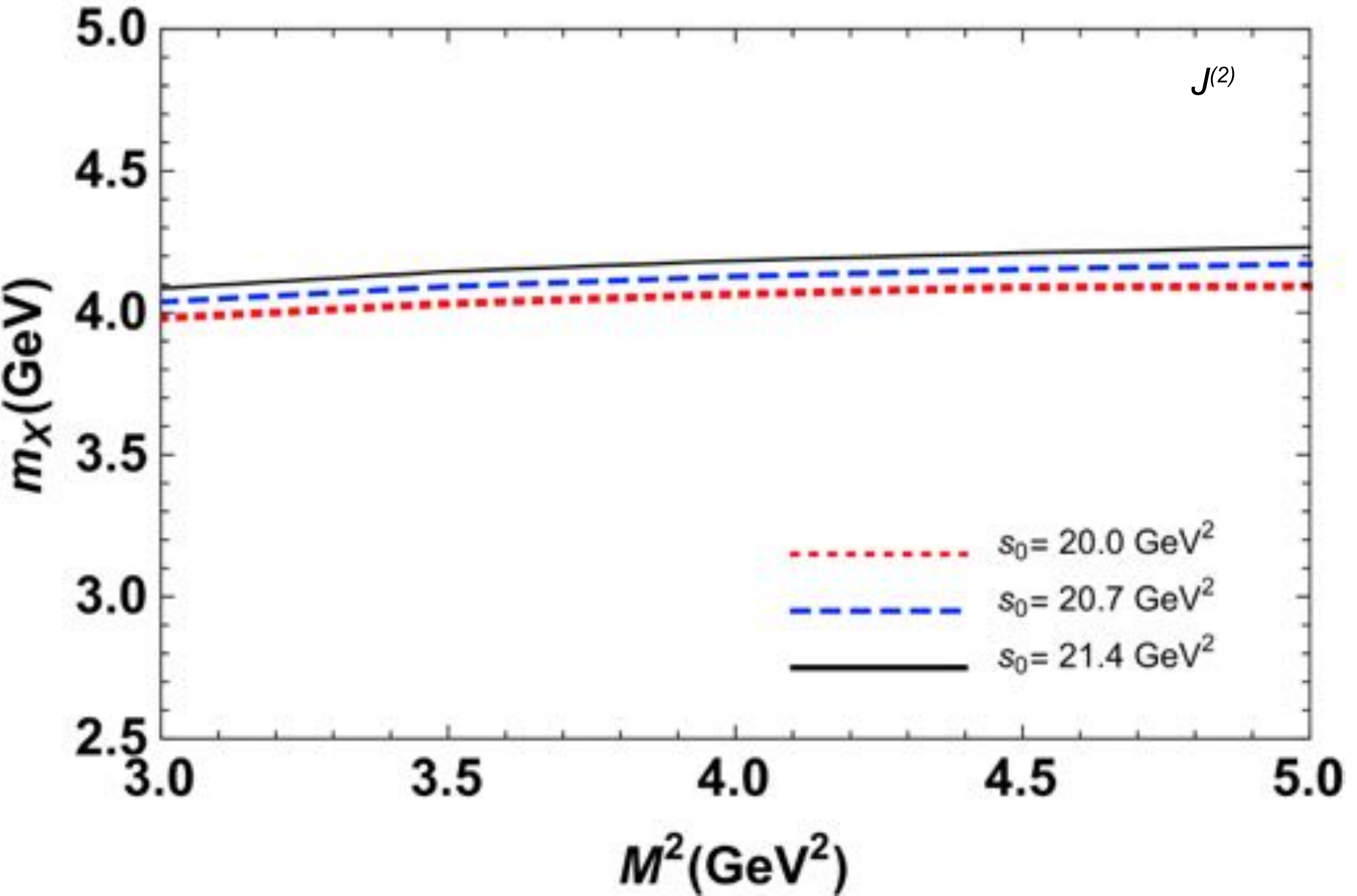}\\
\vskip1.2cm
\includegraphics[width=8.5cm]{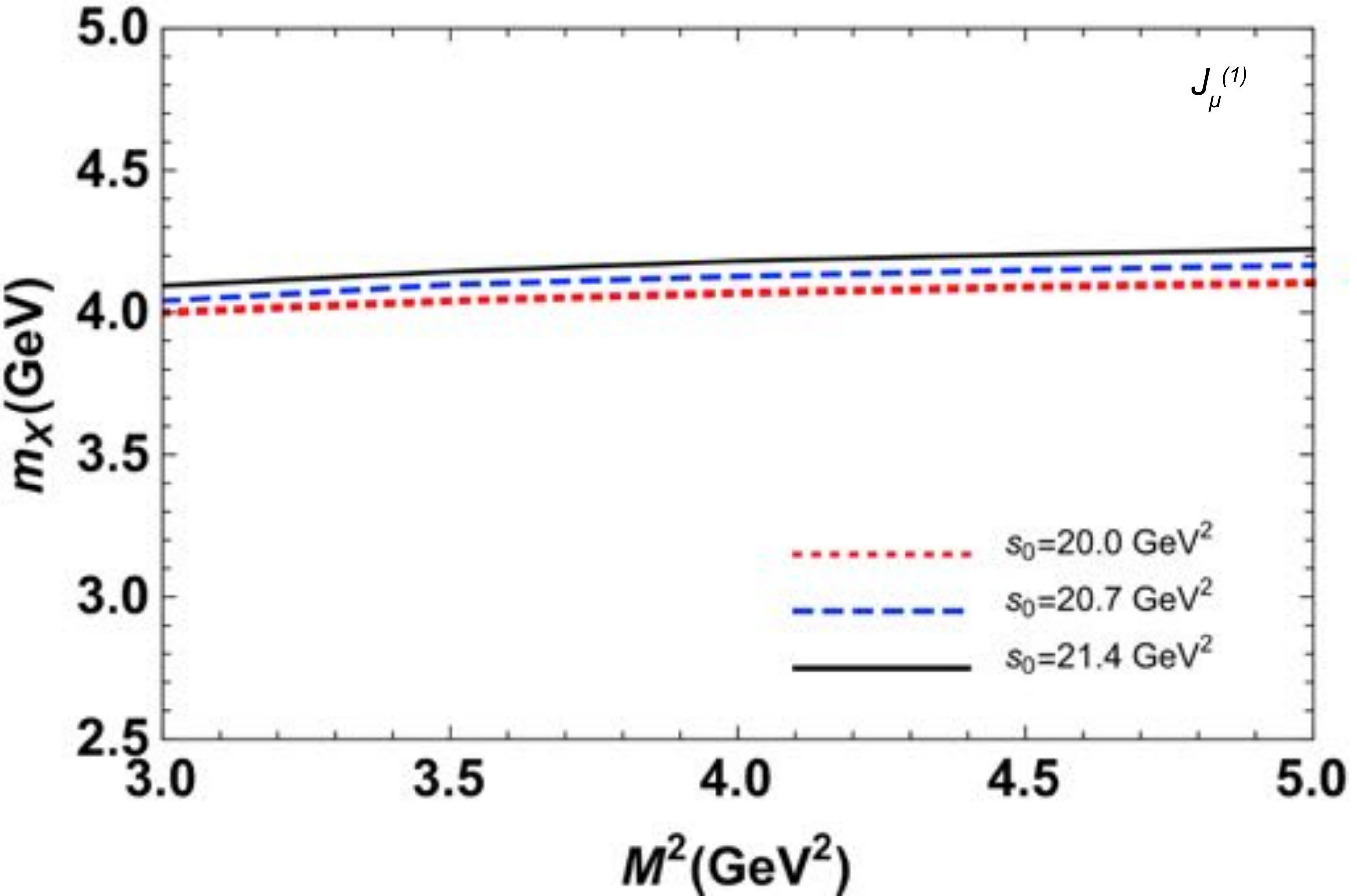}~~~~~~~~\includegraphics[width=8.5cm]{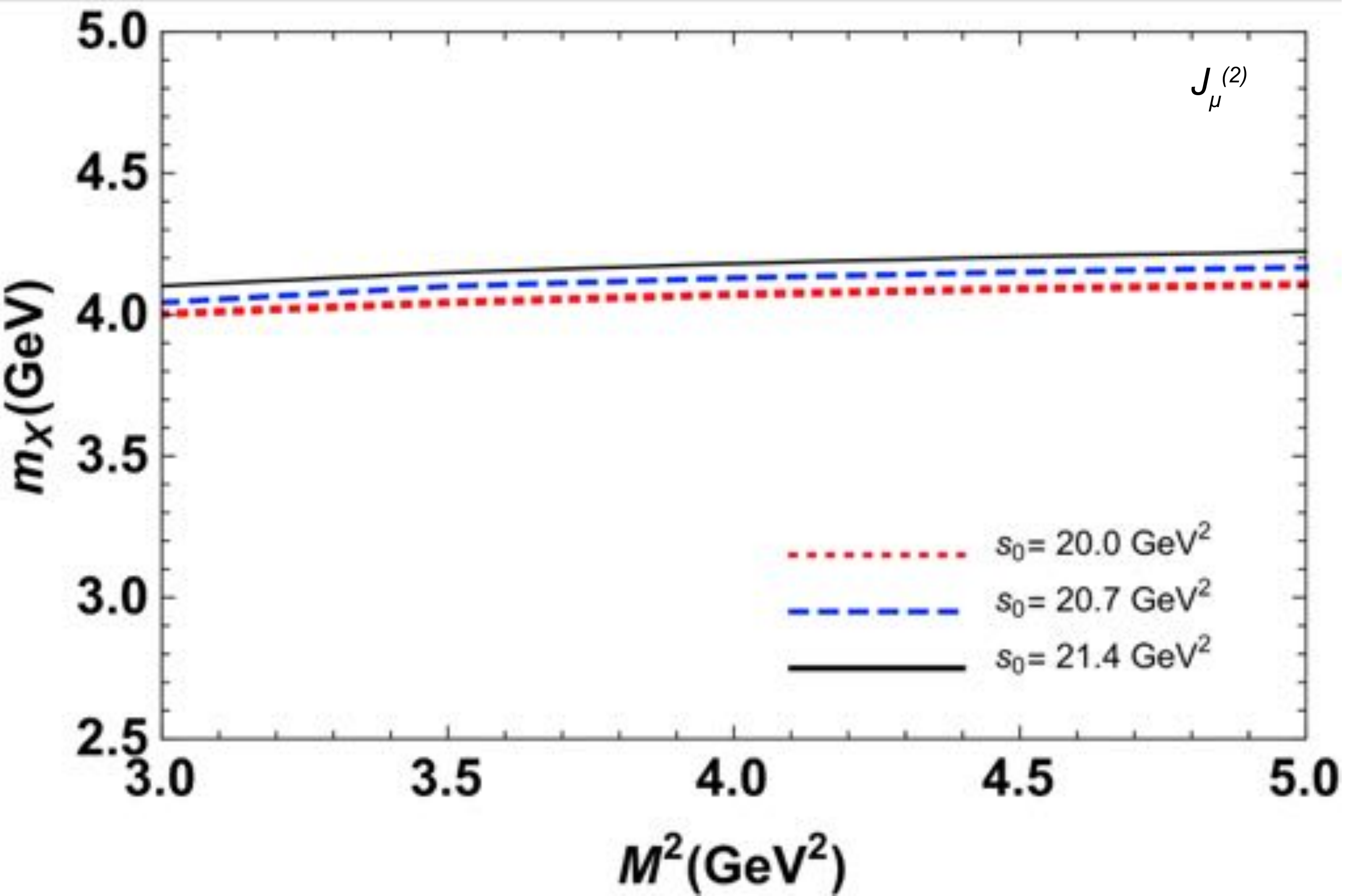}\\
\vskip1.2cm
\includegraphics[width=8.5cm]{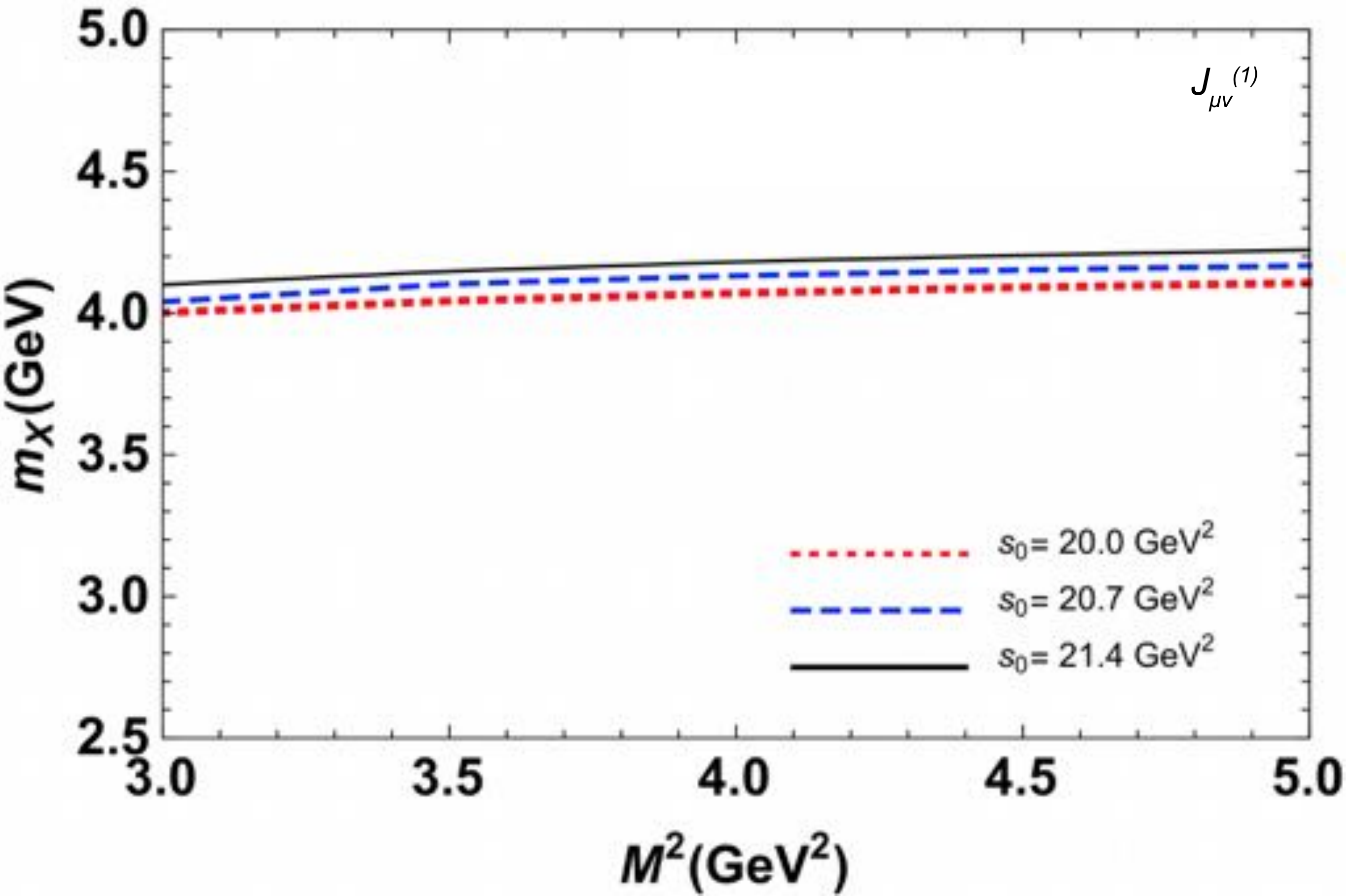}~~~~~~~~\includegraphics[width=8.5cm]{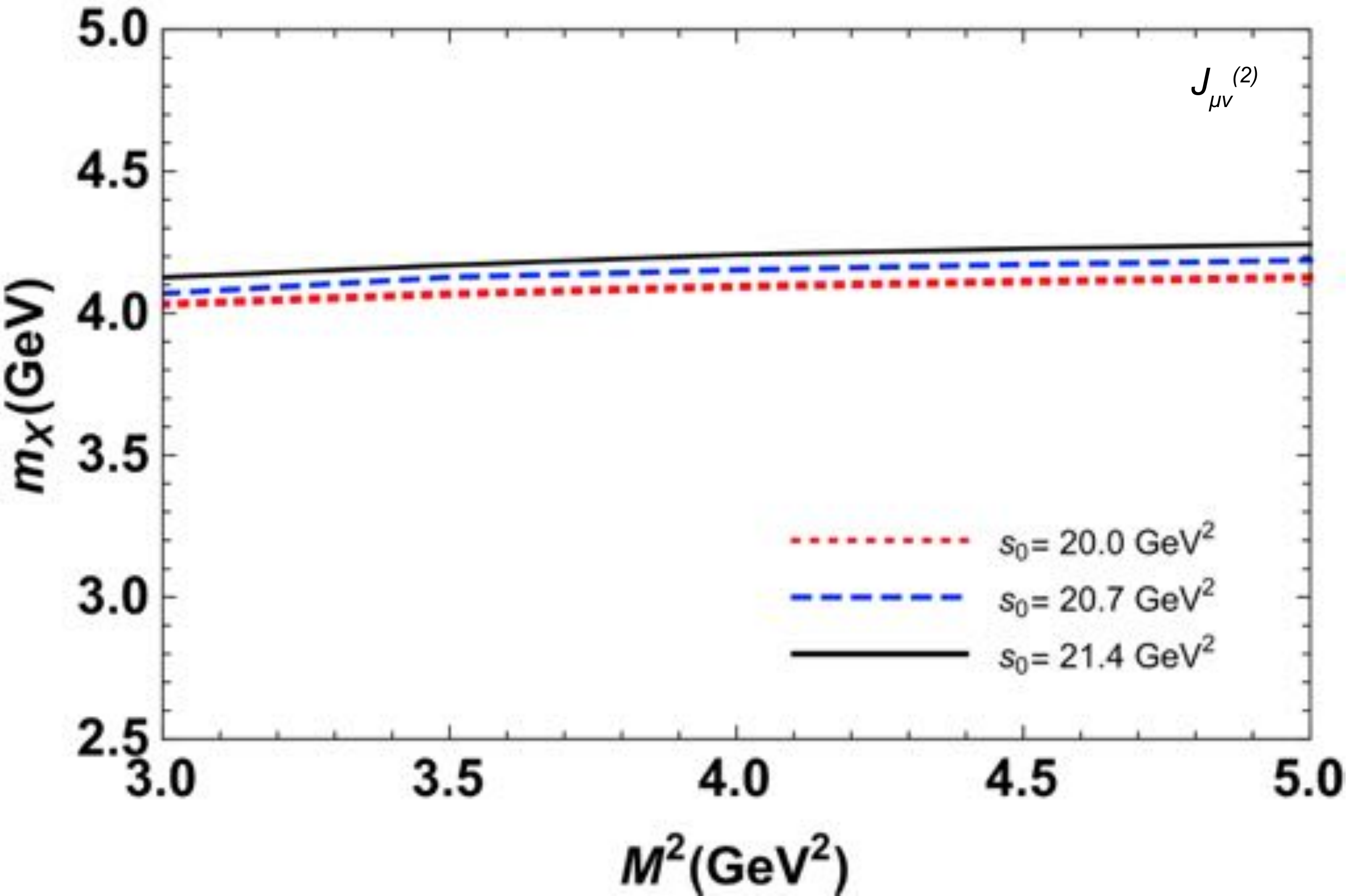}
 \vskip1.2cm \caption{\label{fig::MX} The mass of the ground state coupling to specified current as a function of $M^2$ for different values of $s_0$, for the scalar, axial vector and tensor molecular currents (left panel) and for the scalar, axial vector and tensor diquark-antidiquark currents (right panel). }
\end{center}
\end{figure}

\begin{figure}[h]
\begin{center}
\includegraphics[width=8.5cm]{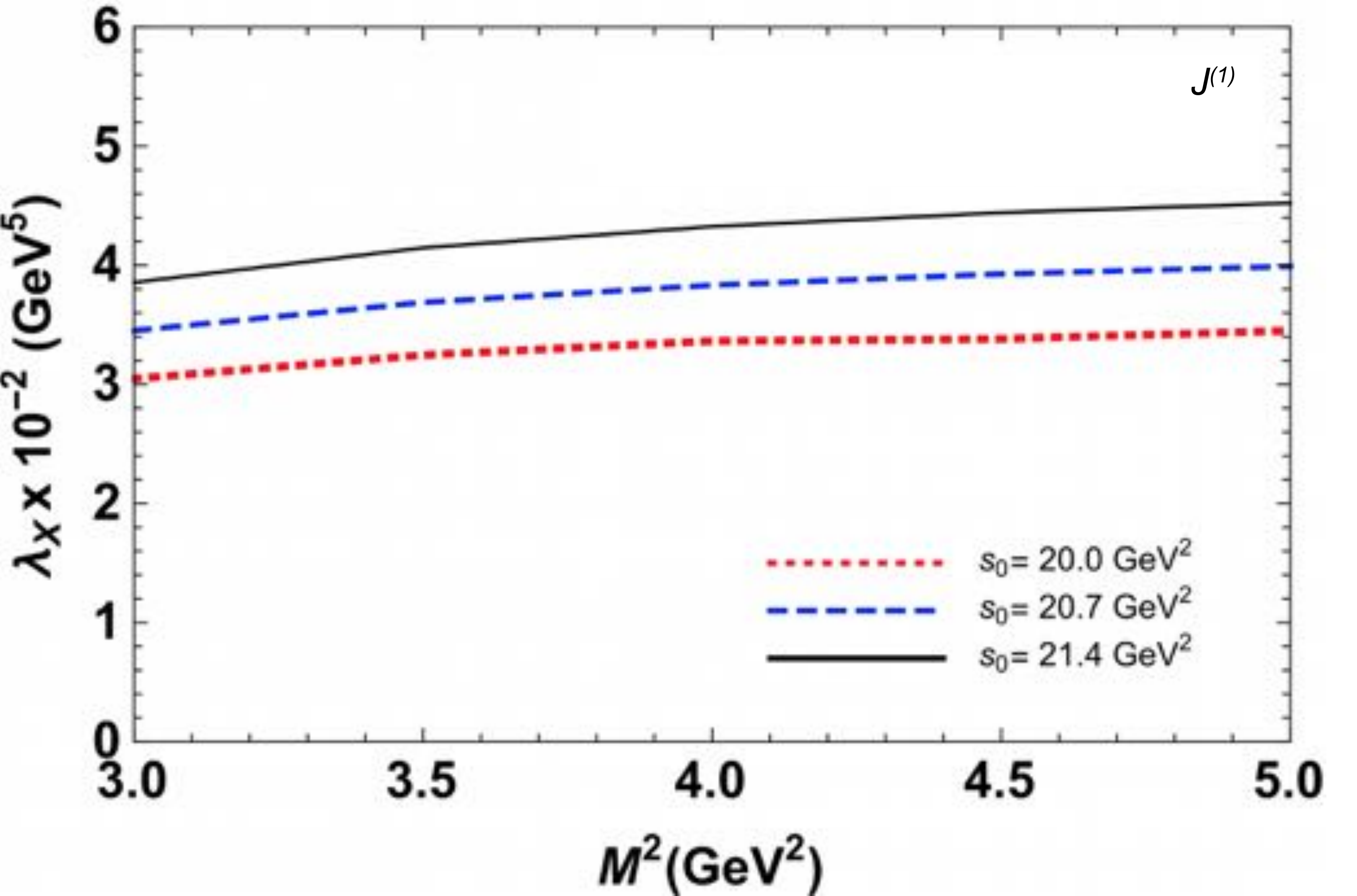}~~~~~~~~\includegraphics[width=8.5cm]{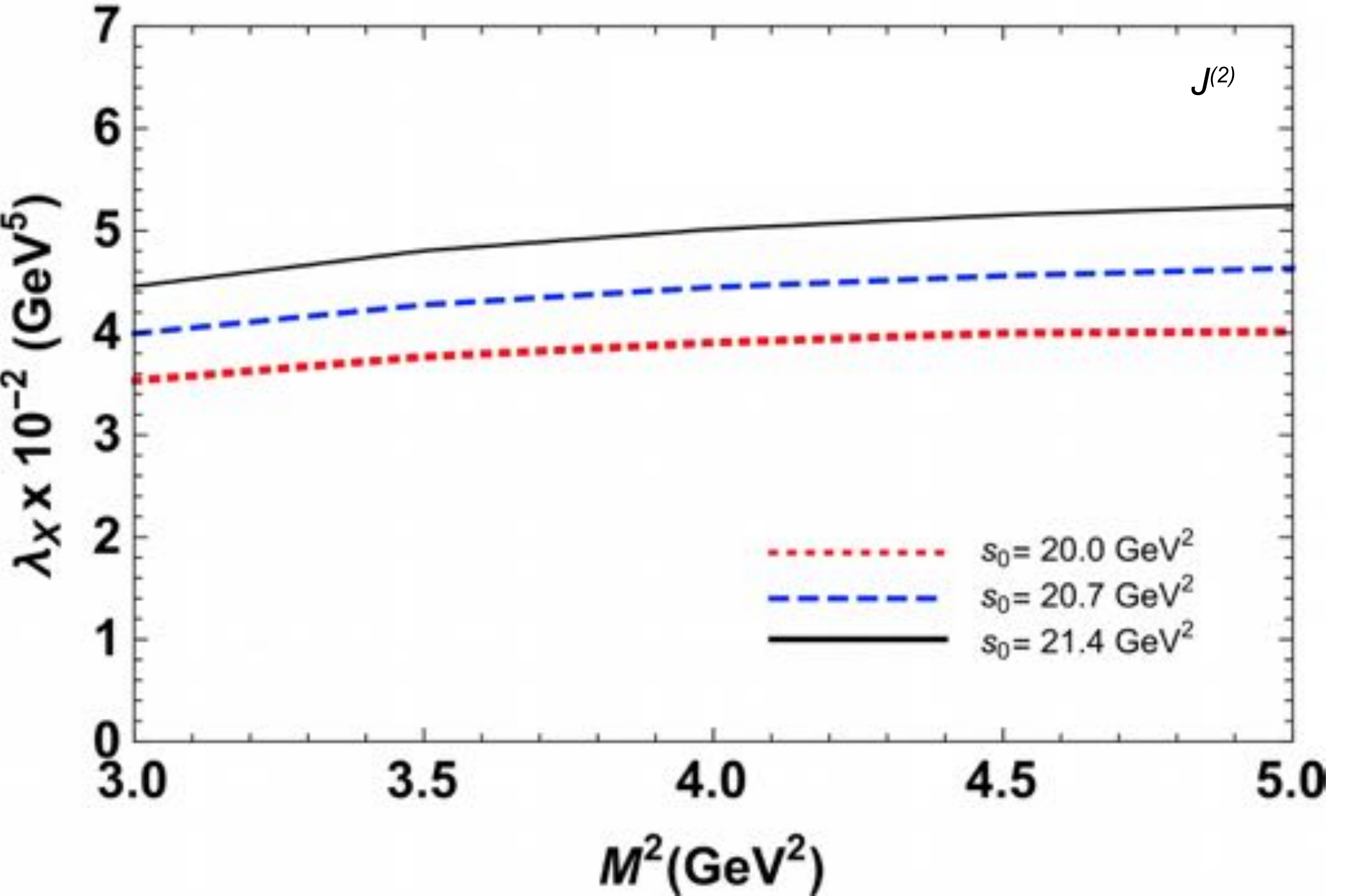}\\
\vskip1.2cm
\includegraphics[width=8.5cm]{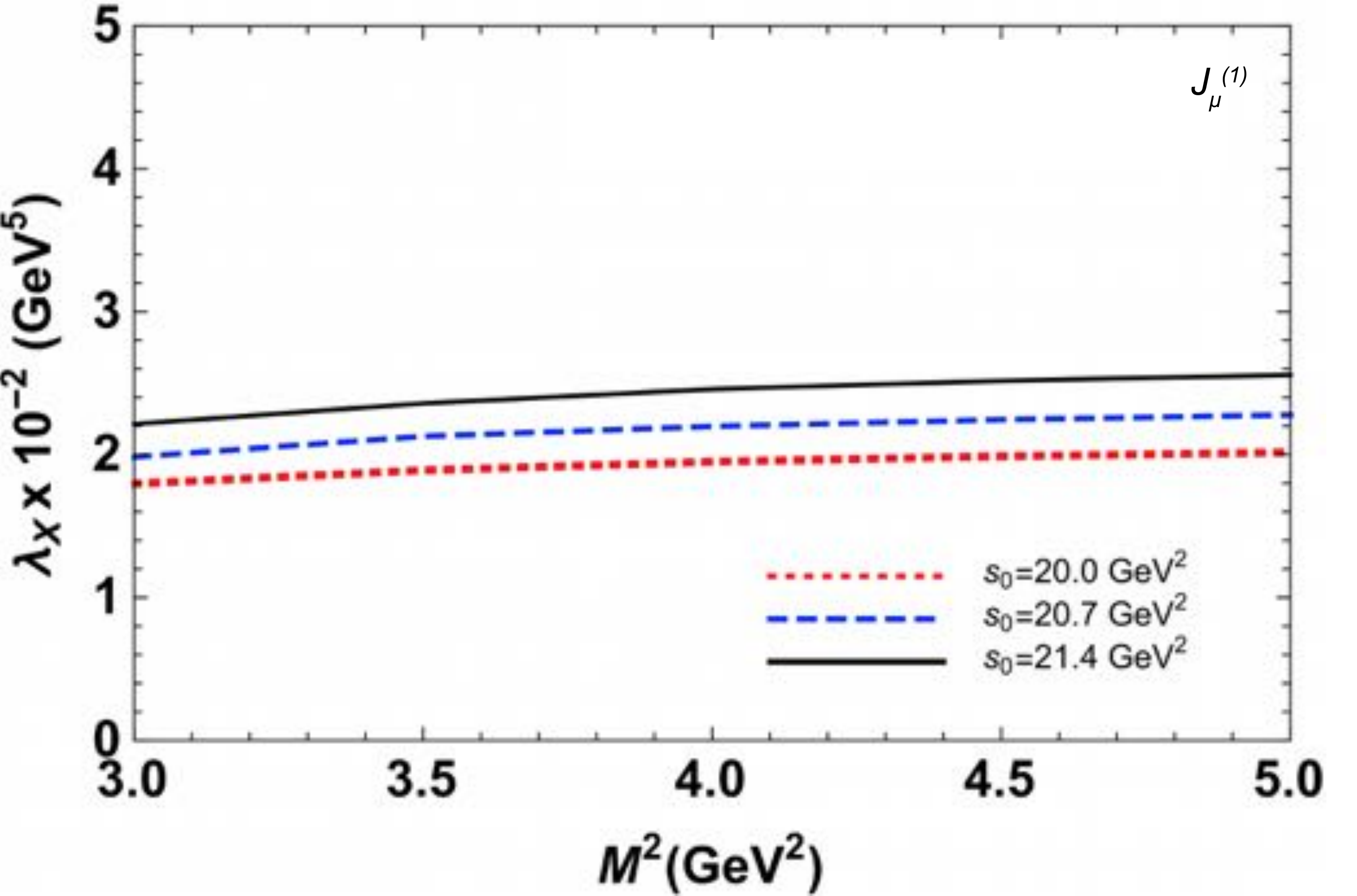}~~~~~~~~\includegraphics[width=8.5cm]{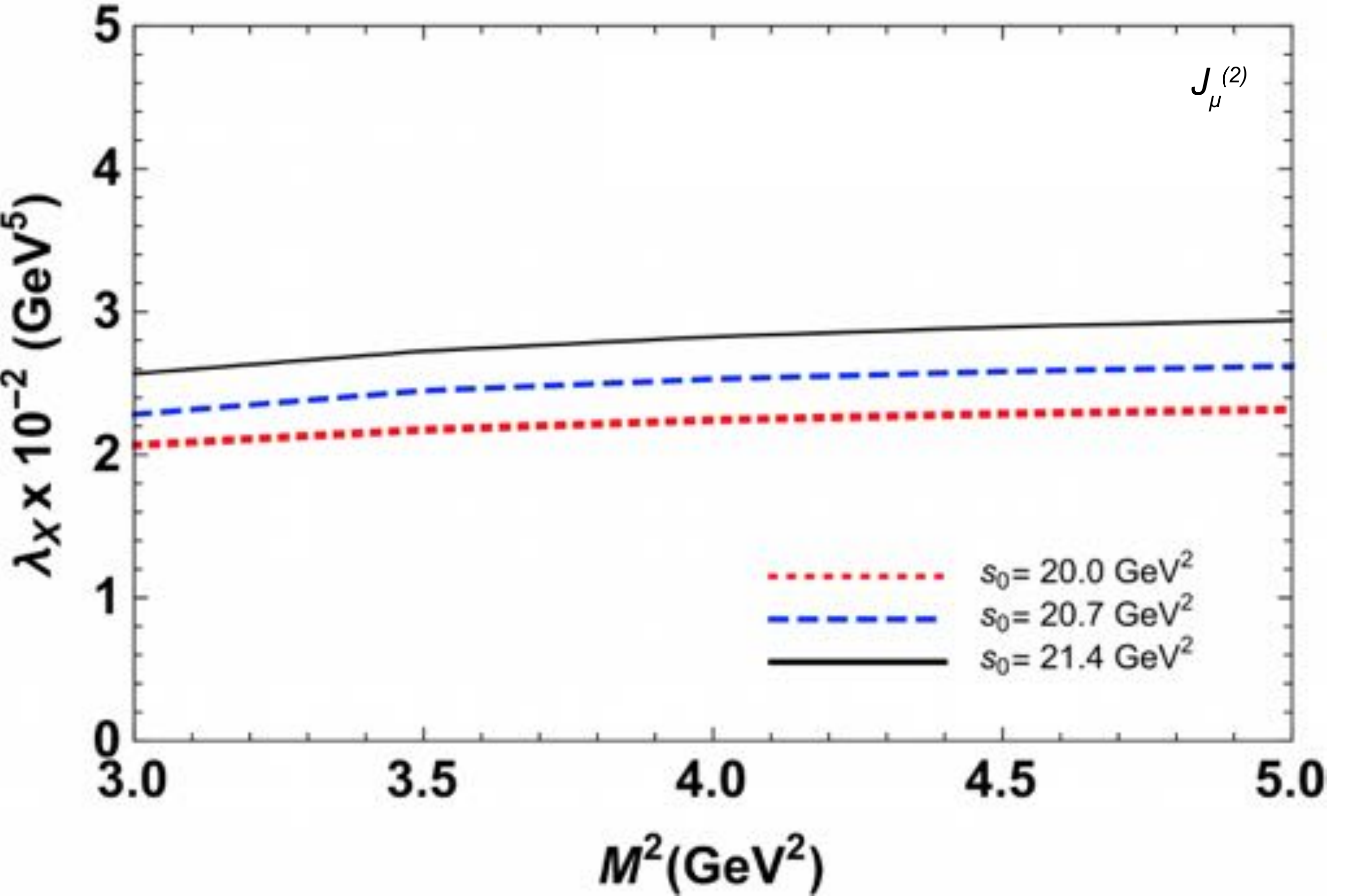}\\
\vskip1.2cm
\includegraphics[width=8.5cm]{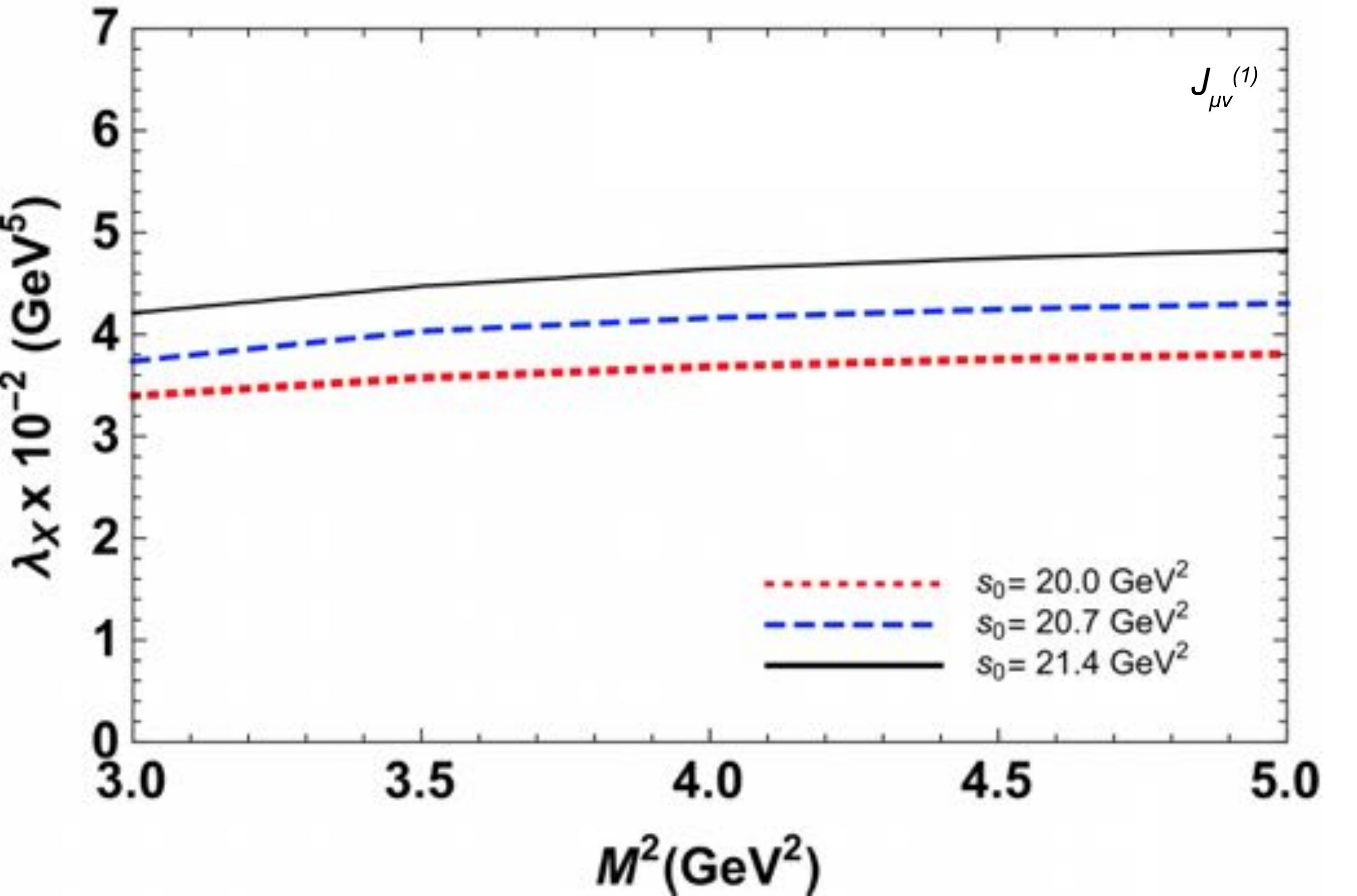}~~~~~~~~\includegraphics[width=8.5cm]{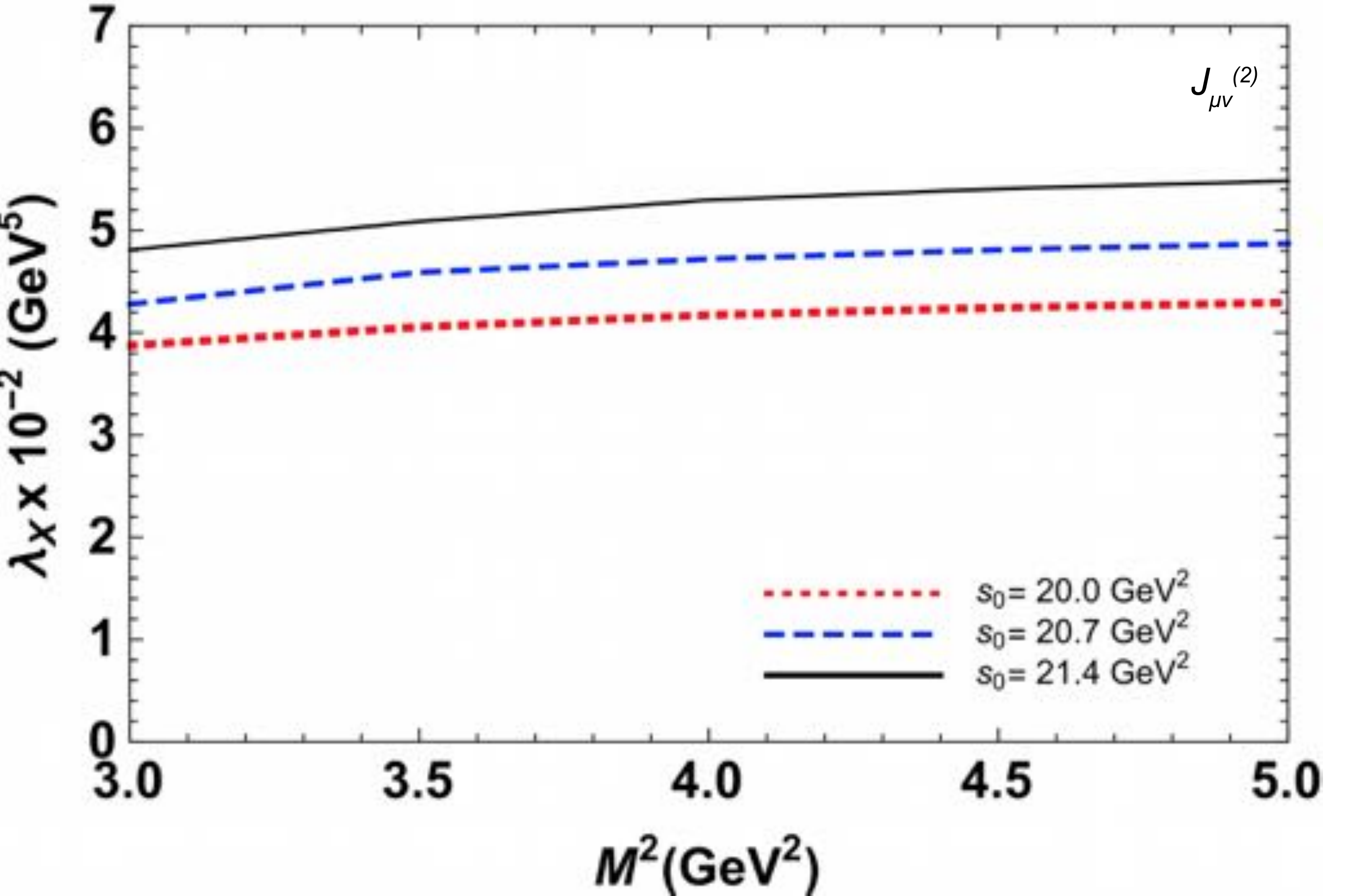}
 \vskip1.2cm \caption{\label{fig::LX} Meson coupling constant ($\lambda$) of the ground state coupling to specified current as a function of $M^2$ for different values of $s_0$, for the scalar, axial vector and tensor molecular currents (left panel) and for the scalar, axial vector and tensor diquark-antidiquark currents (right panel).}
\end{center}
\end{figure}

In traditional sum rules analysis, one last check is necessary to observe the dependence of the physical quantities to parameters $M^2$ and $s_0$. In Fig. \ref{fig::MX} we provide the final results which are obtained for the masses and depict these masses as a function of Borel mass at fixed $s_0$ values. The masses of the ground state hadrons coupling to specified currents are stable with respect to variations of $M^2$ and $s_0$. In Fig. \ref{fig::LX}, the dependence of the meson coupling constants are plotted with respect to $M^2$ at some fixed $s_0$ values within the range of working regions of the sum rules. Even though the dependence of the meson coupling constants to these parameters are within the acceptable limits, the observed variations with respect to continuum thresholds are the main sources of uncertainties in the final results for meson coupling constants.   

Finally, we present our results for the masses and the meson coupling constants of the specified currents in Tables \ref{tab::masses1} and \ref{tab::Residue1}. The uncertainties in the obtained results are due to variations of $s_0$ and $M^2$ within the working regions specified in Table \ref{Tab::WR}. We also considered the errors of the input parameters given in Table \ref{Tab::Param}

\begin{table}[h]
\centering
\caption{Masses obtained in this work, and their comparison with literature.}
\label{tab::masses1}       
\begin{tabular}{lclclclcl}
\hline\hline   Current & $m_X$(GeV) &~ &$m_X$(GeV)  \\
& This Work & & Literature \\
\hline\hline  $ J^{(1)}\ (0^{++}) $  &~ $4.146\pm 0.141$&~   &$4.14\pm0.09$ \cite{Nielsen}  \\
&&~ &$4.13\pm0.10$ \cite{Zhang,Huang}\\
&&~ &$4.48\pm0.17$ \cite{Wang}\\
&&~ &$4.43\pm0.16$ \cite{Wang2}\\
&&~ &$4.14\pm0.08$ \cite{Wang3}\\
\hline$ J^{(1)}_{\mu}\ (1^{++})  $ &~ $4.136\pm0.131$  &~ &  \\
\hline$ J^{(1)}_{\mu\nu}\ (2^{++})  $ &~ $4.138\pm0.130$&~  &   \\
\hline\hline$J^{(2)}\ (0^{++}) $  &~ $4.141\pm0.142$  &~ & $3.98\pm0.08$ \cite{Wang4}   \\
\hline$ J^{(2)}_{\mu}\ (1^{++}) $ &~ $4.138\pm0.130$ &~  &$3.95\pm 0.09$ \cite{Wang5}  \\
  & &~  &$4.07\pm 0.10$ \cite{CZ2011}  \\
  & &~  &$4.183\pm 0.115$ \cite{Agaev}  \\
\hline$ J^{(2)}_{\mu\nu}\ (2^{++})$ &~ $4.162\pm0.125$  &~ &  $4.13\pm0.08$ \cite{Wang4}  \\
\hline\hline
\end{tabular}
\end{table}

\begin{table}[h]
\centering
\caption{Meson coupling constants of the ground state hadron coupling to specified current. The results obtained in this work are given in comparison with literature.}
\label{tab::Residue1}       
\begin{tabular}{lclclclcl}
\hline\hline   Current & $\lambda_X$ &~ &$\lambda_X$ \\
& ($\times 10^{-2}$GeV$^5$) & & ($\times 10^{-2}$GeV$^5$) \\
& This Work & & Literature \\
\hline\hline  $ J^{(1)}\ (0^{++}) $  &~ $3.889\pm 0.951$&~   &$4.20\pm0.96$ \cite{Nielsen}  \\
&& &$6.2\pm1.1$ \cite{Wang}  \\
&& &$5.75\pm0.90$ \cite{Wang3}  \\
\hline $ J^{(1)}_{\mu}\ (1^{++})  $ &~ $2.221\pm0.503$  &~ &  \\
\hline $ J^{(1)}_{\mu\nu}\ (2^{++})  $ &~ $4.199\pm0.948$& &$4.34\pm0.60$ \cite{Wang3}    \\
\hline\hline$J^{(2)}\ (0^{++}) $  &~ $4.510\pm1.099$  &~ & $4.8\pm 0.8$ \cite{Wang4}   \\
\hline$ J^{(2)}_{\mu}\ (1^{++}) $ &~ $2.556\pm0.578$ & &$0.94\pm 0.16$\cite{Agaev}    \\
\hline$ J^{(2)}_{\mu\nu}\ (2^{++})$ &~ $4.775\pm1.085$  &~ &  $5.75\pm0.90$ \cite{Wang4}  \\
\hline\hline
\end{tabular}
\end{table}
%
%
%


\section{Discussion and Concluding Remarks}\label{S4}

In the current work, we performed a QCD sum rules analysis for scalar, axial vector and tensor currents identifying possible $D_s^*\bar{D}_s^*$ molecular states, and scalar, axial vector and tensor, diquark-antidiquark currents. For these currents, the corresponding spectral densities are calculated up to dimension eight, and a careful analysis is done to determine the working regions of the sum rules. The masses of the ground states coupling to these six currents are estimated within 10 MeV vicinity of the mass of X(4140) measured by several experiments\cite{LHCb,CDF1,CDF2,CMS,D0}, which is acceptable with the given uncertainties. Thus we conclude that, if X(4140) is an axial vector state as measured by LHCb \cite{LHCb}, it has scalar and tensor partners with the same mass. In the literature, such scenario was introduced for X(3872) and its possible partners \cite{Nieves}. In addition, existence of states with different spins which couple to $D^*\bar{D}^*$ and $D_s^*\bar{D}_s^*$ were claimed in Ref. \cite{Nielsen2}. We also calculated the meson coupling constants for these ground states, which may be X(4140) and its partners. Since the chosen molecular and diquark currents estimate degenerate masses, we can not favor neither of the $D_s^*\bar{D}_s^*$ molecular nor the diquark-antidiquark structures for these states. 

The masses estimated in this work were presented in Table \ref{tab::masses1}, in comparison with the results of the sum rules analysis which were done using similar currents. Our results are in good agreement with Ref. \cite{Nielsen,Zhang,Huang,Wang3} for the scalar molecular current given in Eq. (\ref{curr_S_m}), and with Ref. \cite{Wang4} for tensor diquark current given in Eq. (\ref{curr_T_t}). There is a discrepancy between this work and Ref. \cite{Wang,Wang2,Wang4,Wang5} for the specified currents in Table \ref{tab::masses1}, in which the authors did not follow the traditional sum rules analysis. Since the results obtained in Ref. \cite{Wang,Wang2} are invalidated by the same authors in Ref. \cite{Wang3}, we conclude that for the scalar molecular and tensor diquark-antidiquark currents, this work confirms literature. For the axial vector and tensor molecular currents, the masses are obtained for the first time  and for the scalar diquark-antidiquark current, the masses are estimated with traditional sum rules analysis for the first time in this work. For the axial vector diquark-antidiquark current, our results are in agreement with the ones found in Ref. \cite{CZ2011} in which the results are less stable with respect to continuum threshold, and with the ones obtained in Ref. \cite{Agaev}, which appeared when this manuscript is being prepared. The axial vector currents analyzed in this work can be associated with the $J^{PC}=1^{++}$, X(4140) exotic meson observed by LHCb \cite{LHCb}.

As is seen from Table \ref{tab::Residue1}, meson coupling constants obtained in this work are in agreement with the sum rules analysis of the similar currents in literature \cite{Nielsen,Wang,Wang3,Wang4}, for the scalar molecular, tensor molecular, scalar diquark and tensor diquark currents. For the diquark-antidiquark axial vector current, our result for the meson coupling constant is bigger than the result obtained in \cite{Agaev}. For molecular axial vector currents, the meson coupling constant of the ground state structure is estimated for the first time.

In summary, we presented a QCD sum rules analysis of the two point correlation function for possible  $D^*_s \bar{D}^*_s$ molecule and diquark-antidiquark currents with $J^{PC}=0^{++},1^{++}$ and $2^{++}$. Our motivation is to search a possible state which can be associated with X(4140) confirmed by several experiments\cite{CDF1,CDF2,CMS,D0}, and measured to be an axial vector state by LHCb\cite{LHCb}. For both molecular and diquark-antidiquark axial vector currents, we obtained a stable mass in agreement with X(4140) observed by LHCb. Thus we confirm the existence and the mass of X(4140), but we can not predict its content. In addition, we also have a subsidiary attempt to reanalyze the scalar molecular, tensor molecular, scalar diquark and tensor diquark currents which were used in studying X(4140) within QCD sum rules. We found that, all of these four currents predict same masses as X(4140), which confirms the traditional sum rules analysis that were done with similar currents. However, interpreting these states as X(4140) contradicts with LHCb measurements. Thus, we conclude that the analyzed  scalar and tensor states might be partners of X(4140) with degenerate masses which can either have a molecular or diquark-antidiquark content, and they have not been observed yet.  In addition to the masses, we also predicted the meson coupling constants for the states coupling to chosen currents. Our findings can be used in further analysis for the decays of X(4140) or its possible partners. Consequently, X(4140) should be investigated more, by studying its decays, and by other approaches as well. Preliminary results of this work was also presented in \cite{A_H_conf12}.    

\section*{Acknowledgements}
This work is supported by TUBITAK under project number 114F215. Authors are thankful to K. Azizi, H. Sundu, A. Ozpineci, Y. Kai, E.A. Yetkin, F. Ozok and G. Erkol for helpful discussions. H. Dag also thanks to B. Isildak and A. Hayreter. 

\onecolumngrid

\clearpage
\onecolumngrid

\appendix*

\section{ Spectral Densities}\label{S4}
\renewcommand{\theequation}{\Alph{section}.\arabic{equation}} \label{sec:App}
In this appendix section, analytic expressions of the spectral densities $\rho^{OPE (a)}_J$, obtained by using currents given in Eqs. (\ref{curr_S_m}-\ref{curr_T_t}) are given. They are obtained from the sum rules given in Eq. (\ref{SR_res}). Spectral densities can be written as 
\begin{equation}\label{app::rho}
\rho^{(a)\text{OPE}}_{S}=\sum_{i}\rho^{(a)\text{OPE}}_{S,i} ,
\end{equation}
where $\rho^{(a)\text{OPE}}_{S,i}$ are the contributions of terms with dimension denoted by  $i=0,2,3,..,8$. In Eq. (\ref{app::rho}), $a=1,2$ denotes the molecular or tetraquark content, and $S=0,1,2$ denotes whether the current is scalar, axial vector or tensor, respectively. In the appendix section, $A=(x_1+x_2-1)$ and $B=(x_2-1)$ are defined for simplicity.
\subsection{Spectral densities of molecular scalar current $(J^{(1)})$}
\begin{eqnarray}
\nonumber \rho^{(1)\text{OPE}}_{0,0}(s)&=&\int_{0}^{1}\int_{0}^{1-x_{1}}dx_{1}~dx_{2}\frac{-3 \left(s x_{1} x_{2}
   A-m_{c}^2
   \left(x_{1}^3+x_{1}^2 (2 x_{2}-1)+2 x_{1}
   B x_{2}+B
   x_{2}^2\right)\right)^2}{1024 \pi
   ^6 A \left(x_{1}^2+x_{1}
   B+B x_{2}\right)^8}\\
 \nn  &\times & \left\lbrace  m_{c}^4
   \left( x_{1}^2-x_{1}+B x_{2}\right)
   \left(x_{1}^3+x_{1}^2 (2 x_{2}-1)+2 x_{1}
   B x_{2}+B x_{2}^2\right)^2\right. \\
 \nn  &- & 4
   m_{c}^3 m_{s} (x_{1}+x_{2})^2
   \left(x_{1}^2+x_{1} B+B
   x_{2}\right)^3 -2 m_{c}^2 \left(x_{1}^2+x_{1}
   B+B x_{2}\right)\\
  \nn & \times & \left(12
   m_{s}^2 \left(x_{1}^2+x_{1}
   B+B x_{2}\right)^3 +s x_{1}
   x_{2} \left(2 x_{1}^4+x_{1}^3 (3
   x_{2}-4)+x_{1}^2 \left(2 x_{2}^2-7
   x_{2}+2\right)\right.\right. \\
  \nn &+& \left. \left. \left.  x_{1} x_{2} \left(3 x_{2}^2-7
   x_{2}+4\right)+2 B^2
   x_{2}^2\right)\right)+ 10 m_{c} m_{s} s x_{1}
   x_{2} \left(x_{1}^2+x_{1}
   B+B x_{2}\right)^2\right. \\
   \nn &\times &\left. 
   \left(x_{1}^2+x_{1} (2 x_{2}-1)+B
   x_{2}\right)+s^2 x_{1}^2 x_{2}^2
   A^2 \left(3 x_{1}^2-x_{1} (4
   x_{2}+3)+3 B x_{2}\right)\right\rbrace 
\end{eqnarray}
\begin{eqnarray}
\nonumber \rho^{(1)\text{OPE}}_{0,3} (s)&=&\int_{0}^{1}\int_{0}^{1-x_{1}}dx_{1}~dx_{2}\frac{-3 \langle
\bar{s}s\rangle}{64 \pi ^4
   \left(x_{1}^2+x_{1} B+B
   x_{2}\right)^6}  \left\lbrace  2 m_{c}^5
   \left(x_{1}^3+x_{1}^2 (2 x_{2}-1)\right. \right. \\   
   \nn &+&\left. \left.  2 x_{1}
   B x_{2}+B x_{2}^2\right)^3+2
   m_{c}^4 m_{s} \left(x_{1}^2+x_{1}
   B+B x_{2}\right)^2 \left(5
   x_{1}^5+x_{1}^4 (13 x_{2}-10)\right. \right. \\
   \nn &+& \left. \left. x_{1}^3 \left(18
   x_{2}^2-28 x_{2}+5\right) +3 x_{1}^2 x_{2}
   \left(6 x_{2}^2-12 x_{2}+5\right)+x_{1}
   x_{2}^2 \left(13 x_{2}^2-28 x_{2}+15\right)\right. \right.\\
   \nn &+& \left. \left. 5
   B^2 x_{2}^3\right) - 2 m_{c}^3
  A \left(x_{1}^3+x_{1}^2 (2
   x_{2}-1)+2 x_{1} B x_{2}+B
   x_{2}^2\right)^2 \left(m_{s}^2
   \left(x_{1}^2+x_{1} B+B
   x_{2}\right)\right. \right. \\
   \nn &+&\left. \left.  3 s x_{1} x_{2}\right)-2 m_{c}^2
   m_{s} s x_{1} x_{2} \left(7
   x_{1}^7+x_{1}^6 (19 x_{2}-28)+6 x_{1}^5
   \left(4 x_{2}^2-13 x_{2}+7\right)+2 x_{1}^4 \right. \right. \\
   \nn &\times &\left. \left. 
   \left(11 x_{2}^3-57 x_{2}^2+60
   x_{2}-14\right)+x_{1}^3 \left(22 x_{2}^4-124
   x_{2}^3+177 x_{2}^2-82 x_{2}+7\right)+3
   x_{1}^2 B^2 x_{2}\right. \right. \\
   \nn &\times & \left. \left.  \left(8 x_{2}^2-22
   x_{2}+7\right)+ x_{1} B^3 x_{2}^2 (19
   x_{2}-21)+7 B^4 x_{2}^3\right)+m_{c}
   s x_{1} x_{2}A^2
   \left(x_{1}^3 \right. \right. \\
   \nn &+&  \left. \left. x_{1}^2 (2 x_{2}-1)+ 2 x_{1}
   B x_{2}+B x_{2}^2\right)
   \left(3 m_{s}^2 \left(x_{1}^2+x_{1}
   B+B x_{2}\right)+4 s x_{1}
   x_{2}\right) \right.  \\
   &+&  \left. 4 m_{s} s^2 x_{1}^2 x_{2}^2
  A^3 \left(x_{1}^2- x_{1} (4
   x_{2}+1)+B x_{2}\right)\right\rbrace 
\end{eqnarray}
\begin{eqnarray}
\nonumber \rho^{(1)\text{OPE}}_{0,4}(s)&=&\int_{0}^{1}\int_{0}^{1-x_{1}}dx_{1}~dx_{2}\frac{\langle
\alpha_{s} \frac{G^{2}}{\pi}\rangle}{512
   \pi ^4 A \left(x_{1}^2+x_{1}
   B+B x_{2}\right)^6}\\
   \nn &\times & \left\lbrace m_{c}^4
   \left(x_{1}^3+x_{1}^2 (2 x_{2}-1)+2 x_{1}
   B x_{2}+B x_{2}^2\right)^2
   \left(4 x_{1}^4+x_{1}^3 (29 x_{2}-10)\right.\right.\\
   \nn &+&  \left. x_{1}^2
   \left(38 x_{2}^2-39 x_{2}+6\right)+x_{1} x_{2}
   \left(29 x_{2}^2-39 x_{2}+12\right)+2 x_{2}^2
   \left(2 x_{2}^2-5 x_{2}+3\right)\right)\\
   \nn &+&\left. 2 m_{c}^3
   m_{s} \left(x_{1}^2+x_{1}
   B+B x_{2}\right)^2 \left(2
   x_{1}^6+2 x_{1}^5 (7 x_{2}-1)+x_{1}^4
   x_{2} (35 x_{2}-26)\right.\right.\\
   \nn &+& \left. \left. 12 x_{1}^3 x_{2} \left(4
   x_{2}^2-5 x_{2}+1\right)+x_{1}^2 x_{2}^2
   \left(35 x_{2}^2-60 x_{2}+24\right)+2 x_{1}
   x_{2}^3 \left(7 x_{2}^2-13 x_{2}+6\right)\right. \right. \\
   \nn &+& \left. \left. 2
   B x_{2}^5\right)+m_{c}^2
   \left(x_{1}^2+x_{1} B+B
   x_{2}\right) \left(8 m_{s}^2
   \left(x_{1}^2+x_{1} B+B
   x_{2}\right)^2 \right.\right. \\
   \nn & \times &\left.\left. \left(x_{1}^4-x_{1}^3+B
   x_{2}^3\right)-s x_{1} x_{2} \left(15
   x_{1}^6+4 x_{1}^5 (31 x_{2}-12)+x_{1}^4
   \left(340 x_{2}^2-301 x_{2}+51\right)\right.\right.\right.\\
   \nn &+& \left.\left.\left. 3 x_{1}^3
   \left(154 x_{2}^3-214 x_{2}^2+77 x_{2}-6\right)+2
   x_{1}^2 x_{2} \left(170 x_{2}^3-321
   x_{2}^2+180 x_{2}-27\right)\right.\right.\right.\\
   \nn &+&\left.\left.\left.  x_{1} x_{2}^2
   \left(124 x_{2}^3-301 x_{2}^2+231
   x_{2}-54\right)+3 B^2 x_{2}^3 (5
   x_{2}-6)\right)\right)-6 m_{c} m_{s} s x_{1}
   x_{2} \right.\\
   \nn & \times & \left.  \left(x_{1}^8+x_{1}^7 (7
   x_{2}-3)+x_{1}^6 \left(23 x_{2}^2-26
   x_{2}+3\right)+x_{1}^5 \left(44 x_{2}^3-81
   x_{2}^2+37 x_{2}-1\right)\right.\right.\\
   \nn &+& \left.\left. 3 x_{1}^4 x_{2}
   \left(18 x_{2}^3-45 x_{2}^2+35
   x_{2}-8\right)+x_{1}^3 B^2 x_{2}
   \left(44 x_{2}^2-47 x_{2}+6\right)\right.\right.\\
   \nn &+& \left.\left. x_{1}^2
   B^3 x_{2}^2 (23 x_{2}-12)+x_{1}
   B^2 x_{2}^3 \left(7 x_{2}^2-12
   x_{2}+6\right)+B^3 x_{2}^5\right)\right.\\
    &+& \left. 6 s^2
   x_{1}^2 x_{2}^2 A^3 \left(2
   x_{1}^3+x_{1}^2 (9 x_{2}-2)+x_{1} x_{2} (9
   x_{2}-4)+2 B x_{2}^2\right)\right\rbrace 
\end{eqnarray}
\begin{eqnarray}
\nonumber \rho^{(1)\text{OPE}}_{0,5} (s)&=&\int_{0}^{1} dx \frac{m_{0}^2 m_{s} \langle
\bar{s}s\rangle \left(-8
   m_{c}^2+m_{c} m_{s}-3 s (x-1) x\right)}{64 \pi
   ^4} \\
   \nn & +& \int_{0}^{1}\int_{0}^{1-x_{1}}dx_{1}~dx_{2}
\frac{m_{0}^2 \langle
\bar{s}s\rangle A}{128 \pi ^4
   \left(x_{1}^2+x_{1} B+B
   x_{2}\right)^5}\\
   \nn & \times & \left\lbrace 6
   m_{c}^3 \left(x_{1}^3+x_{1}^2 (2 x_{2}-1)+2
   x_{1} B x_{2}+B
   x_{2}^2\right)^2-8 m_{c}^2 m_{s} x_{1}
   x_{2} \left(x_{1}^4+x_{1}^3 (3
   x_{2}-2)\right.\right.\\
   \nn &+& \left. \left. x_{1}^2 \left(4 x_{2}^2-5
   x_{2}+1\right)+x_{1} x_{2} \left(3 x_{2}^2-5
   x_{2}+2\right)+B^2 x_{2}^2\right)-9
   m_{c} s x_{1} x_{2} \right. \\
   \nn &\times & \left. \left(x_{1}^4+x_{1}^3
   (3 x_{2}-2)+ x_{1}^2 \left(4 x_{2}^2-5
   x_{2}+1\right)+x_{1} x_{2} \left(3 x_{2}^2-5
   x_{2}+2\right)+B^2 x_{2}^2\right) \right. \\
   &+&\left.  18
   m_{s} s x_{1}^2 x_{2}^2
   A^2\right\rbrace
\end{eqnarray}
\begin{eqnarray}
\nonumber \rho^{(1)\text{OPE}}_{0,6} (s)&=&\int_{0}^{1} dx \frac{\langle
\bar{s}s\rangle^2 }{432 \pi ^4} \left\lbrace g_{s}^2 \left(2
   m_{c}^2-m_{c} m_{s}+3 s (x-1) x\right)- 54 \pi
   ^2 \left(2 m_{c}^2-m_{c} m_{s} \right. \right. \\
   \nn &-&\left.\left. 3 m_{s}^2
   (x-1) x\right)\right\rbrace +\int_{0}^{1}\int_{0}^{1-x_{1}}dx_{1}~dx_{2} \frac{g_{s}^2 \langle
\bar{s}s\rangle^2 x_{1} x_{2}
   A^2}{432 \pi ^4
   \left(x_{1}^2+x_{1} B+B
   x_{2}\right)^5}\\
    & \times & \left\lbrace 4 m_{c}^2
   \left(x_{1}^3+x_{1}^2 (2 x_{2}-1)+2 x_{1}
   B x_{2}+B x_{2}^2\right)-9 s
   x_{1} x_{2} A\right\rbrace
\end{eqnarray}
\begin{eqnarray}
\nonumber \rho^{(1)\text{OPE}}_{0,7} (s)&=&\int_{0}^{1} dx \frac{\langle
\alpha \frac{G^{2}}{\pi}\rangle  \langle
\bar{s}s\rangle (m_{c}-6
   m_{s})}{192 \pi ^2} +\int_{0}^{1}\int_{0}^{1-x_{1}}dx_{1}~dx_{2} \frac{\langle
\alpha \frac{G^{2}}{\pi}\rangle m_{c} \langle
\bar{s}s\rangle}{96 \pi ^2 B
   \left(x_{1}^2+x_{1} B+B
   x_{2}\right)^4}\\
   \nn & \times &\left\lbrace 2
   x_{1}^6+4 x_{1}^5 B+x_{1}^4 \left(-5
   x_{2}^2+3 x_{2}+2\right)-13 x_{1}^3
   B^2 x_{2} +x_{1}^2 x_{2} \left(-13
   x_{2}^3+31 x_{2}^2 \right. \right.\\
   &-& \left. \left. 24 x_{2}+6\right) -  x_{1}
   B^2 x_{2}^2 (7 x_{2}-6)+2 B^2
   x_{2}^4\right\rbrace
\end{eqnarray}
\begin{equation}
 \rho^{(1)\text{OPE}}_{0,8}(s)=0
\end{equation}
\subsection{Spectral densities of molecular axialvector current $(J^{(1)}_\mu)$}
\begin{eqnarray}
\nonumber \rho^{(1)\text{OPE}}_{1,0}(s)&=&\int_{0}^{1}\int_{0}^{1-x_{1}}dx_{1}~dx_{2}\frac{-1}{4096 \pi ^6
   A \left(x_{1}^2+x_{1}
   B+B x_{2}\right)^8}  \left\lbrace s x_{1} x_{2}
   A \right. \\
   \nn &-& \left. m_{c}^2
   \left(x_{1}^3+x_{1}^2 (2 x_{2}-1)+2 x_{1}
   B x_{2}+B
   x_{2}^2\right)\right)^2  \left(3 m_{c}^4 x_{1}
   x_{2} \left(x_{1}^3+x_{1}^2 (2 x_{2}-1) \right. \right. \\
   \nn &+& \left. \left. 2
   x_{1} B x_{2}+B
   x_{2}^2\right)^2+18 m_{c}^3 m_{s}
   (x_{1}+x_{2})^2 \left(x_{1}^2+x_{1}
   B+B x_{2}\right)^3+2
   m_{c}^2 \left(x_{1}^2 \right. \right. \\
   \nn &+& \left. \left.  x_{1}
   B+B x_{2}\right)  \left(36
   m_{s}^2 \left(x_{1}^2+x_{1}
   B+B x_{2}\right)^3-13 s
   x_{1}^2 x_{2}^2 \left(x_{1}^2+x_{1} (2
   x_{2}-1) \right. \right. \right. \\
   \nn &+& \left. \left. \left. B x_{2}\right)\right)-54
   m_{c} m_{s} s x_{1} x_{2}
   \left(x_{1}^2+x_{1} B+B
   x_{2}\right)^2 \left(x_{1}^2+x_{1} (2
   x_{2}-1)+B x_{2}\right)\right.\\
    &+& \left. 35 s^2
   x_{1}^3 x_{2}^3
   A^2\right\rbrace
\end{eqnarray}
\begin{eqnarray}
\nonumber \rho^{(1)\text{OPE}}_{1,3}(s)&=&\int_{0}^{1}\int_{0}^{1-x_{1}}dx_{1}~dx_{2} \frac{3 \langle
\bar{s}s\rangle}{256 \pi ^4
   \left(x_{1}^2+x_{1} B+B
   x_{2}\right)^6} \left\lbrace 3 m_{c}^5
   \left(x_{1}^3+x_{1}^2 (2 x_{2}-1)\right.\right.\\
   \nn &+&\left.\left.  2 x_{1}
   B x_{2} B x_{2}^2\right)^3+2
   m_{c}^4 m_{s} \left(x_{1}^2+x_{1}
   B+B x_{2}\right)^2 \left(4
   x_{1}^5+x_{1}^4 (9 x_{2}-8) \right. \right.\\
   \nn &+&\left. \left.  x_{1}^3 \left(11
   x_{2}^2   - 21 x_{2}+4\right)+x_{1}^2 x_{2}
   \left(11 x_{2}^2-26 x_{2}+12\right)+3 x_{1}
   x_{2}^2 \left(3 x_{2}^2-7 x_{2}+4\right)\right.\right.\\
   \nn &+& \left.\left. 4
   B^2 x_{2}^3\right)-m_{c}^3
   A \left(x_{1}^3+x_{1}^2 (2
   x_{2}-1)+2 x_{1} B
   x_{2}+B x_{2}^2\right)^2\left(3
   m_{s}^2 \left(x_{1}^2+x_{1}
   B + B x_{2}\right) \right. \right. \\
   \nn &+& \left. \left. 10 s x_{1}
   x_{2}\right) -8 m_{c}^2 m_{s} s x_{1}
   x_{2} \left(x_{1}^7-4 x_{1}^6+x_{1}^5
   \left(-7 x_{2}^2-3 x_{2}+6\right) \right. \right. \\
   \nn &+& \left. \left. x_{1}^4
   \left(-15 x_{2}^3+10 x_{2}^2+9
   x_{2}-4\right) +x_{1}^3 \left(-15 x_{2}^4+19
   x_{2}^3+4 x_{2}^2-9 x_{2}+1\right) \right. \right. \\
   \nn &-& \left. \left. x_{1}^2
   B^2 x_{2} \left(7 x_{2}^2+4
   x_{2}-3\right) -3 x_{1} B^3
   x_{2}^2+B^4 x_{2}^3\right)+m_{c} s
   x_{1} x_{2} A^2
   \left(x_{1}^3+x_{1}^2 (2 x_{2}-1)\right.\right.\\
    &+& \left.\left. 2 x_{1}
   B x_{2}+B x_{2}^2\right)
   \left(5 m_{s}^2 \left(x_{1}^2+x_{1}
   B+B x_{2}\right)+7 s x_{1}
   x_{2}\right) -30 m_{s} s^2 x_{1}^3 x_{2}^3
   A^3\right\rbrace
\end{eqnarray}
\begin{eqnarray}
\nonumber \rho^{(1)\text{OPE}}_{1,4}(s)&=&\int_{0}^{1}\int_{0}^{1-x_{1}}dx_{1}~dx_{2} \frac{-\langle
\alpha_{s} \frac{G^{2}}{\pi}\rangle}{6144 \pi ^4
   A \left(x_{1}^2+x_{1}
   B+B x_{2}\right)^6} \left\lbrace 6 m_{c}^4 x_{1}
   x_{2} \left(x_{1}^2-x_{1}
   x_{2}+x_{2}^2\right)\right.  \\
   \nn &\times &\left. \left(x_{1}^3+x_{1}^2
   (2 x_{2}-1)+2 x_{1} B
   x_{2}+B x_{2}^2\right)^2 +9 m_{c}^3
   m_{s} \left(x_{1}^2+x_{1}
   B+B x_{2}\right)^2 \right.\\
   \nn &\times & \left. \left(2
   x_{1}^6-2 x_{1}^5 (x_{2}+1) +x_{1}^4 (6-13
   x_{2}) x_{2} -4 x_{1}^3 x_{2} \left(4
   x_{2}^2-5 x_{2}+1\right)\right. \right. \\
   \nn &+& \left. \left. x_{1}^2 x_{2}^2
   \left(-13 x_{2}^2+20 x_{2}-8\right)-2 x_{1}
   x_{2}^3 \left(x_{2}^2-3 x_{2}+2\right) +2
   B x_{2}^5\right)\right.\\
   \nn &+& \left. 8 m_{c}^2
   \left(x_{1}^2+x_{1} B+B
   x_{2}\right) \left(3 m_{s}^2
   \left(x_{1}^2+x_{1} B+B
   x_{2}\right)^2
   \left(x_{1}^4-x_{1}^3+B
   x_{2}^3\right) \right.\right.\\
   \nn &+& \left. \left. s x_{1}^2 x_{2}^2 \left(x_{1}^4+2 x_{1}^3 (5 x_{2}-2)+3
   x_{1}^2 \left(6 x_{2}^2-6 x_{2}+1\right)+2
   x_{1} x_{2} \left(5 x_{2}^2-9
   x_{2}+3\right)\right.\right.\right.\\
   \nn &+& \left.\left. \left. x_{2}^2 \left(x_{2}^2-4
   x_{2}+3\right)\right)\right)-30 m_{c} m_{s} s
   x_{1} x_{2} \left(x_{1}^8-x_{1}^7
   (x_{2}+3)+x_{1}^6 \left(-9 x_{2}^2+6
   x_{2}+3\right)\right.\right.\\
   \nn &-&  \left.\left. x_{1}^5 \left(20 x_{2}^3-31
   x_{2}^2+11 x_{2}+1\right)+x_{1}^4 x_{2}
   \left(-26 x_{2}^3+57 x_{2}^2-39
   x_{2}+8\right)\right.\right.\\
   \nn &-& \left.\left. x_{1}^3 B^2 x_{2}
   \left(20 x_{2}^2-17 x_{2}+2\right)-x_{1}^2
   B^3 x_{2}^2 (9 x_{2}-4)-x_{1}
   B^2 x_{2}^3 \left(x_{2}^2-4
   x_{2}+2\right) \right. \right.\\
   &+&\left. \left.  B^3 x_{2}^5\right)-30 s^2
   x_{1}^3 x_{2}^3 A^3
   (x_{1}+x_{2})\right\rbrace
\end{eqnarray}
\begin{eqnarray}
\nonumber \rho^{(1)\text{OPE}}_{1,5}(s)&=&\int_{0}^{1} dx \frac{3 m_{0}^2 m_{c} m_{s} \langle
\bar{s}s\rangle (4
   m_{c}-m_{s})}{512 \pi ^4} +\int_{0}^{1}\int_{0}^{1-x_{1}}dx_{1}~dx_{2} \frac{-m_{0}^2 \langle
\bar{s}s\rangle A}{512 \pi ^4
   \left(x_{1}^2+x_{1} B+B
   x_{2}\right)^5}\\
   \nn & \times & \left\lbrace 9
   m_{c}^3 \left(x_{1}^3+x_{1}^2 (2 x_{2}-1)+2
   x_{1} B x_{2}+B
   x_{2}^2\right)^2 -8 m_{c}^2 m_{s} x_{1}
   x_{2} \left(x_{1}^4+x_{1}^3 (3
   x_{2}-2)\right. \right. \\
   \nn &+& \left. \left. x_{1}^2 \left(4 x_{2}^2-5
   x_{2}+1\right)+x_{1} x_{2} \left(3 x_{2}^2-5
   x_{2}+2\right) +B^2 x_{2}^2\right)-15
   m_{c} s x_{1} x_{2}
   \left(x_{1}^4\right. \right. \\
   \nn &+& \left. \left.  x_{1}^3 (3 x_{2}-2) + x_{1}^2
   \left(4 x_{2}^2-5 x_{2}+1\right)+x_{1} x_{2}
   \left(3 x_{2}^2-5 x_{2}+2\right)+B^2
   x_{2}^2\right) \right. \\
   &+& \left. 24 m_{s} s x_{1}^2 x_{2}^2
   A^2\right\rbrace
\end{eqnarray}
\begin{eqnarray}
\nonumber \rho^{(1)\text{OPE}}_{1,6}(s)&=&\int_{0}^{1} dx \frac{\langle
\bar{s}s\rangle^2 \left\lbrace g_{s}^2 m_{c} m_{s}+18 \pi ^2 \left(4 m_{c}^2-3
   m_{c} m_{s}-3 m_{s}^2 (x-1) x\right)\right\rbrace}{1152 \pi ^4}\\
   \nn & +& \int_{0}^{1}\int_{0}^{1-x_{1}}dx_{1}~dx_{2} \frac{-g_{s}^2 \langle
\bar{s}s\rangle^2 x_{1} x_{2} A^2 }{432 \pi ^4
   \left(x_{1}^2+x_{1} B+B x_{2}\right)^5} \left\lbrace m_{c}^2
   \left(x_{1}^3+x_{1}^2 (2 x_{2}-1)\right. \right. \\
   &+& \left. \left.  2 x_{1} B x_{2}+B
   x_{2}^2\right)-3 s x_{1} x_{2} A\right\rbrace
\end{eqnarray}
\begin{eqnarray}
\nonumber \rho^{(1)\text{OPE}}_{1,7}(s)&=&\int_{0}^{1} dx -\frac{\langle
\alpha_{s} \frac{G^{2}}{\pi}\rangle m_{c} \langle
\bar{s}s\rangle}{512 \pi ^2} +\int_{0}^{1}\int_{0}^{1-x_{1}}dx_{1}~dx_{2} \frac{-\langle
\alpha_{s} \frac{G^{2}}{\pi}\rangle m_{c} \langle
\bar{s}s\rangle}{256 \pi ^2
   B \left(x_{1}^2+x_{1} B+B x_{2}\right)^4}\\
   \nn & \times & \left\lbrace 2 x_{1}^6+4 x_{1}^5
   B+x_{1}^4 \left(3 x_{2}^2-5 x_{2}+2\right)+3 x_{1}^3 B^2
   x_{2} +x_{1}^2 x_{2} \left(3 x_{2}^3-9 x_{2}^2+8 x_{2}-2\right) \right. \\
   &+& \left. x_{1}
   (x_{2}-2) B^2 x_{2}^2+2 B^2 x_{2}^4\right\rbrace
\end{eqnarray}
\begin{equation}
 \rho^{(1)\text{OPE}}_{1,8}(s)=0
\end{equation}
\subsection{Spectral densities of molecular tensor current $(J^{(1)}_{\mu\nu})$}
\begin{eqnarray}
\nonumber \rho^{(1)\text{OPE}}_{2,0}(s)&=&\int_{0}^{1}\int_{0}^{1-x_{1}}dx_{1}~dx_{2}\frac{1}{512 \pi ^6
   A \left(x_{1}^2+x_{1}
   B+B x_{2}\right)^8} \left\lbrace s x_{1} x_{2}
   A \right. \\
   \nn &-& \left. m_{c}^2
   \left(x_{1}^3+x_{1}^2 (2 x_{2}-1)+2
   x_{1} B x_{2}+B
   x_{2}^2\right)\right)^2  \left(m_{c}^4
   x_{1} x_{2} \left(x_{1}^3+x_{1}^2 (2
   x_{2}-1) \right. \right. \\
   \nn &+& \left. \left. 2 x_{1} B
   x_{2}+B x_{2}^2\right)^2+6
   m_{c}^3 m_{s}  (x_{1}+x_{2})^2
   \left(x_{1}^2+x_{1}
   B+B x_{2}\right)^3 \right.  \\
   \nn &+& \left.  2
   m_{c}^2 \left(x_{1}^2+x_{1}
   B+B x_{2}\right)
   \left(18 m_{s}^2 \left(x_{1}^2+x_{1}
   B+B x_{2}\right)^3-5 s
   x_{1}^2 x_{2}^2 \right. \right. \\
   \nn & \times & \left. \left. \left(x_{1}^2+x_{1}
   (2 x_{2}-1)+B
   x_{2}\right)\right) -24 m_{c} m_{s} s
   x_{1} x_{2} \left(x_{1}^2+x_{1}
   B+B x_{2}\right)^2 \right. \\
&\times & \left.   \left(x_{1}^2+x_{1} (2
   x_{2}-1)+B x_{2}\right)  +15 s^2
   x_{1}^3 x_{2}^3
   A^2\right\rbrace
\end{eqnarray}
\begin{eqnarray}
\nonumber \rho^{(1)\text{OPE}}_{2,3}(s)&=&\int_{0}^{1}\int_{0}^{1-x_{1}}dx_{1}~dx_{2}\frac{-3 \langle
\bar{s}s\rangle}{32 \pi ^4
   \left(x_{1}^2+x_{1}
   B+B x_{2}\right)^6}  \left\lbrace m_{c}^5
   \left(x_{1}^3+x_{1}^2 (2 x_{2}-1) \right. \right.\\
   \nn &+& \left. \left.  2
   x_{1} B x_{2}+B
   x_{2}^2\right)^3+2 m_{c}^4 m_{s}
   \left(x_{1}^2+x_{1}
   B+B x_{2}\right)^2
   \left(2 x_{1}^5+x_{1}^4 (5
   x_{2}-4) \right. \right. \\
   \nn &+& \left. \left.  x_{1}^3 \left(7 x_{2}^2-11
   x_{2}+2\right) +x_{1}^2 x_{2} \left(7
   x_{2}^2-14 x_{2}+6\right)   +x_{1}
   x_{2}^2 \left(5 x_{2}^2-11
   x_{2}+6\right) \right. \right. \\
   \nn &+& \left. \left. 2 B^2
   x_{2}^3\right) -m_{c}^3
   A
   \left(x_{1}^3+x_{1}^2 (2 x_{2}-1)+2
   x_{1} B x_{2}+B
   x_{2}^2\right)^2 \right. \\
   \nn & \times & \left.  \left(m_{s}^2
   \left(x_{1}^2+x_{1}
   B+B x_{2}\right)+4 s
   x_{1} x_{2}\right)-4 m_{c}^2 m_{s}
   s x_{1} x_{2} \left(x_{1}^7+x_{1}^6
   (x_{2}-4)\right. \right. \\
   \nn &-& \left. \left. 3 x_{1}^5 \left(x_{2}^2+2
   x_{2}-2\right)-4 x_{1}^4 \left(2
   x_{2}^3-3 x_{2}+1\right)+x_{1}^3
   \left(-8 x_{2}^4+5 x_{2}^3+12 x_{2}^2 \right. \right.\right. \\
   \nn &-& \left. \left. \left.  10
   x_{2}+1\right)- 3 x_{1}^2 x_{2}
   \left(x_{2}^4-4 x_{2}^2+4
   x_{2}-1\right)+x_{1} (x_{2}-3)
   B^3 x_{2}^2+B^4
   x_{2}^3\right)\right. \\
   \nn &+& \left.  m_{c} s x_{1} x_{2}
   A^2
   \left(x_{1}^3+x_{1}^2 (2 x_{2}-1)+2
   x_{1} B x_{2}+B
   x_{2}^2\right) \right. \\
    & \times & \left. \left(2 m_{s}^2
   \left(x_{1}^2+x_{1}
   B+B x_{2}\right)+3 s
   x_{1} x_{2}\right)-12 m_{s} s^2
   x_{1}^3 x_{2}^3
   A^3\right\rbrace
\end{eqnarray}
\begin{eqnarray}
\nonumber \rho^{(1)\text{OPE}}_{2,4}(s)&=&\int_{0}^{1}\int_{0}^{1-x_{1}}dx_{1}~dx_{2} \frac{\langle
\alpha_{s} \frac{G^{2}}{\pi}\rangle}{768 \pi ^4
   A \left(x_{1}^2+x_{1}
   B+B x_{2}\right)^6} \left\lbrace 2 m_{c}^4 x_{1}
   x_{2} \left(x_{1}^3+x_{1}^2 (2
   x_{2}-1) \right. \right. \\
   \nn &+&\left. \left.  2 x_{1} B
   x_{2}+B x_{2}^2\right)^2 \left(7
   x_{1}^2+x_{1} (2 x_{2}-3)+x_{2} (7
   x_{2}-3)\right)+6 m_{c}^3 m_{s}
   \left(x_{1}^2+x_{1}
   B+B x_{2}\right)^2 \right. \\
   \nn & \times &\left.  \left(2 x_{1}^6+x_{1}^5 (6
   x_{2}-2)+x_{1}^4 x_{2} (11
   x_{2}-10)+4 x_{1}^3 x_{2} \left(4
   x_{2}^2-5 x_{2}+1\right)\right. \right. \\
   \nn &+& \left. \left.  x_{1}^2
   x_{2}^2 \left(11 x_{2}^2-20
   x_{2}+8\right)+2 x_{1} x_{2}^3  \left(3
   x_{2}^2-5 x_{2}+2\right)+2 B
   x_{2}^5\right) \right. \\
   \nn &+& \left. 2 m_{c}^2
   \left(x_{1}^2+x_{1}
   B+B x_{2}\right) \left(6
   m_{s}^2 \left(x_{1}^2+x_{1}
   B+B x_{2}\right)^2
   \left(x_{1}^4-x_{1}^3+B
   x_{2}^3\right)-s x_{1}^2 x_{2}^2\right. \right. \\
   \nn & \times & \left. \left.  \left(22 x_{1}^4+x_{1}^3 (67
   x_{2}-37)+15 x_{1}^2 \left(6 x_{2}^2-6
   x_{2}+1\right)+x_{1} x_{2} \left(67
   x_{2}^2-90 x_{2}+30\right)\right. \right. \right. \\
   \nn &+& \left. \left. \left.  x_{2}^2
   \left(22 x_{2}^2-37
   x_{2}+15\right)\right)\right)-18 m_{c}
   m_{s} s x_{1} x_{2} \left(x_{1}^8+3
   x_{1}^7 B+x_{1}^6 \left(7
   x_{2}^2-10 x_{2}+3\right)\right. \right. \\
   \nn &+& \left. \left. x_{1}^5
   \left(12 x_{2}^3-25 x_{2}^2+13
   x_{2}-1\right)+x_{1}^4 x_{2} \left(14
   x_{2}^3-39 x_{2}^2+33
   x_{2}-8\right)\right. \right. \\
   \nn &+& \left. \left.  x_{1}^3 B^2
   x_{2} \left(12 x_{2}^2-15
   x_{2}+2\right)+x_{1}^2 B^3
   x_{2}^2 (7 x_{2}-4)+x_{1}
   B^2 x_{2}^3 \right. \right. \\
    &\times & \left. \left.  \left(3 x_{2}^2-4
   x_{2}+2\right)+B^3
   x_{2}^5\right)+27 s^2 x_{1}^3 x_{2}^3
   A^3
   (x_{1}+x_{2})\right\rbrace
\end{eqnarray}
\begin{eqnarray}
\nonumber \rho^{(1)\text{OPE}}_{2,5}(s)&=&\int_{0}^{1} dx \frac{- m_{0}^2 m_{c} m_{s} \langle
\bar{s}s\rangle (6
   m_{c}-m_{s})}{64 \pi ^4} +\int_{0}^{1}\int_{0}^{1-x_{1}}dx_{1}~dx_{2} \frac{m_{0}^2 \langle
\bar{s}s\rangle A}{64 \pi
   ^4 \left(x_{1}^2+x_{1}
   B+B x_{2}\right)^5}\\
   \nn &\times &  \left\lbrace 3 m_{c}^3 \left(x_{1}^3+x_{1}^2
   (2 x_{2}-1)+2 x_{1} B
   x_{2}+B x_{2}^2\right)^2-2
   m_{c}^2 m_{s} x_{1} x_{2} \right. \\
   \nn &\times & \left.   \left(x_{1}^4+x_{1}^3 (3
   x_{2}-2)+x_{1}^2 \left(4 x_{2}^2-5
   x_{2}+1\right)+x_{1} x_{2} \left(3
   x_{2}^2-5 x_{2}+2\right)+B^2
   x_{2}^2\right)\right. \\
   \nn &-& \left.  6 m_{c} s x_{1}
   x_{2} \left(x_{1}^4+x_{1}^3 (3
   x_{2}-2)+x_{1}^2 \left(4 x_{2}^2-5
   x_{2}+1\right)+x_{1} x_{2} \left(3
   x_{2}^2-5 x_{2}+2\right)\right. \right. \\
    &+& \left. \left.  B^2
   x_{2}^2\right)+8 m_{s} s x_{1}^2
   x_{2}^2 A^2\right\rbrace
\end{eqnarray}
\begin{eqnarray}
\nonumber \rho^{(1)\text{OPE}}_{2,6}(s)&=&\int_{0}^{1} dx \frac{\langle
\bar{s}s\rangle^2 \left\lbrace 2 g_{s}^2 m_{c}
   m_{s}+27 \pi ^2 \left(8 m_{c}^2-4
   m_{c} m_{s}-9 m_{s}^2 (x-1)
   x\right)\right\rbrace}{864 \pi ^4}\\
   \nn &+& \int_{0}^{1}\int_{0}^{1-x_{1}}dx_{1}~dx_{2} \frac{g_{s}^2 \langle
\bar{s}s\rangle^2 x_{1} x_{2}
   A^2 }{216 \pi ^4
   \left(x_{1}^2+x_{1}
   B+B x_{2}\right)^5}  \left\lbrace m_{c}^2
   \left(x_{1}^3+x_{1}^2 (2 x_{2}-1) \right. \right. \\
    &+&\left. \left.  2
   x_{1} B x_{2}+B
   x_{2}^2\right)-4 s x_{1} x_{2}
   A\right\rbrace
\end{eqnarray}
\begin{eqnarray}
\nonumber \rho^{(1)\text{OPE}}_{2,7}(s)&=&\int_{0}^{1} dx \frac{\langle
\alpha_{s} \frac{G^{2}}{\pi}\rangle m_{c} \langle
\bar{s}s\rangle}{192 \pi
   ^2} +\int_{0}^{1}\int_{0}^{1-x_{1}}dx_{1}~dx_{2} \frac{\langle
\alpha_{s} \frac{G^{2}}{\pi}\rangle m_{c} \langle
\bar{s}s\rangle}{96 \pi ^2
   B \left(x_{1}^2+x_{1}
   B+B x_{2}\right)^4}\\
   \nn &\times &   \left\lbrace m_{c}
   \left(2 x_{1}^6+4 x_{1}^5
   B+x_{1}^4 \left(-5 x_{2}^2+3
   x_{2}+2\right)-13 x_{1}^3 B^2
   x_{2}+x_{1}^2 x_{2} \left(-13
   x_{2}^3 \right. \right.\right.  \\
    &+& \left. \left. \left.  31 x_{2}^2-24
   x_{2}+6\right)-x_{1} B^2
   x_{2}^2 (7 x_{2}-6) +2 B^2
   x_{2}^4\right)+m_{s} x_{1}
   B x_{2} (x_{1}+x_{2})
   A^2\right\rbrace
\end{eqnarray}
\begin{equation}
 \rho^{(1)\text{OPE}}_{2,8}(s)=0
\end{equation}
\subsection{Spectral densities of tetraquark scalar current $(J^{(2)})$}
\begin{eqnarray}
\nonumber \rho^{(2)\text{OPE}}_{0,0}(s)&=&\int_{0}^{1}\int_{0}^{1-x_{1}}dx_{1}~dx_{2}\frac{-1}{256 \pi ^6 A
   \left(x_{1}^2+x_{1} B+B
   x_{2}\right)^8} \left\lbrace s x_{1} x_{2}
   A\right. \\
   \nn &-&\left.  m_{c}^2
   \left(x_{1}^3+x_{1}^2 (2 x_{2}-1)+2 x_{1}
   B x_{2}+B x_{2}^2\right)\right)^2 \left(m_{c}^4 \left(x_{1}^2-x_{1}+B
   x_{2}\right)\right. \\
   \nn & \times & \left. \left(x_{1}^3+x_{1}^2 (2 x_{2}-1)+2
   x_{1} B x_{2}+B
   x_{2}^2\right)^2 -4 m_{c}^3 m_{s}
   (x_{1}+x_{2})^2 \right. \\
   \nn &\times & \left. \left(x_{1}^2+x_{1}
   B+B x_{2}\right)^3 - 2 m_{c}^2
   \left(x_{1}^2+x_{1} B   +B
   x_{2}\right) \left(12 m_{s}^2
   \left(x_{1}^2+x_{1} B+B
   x_{2}\right)^3\right. \right. \\
   \nn &+&  \left. \left. s x_{1} x_{2} \left(2
   x_{1}^4+x_{1}^3 (3 x_{2}-4)  +x_{1}^2 \left(2
   x_{2}^2-7 x_{2}+2\right)+x_{1} x_{2} \left(3
   x_{2}^2-7 x_{2}+4\right)\right. \right. \right. \\
   \nn &+& \left. \left. \left. 2 B^2
   x_{2}^2\right)\right) +10 m_{c} m_{s} s x_{1}
   x_{2} \left(x_{1}^2+x_{1} B+B
   x_{2}\right)^2 \left(x_{1}^2+x_{1} (2
   x_{2}-1)+B x_{2}\right)\right. \\
   &+&\left.  s^2 x_{1}^2
   x_{2}^2 A^2 \left(3
   x_{1}^2-x_{1} (4 x_{2}+3)+3 B
   x_{2}\right)\right\rbrace
\end{eqnarray}
\begin{eqnarray}
\nonumber \rho^{(2)\text{OPE}}_{0,3}(s)&=&\int_{0}^{1}\int_{0}^{1-x_{1}}dx_{1}~dx_{2}\frac{-\langle
\bar{s}s\rangle}{16 \pi ^4 \left(x_{1}^2+x_{1}
   B+B x_{2}\right)^6} \left\lbrace 2 m_{c}^5 \left(x_{1}^3+x_{1}^2
   (2 x_{2}-1) \right. \right. \\
   \nn &+& \left. \left. 2 x_{1} B x_{2}+B
   x_{2}^2\right)^3+2 m_{c}^4 m_{s}
   \left(x_{1}^2+x_{1} B+B
   x_{2}\right)^2  \left(5 x_{1}^5+x_{1}^4 (13
   x_{2}-10) \right. \right. \\
   \nn &+& \left. \left.   x_{1}^3 \left(18 x_{2}^2-28
   x_{2}+5+ 3 x_{1}^2 x_{2} \left(6 x_{2}^2-12
   x_{2}+5\right)+x_{1} x_{2}^2 \left(13 x_{2}^2-28
   x_{2}+15\right) \right.\right. \right. \\
   \nn &+& \left. \left. \left.  5 B^2 x_{2}^3\right)-2
   m_{c}^3 A
   \left(x_{1}^3+x_{1}^2 (2 x_{2}-1)+2 x_{1}
   B x_{2}+B x_{2}^2\right)^2  \left(m_{s}^2 \left(x_{1}^2+x_{1}
   B+B x_{2}\right) \right. \right. \right. \\
   \nn &+& \left.\left. \left.  3 s x_{1}
   x_{2}\right)-2 m_{c}^2 m_{s} s x_{1} x_{2}
   \left(7 x_{1}^7+x_{1}^6 (19 x_{2}-28)  +6 x_{1}^5
   \left(4 x_{2}^2-13 x_{2}+7\right)\right.\right.\right. \\  
   \nn &+& \left. \left. \left. 2 x_{1}^4 \left(11
   x_{2}^3-57 x_{2}^2+60 x_{2}-14\right)+x_{1}^3
   \left(22 x_{2}^4-124 x_{2}^3  +177 x_{2}^2-82
   x_{2}+7\right)\right. \right.\right. \\
   \nn &+& \left. \left. \left. 3 x_{1}^2 B^2 x_{2}
   \left(8 x_{2}^2-22 x_{2}+7\right)+x_{1}
   B^3 x_{2}^2 (19 x_{2}-21)+7 B^4
   x_{2}^3\right)\right. \right. \\
   \nn &+&\left.\left.  m_{c} s x_{1} x_{2}
   A^2 \left(x_{1}^3+x_{1}^2 (2
   x_{2}-1)  +2 x_{1} B x_{2}+B
   x_{2}^2\right) \left(3 m_{s}^2
   \left(x_{1}^2+x_{1} B+B
   x_{2}\right)\right. \right. \right. \\
   &+& \left. \left. \left. 4 s x_{1} x_{2}\right) +4 m_{s} s^2
   x_{1}^2 x_{2}^2 A^3
   \left(x_{1}^2-x_{1} (4 x_{2}+1)+B
   x_{2}\right)\right\rbrace\right.
\end{eqnarray}
\begin{eqnarray}
\nonumber \rho^{(2)\text{OPE}}_{0,4}(s)&=&\int_{0}^{1}\int_{0}^{1-x_{1}}dx_{1}~dx_{2} \frac{\langle
\alpha_{s} \frac{G^{2}}{\pi}\rangle}{1536 \pi ^4 B
   A \left(x_{1}^2+x_{1}
   B+B x_{2}\right)^6} \left\lbrace m_{c}^4 B
   \left(x_{1}^3+x_{1}^2 (2 x_{2}-1) \right. \right. \\
   \nn &+& \left. \left. 2 x_{1}
   B x_{2}+B x_{2}^2\right)^2  \left(4 x_{1}^4+x_{1}^3 (53 x_{2}-16)+x_{1}^2
   \left(50 x_{2}^2-69 x_{2}+12\right) \right. \right. \\
   \nn &+& \left. \left. x_{1} x_{2}
   \left(53 x_{2}^2-69 x_{2}+24\right)  +4 x_{2}^2
   \left(x_{2}^2-4 x_{2}+3\right)\right) \right.\\
   \nn &+& \left. 4 m_{c}^3
   m_{s} \left(x_{1}^2+x_{1}
   B+B x_{2}\right)^2 \left(4
   x_{1}^7+x_{1}^6 (9 x_{2}-5)+x_{1}^5 \left(15
   x_{2}^2-13 x_{2}-2\right)\right. \right. \\
   \nn &+&\left. \left.  x_{1}^4 \left(23
   x_{2}^3-36 x_{2}^2+10 x_{2}+3\right) +3 x_{1}^3
   x_{2} \left(10 x_{2}^3-23 x_{2}^2+15
   x_{2}-2\right) \right. \right. \\
   \nn &+& \left.\left. x_{1}^2 x_{2}^2 \left(19
   x_{2}^3-58 x_{2}^2+57 x_{2}-18\right) +x_{1}
   B^2 x_{2}^3 (7 x_{2}-6)+B^2
   x_{2}^4 (x_{2}+3)\right)\right. \\
   \nn &+&\left.  4 m_{c}^2
   \left(x_{1}^2+x_{1} B+B
   x_{2}\right) \left(8 m_{s}^2 \left(x_{1}^2+x_{1}
   B+B x_{2}\right)^2
   \left(x_{1}^5+2 x_{1}^4 B+x_{1}^3
   B^2 \right. \right. \right.\\
   \nn &+& \left. \left. \left. B^2 x_{2}^3\right)-s x_{1}
   B x_{2} \left(6 x_{1}^6+x_{1}^5 (52
   x_{2}-21)+x_{1}^4 \left(133 x_{2}^2-130
   x_{2}+24\right) \right. \right. \right.\\
   \nn &+& \left. \left. \left. 3 x_{1}^3 \left(58 x_{2}^3-88
   x_{2}^2+35 x_{2}-3\right)+x_{1}^2 x_{2} \left(133
   x_{2}^3-264 x_{2}^2+162 x_{2}-27\right) \right. \right. \right. \\
   \nn &+& \left. \left. \left.  x_{1}
   x_{2}^2 \left(52 x_{2}^3-130 x_{2}^2+105
   x_{2}-27\right)+3 B^2 x_{2}^3 (2
   x_{2}-3)\right)\right) \right. \\
   \nn &-& \left. 6 m_{c} m_{s} s x_{1}
   x_{2} \left(4 x_{1}^9+13 x_{1}^8 B+12
   x_{1}^7 \left(2 x_{2}^2-3 x_{2}+1\right)+x_{1}^6
   \left(37 x_{2}^3 \right. \right. \right. \\
   \nn &-& \left. \left. \left. 75 x_{2}^2+36 x_{2}+2\right)+2
   x_{1}^5 \left(25 x_{2}^4-71 x_{2}^3+61 x_{2}^2-11
   x_{2}-4\right)+x_{1}^4 B^2 \right. \right. \\
   \nn &\times & \left. \left.   \left(52
   x_{2}^3-94 x_{2}^2+21 x_{2}+3\right)+2 x_{1}^3
   B^3 x_{2} \left(19 x_{2}^2-31
   x_{2}+3\right)+x_{1}^2 B^3 x_{2}^2 \right. \right. \\
   \nn &\times & \left. \left.  \left(17 x_{2}^2-38 x_{2}+18\right)+2 x_{1}
   B^3 x_{2}^3 \left(2 x_{2}^2-3
   x_{2}+3\right)+B^4 x_{2}^4
   (x_{2}+3)\right)\right. \\
    &+& \left. 6 s^2 x_{1}^2 B x_{2}^2
   A^3 \left(4 x_{1}^3+x_{1}^2 (13
   x_{2}-4)+x_{1} x_{2} (13 x_{2}-8)+4 B
   x_{2}^2\right)\right\rbrace
\end{eqnarray}
\begin{eqnarray}
\nonumber \rho^{(2)\text{OPE}}_{0,5}(s)&=&\int_{0}^{1} dx  \frac{ m_{0}^2 m_{s} \langle
\bar{s}s\rangle \left(-8
   m_{c}^2+m_{c} m_{s}-3 s (x-1) x\right)}{48 \pi ^4} \\
   \nn &+& \int_{0}^{1}\int_{0}^{1-x_{1}}dx_{1}~dx_{2} \frac{m_{0}^2 \langle
\bar{s}s\rangle A}{96 \pi ^4
   \left(x_{1}^2+x_{1} B+B
   x_{2}\right)^5} \left\lbrace 6
   m_{c}^3 \left(x_{1}^3+x_{1}^2 (2 x_{2}-1) \right. \right. \\
   \nn &+& \left. \left. 2
   x_{1} B x_{2}+B
   x_{2}^2\right)^2-8 m_{c}^2 m_{s} x_{1}
   x_{2} \left(x_{1}^4+x_{1}^3 (3
   x_{2}-2)+x_{1}^2 \left(4 x_{2}^2-5
   x_{2}+1\right)\right. \right. \\
   \nn &+& \left. \left. x_{1} x_{2} \left(3 x_{2}^2-5
   x_{2}+2\right)+B^2 x_{2}^2\right)  -9
   m_{c} s x_{1} x_{2} \left(x_{1}^4+x_{1}^3 (3
   x_{2}-2)\right. \right.  \\
   \nn &+& \left. \left.  x_{1}^2 \left(4 x_{2}^2-5
   x_{2}+1\right)+ x_{1} x_{2} \left(3 x_{2}^2-5
   x_{2}+2\right) +B^2 x_{2}^2\right)\right. \\
    &+& \left. 18
   m_{s} s x_{1}^2 x_{2}^2
   A^2\right\rbrace
\end{eqnarray}
\begin{eqnarray}
\nonumber \rho^{(2)\text{OPE}}_{0,6}(s)&=&\int_{0}^{1} dx \frac{\langle
\bar{s}s\rangle^2}{324 \pi ^4}  \left\lbrace g_{s}^2 \left(2 m_{c}^2-m_{c}
   m_{s}+3 s (x-1) x\right)-  54 \pi ^2 \left(2
   m_{c}^2-m_{c} m_{s}\right. \right. \\
   \nn &-& \left. \left. 3 m_{s}^2 (x-1)
   x\right)\right\rbrace +\int_{0}^{1}\int_{0}^{1-x_{1}}dx_{1}~dx_{2} \frac{g_{s}^2 \langle
\bar{s}s\rangle^2 x_{1} x_{2}
   A^2}{324 \pi ^4
   \left(x_{1}^2+x_{1} B+B
   x_{2}\right)^5}\\
    &\times &  \left\lbrace 4 m_{c}^2
   \left(x_{1}^3+x_{1}^2 (2 x_{2}-1)+2 x_{1}
   B x_{2}+B x_{2}^2\right)-9 s
   x_{1} x_{2} A\right\rbrace
\end{eqnarray}
\begin{eqnarray}
\nonumber \rho^{(2)\text{OPE}}_{0,7}(s)&=&\int_{0}^{1} dx \frac{\langle
\alpha_{s} \frac{G^{2}}{\pi}\rangle  \langle
\bar{s}s\rangle (m_{c}-3 m_{s})}{144
   \pi ^2} +\int_{0}^{1}\int_{0}^{1-x_{1}}dx_{1}~dx_{2} \frac{\langle
\alpha_{s} \frac{G^{2}}{\pi}\rangle m_{c} \langle
\bar{s}s\rangle}{288 \pi ^2 B
   \left(x_{1}^2+x_{1} B+B
   x_{2}\right)^4}\\
   \nn &\times &   \left\lbrace m_{c} \left(8
   x_{1}^6+19 x_{1}^5 B-2 x_{1}^4
   \left(x_{2}^2+6 x_{2}-7\right)-x_{1}^3 \left(19
   x_{2}^3-38 x_{2}^2+16 x_{2}+3\right) \right. \right.\\
   \nn &+&  \left. \left.  x_{1}^2
   x_{2} \left(-19 x_{2}^3+43 x_{2}^2-33
   x_{2}+9\right)-x_{1} B^2 x_{2}^2 (10
   x_{2}-9)+B^2 x_{2}^3 (11
   x_{2}-3)\right) \right. \\
    &-& \left. 3 m_{s} x_{1} B
   x_{2} A^2
   (x_{1}+x_{2})\right\rbrace
\end{eqnarray}
\begin{equation}
 \rho^{(2)\text{OPE}}_{0,8}(s)=0
\end{equation}
\subsection{Spectral densities of tetraquark axialvector current $(J^{(2)}_{\mu})$}
\begin{eqnarray}
\nonumber \rho^{(2)\text{OPE}}_{1,0}(s)&=&\int_{0}^{1}\int_{0}^{1-x_{1}}dx_{1}~dx_{2}\frac{-1}{3072 \pi ^6
   A \left(x_{1}^2+x_{1}
   B+B x_{2}\right)^8}  \left\lbrace s x_{1} x_{2}
   A \right. \\
   \nn &-& \left. m_{c}^2 \left(x_{1}^3+x_{1}^2
   (2 x_{2}-1)+2 x_{1} B x_{2}+B
   x_{2}^2\right)\right)^2  \left(3 m_{c}^4 x_{1}
   x_{2} \right. \\
   \nn &\times & \left.  \left(x_{1}^3+x_{1}^2 (2 x_{2}-1)+2 x_{1}
   B x_{2}+B x_{2}^2\right)^2 +18
   m_{c}^3 m_{s} (x_{1}+x_{2})^2\right.
   \\
   \nn &\times & \left. \left(x_{1}^2+x_{1} B+B
   x_{2}\right)^3+2 m_{c}^2 \left(x_{1}^2+x_{1}
   B  +B x_{2}\right) \right. \\
   \nn &\times & \left. \left(36 m_{s}^2
   \left(x_{1}^2+x_{1} B+B
   x_{2}\right)^3-13 s x_{1}^2 x_{2}^2
   \left(x_{1}^2+x_{1} (2 x_{2}-1) +B
   x_{2}\right)\right)\right. \\
   \nn &-& \left. 54 m_{c} m_{s} s x_{1}
   x_{2} \left(x_{1}^2+x_{1} B+B
   x_{2}\right)^2 \left(x_{1}^2+x_{1} (2
   x_{2}-1) +B x_{2}\right)\right. \\
    &+&\left.  35 s^2 x_{1}^3
   x_{2}^3 A^2\right\rbrace
\end{eqnarray}
\begin{eqnarray}
\nonumber \rho^{(2)\text{OPE}}_{1,3}(s)&=&\int_{0}^{1}\int_{0}^{1-x_{1}}dx_{1}~dx_{2}\frac{\langle
\bar{s}s\rangle}{64 \pi ^4
   \left(x_{1}^2+x_{1} B+B
   x_{2}\right)^6} \left\lbrace 3 m_{c}^5 \left(x_{1}^3+x_{1}^2 (2
   x_{2}-1)\right.\right. \\
   \nn &+&\left.  \left. 2 x_{1} B x_{2}+B
   x_{2}^2\right)^3  +2 m_{c}^4 m_{s}
   \left(x_{1}^2+x_{1} B+B
   x_{2}\right)^2 \left(4 x_{1}^5+x_{1}^4 (9
   x_{2}-8)\right. \right. \\
   \nn &+& \left. \left.  x_{1}^3 \left(11 x_{2}^2-21
   x_{2}+4\right)+x_{1}^2 x_{2} \left(11 x_{2}^2-26
   x_{2}+12\right)+3 x_{1} x_{2}^2 \left(3 x_{2}^2-7
   x_{2}+4\right) \right. \right. \\
   \nn &+& \left. \left.  4 B^2
   x_{2}^3\right)-m_{c}^3 A
   \left(x_{1}^3+x_{1}^2 (2 x_{2}-1)+2 x_{1}
   B x_{2}+B x_{2}^2\right)^2  \right. \\
   \nn &\times & \left.  \left(3
   m_{s}^2 \left(x_{1}^2+x_{1}
   B+B x_{2}\right)+10 s x_{1}
   x_{2}\right)-8 m_{c}^2 m_{s} s x_{1} x_{2}
   \left(x_{1}^7-4 x_{1}^6+x_{1}^5 \right. \right. \\
   \nn &\times & \left. \left.   \left(-7 x_{2}^2-3
   x_{2}+6\right)+x_{1}^4 \left(-15 x_{2}^3+10
   x_{2}^2+9 x_{2}-4\right)+x_{1}^3 \left(-15
   x_{2}^4+19 x_{2}^3 \right. \right. \right. \\
   \nn &+& \left. \left. \left. 4 x_{2}^2 - 9
   x_{2}+1\right)-x_{1}^2 B^2 x_{2} \left(7
   x_{2}^2+4 x_{2}-3\right)-3 x_{1} B^3
   x_{2}^2+B^4 x_{2}^3\right) \right. \\
   \nn &+& \left. m_{c} s
   x_{1} x_{2} A^2
   \left(x_{1}^3+x_{1}^2 (2 x_{2}-1)+2 x_{1}
   B x_{2}+B x_{2}^2\right) \right.  \\
    &\times & \left.  \left(5
   m_{s}^2 \left(x_{1}^2+x_{1}
   B+B x_{2}\right)+7 s x_{1}
   x_{2}\right)-30 m_{s} s^2 x_{1}^3 x_{2}^3
   A^3\right\rbrace
\end{eqnarray}
\begin{eqnarray}
\nonumber \rho^{(2)\text{OPE}}_{1,4}(s)&=&\int_{0}^{1}\int_{0}^{1-x_{1}}dx_{1}~dx_{2} \frac{-\langle
\alpha_{s} \frac{G^{2}}{\pi}\rangle}{18432 \pi ^4
   A^2 \left(x_{1}^2+x_{1}
   B+B x_{2}\right)^6}\\
   \nn &\times &  \left\lbrace m_{c}^4
   \left(x_{1}^2+x_{1} B+B
   x_{2}\right)^2 \left(72 x_{1}^7+x_{1}^6 (481
   x_{2}-252) +x_{1}^5 \left(1277 x_{2}^2\right. \right. \right. \\
   \nn &-&  \left. \left. \left. 1420
   x_{2}+324\right) +x_{1}^4 \left(1939 x_{2}^3-3274
   x_{2}^2+1533 x_{2}-180\right) \right. \right. \\
   \nn &+& \left. \left. x_{1}^3 \left(1939
   x_{2}^4-4212 x_{2}^3+2961 x_{2}^2-702
   x_{2}+36\right)+x_{1}^2 x_{2} \left(1277
   x_{2}^4-3274 x_{2}^3 \right. \right. \right. \\
   \nn &+& \left. \left. \left.  2961 x_{2}^2-1044
   x_{2}+108\right)+x_{1} x_{2}^2 \left(481
   x_{2}^4-1420 x_{2}^3+1533 x_{2}^2-702
   x_{2}+108\right) \right. \right. \\
   \nn &+& \left. \left. 36 B^3 x_{2}^3 (2
   x_{2}-1)\right)+9 m_{c}^3 m_{s}
   \left(x_{1}^2+x_{1} B+B
   x_{2}\right)^2  \left(8 x_{1}^7+x_{1}^6 (13
   x_{2}-16) \right. \right. \\
   \nn &-& \left. \left. 2 x_{1}^5 \left(4 x_{2}^2+7
   x_{2}-4\right)+x_{1}^4 x_{2} \left(-25 x_{2}^2+29
   x_{2}-3\right) \right. \right. \\
   \nn &+& \left. \left. x_{1}^3 x_{2} \left(-25 x_{2}^3+46
   x_{2}^2-29 x_{2}+4\right)+x_{1}^2 x_{2}^2 \left(-8
   x_{2}^3+29 x_{2}^2-29 x_{2}+8\right) \right. \right. \\
   \nn &+& \left. \left. x_{1}
   x_{2}^3 \left(13 x_{2}^3-14 x_{2}^2-3
   x_{2}+4\right)+8 B^2
   x_{2}^5\right)+m_{c}^2 \left(x_{1}^3+2 x_{1}^2
   B \right. \right. \\
   \nn &+& \left. \left. x_{1} \left(2 x_{2}^2-3
   x_{2}+1\right)+B^2 x_{2}\right) \left(6
   m_{s}^2 \left(x_{1}^2+x_{1}
   B+B x_{2}\right)^2 \right. \right.  \\
   \nn &\times  & \left. \left. \left(16
   x_{1}^4-x_{1}^3 (3 x_{2}+16)+3 x_{1}^2 (1-2
   x_{2}) x_{2}-3 x_{1} B x_{2}^2+16
   B x_{2}^3\right) \right. \right. \\
   \nn &-& \left. \left. s x_{1} x_{2} \left(72
   x_{1}^6+x_{1}^5 (431 x_{2}-252)+4 x_{1}^4 \left(201
   x_{2}^2-299 x_{2}+81\right)+x_{1}^3 \left(895
   x_{2}^3 \right. \right. \right. \right.\\
   \nn &-& \left. \left. \left. \left. 2014 x_{2}^2+1239 x_{2}-180\right)+2
   x_{1}^2 \left(402 x_{2}^4-1007 x_{2}^3+894
   x_{2}^2-273 x_{2}+18\right) \right. \right. \right. \\
   \nn &+& \left. \left. \left. x_{1} x_{2} \left(431
   x_{2}^4-1196 x_{2}^3+1239 x_{2}^2-546
   x_{2}+72\right)+36 B^3 x_{2}^2 (2
   x_{2}-1)\right)\right) \right.\\
   \nn &-&\left.  15 m_{c} m_{s} s x_{1}
   x_{2} A^2 \left(8 x_{1}^7-x_{1}^6
   (3 x_{2}+16)+x_{1}^5 \left(-17 x_{2}^2+10
   x_{2}+8\right) \right. \right. \\
   \nn &+& \left. \left. x_{1}^4 x_{2} \left(-39 x_{2}^2+42
   x_{2}-11\right)+x_{1}^3 x_{2} \left(-39 x_{2}^3+68
   x_{2}^2-33 x_{2}+4\right) \right. \right. \\
   \nn &-& \left. \left. x_{1}^2 B^2
   x_{2}^2 (17 x_{2}-8)-x_{1} B^2 x_{2}^3
   (3 x_{2}-4)+8 B^2 x_{2}^5\right) \right. \\
   \nn &+& \left. 6 s
   x_{1}^2 x_{2}^2 A^3 \left(5
   m_{s}^2 \left(x_{1}^2+x_{1}
   B+B x_{2}\right)^2 \right. \right. \\
    &-& \left. \left. 11 s x_{1}
   x_{2} \left(x_{1}^2+x_{1} (2 x_{2}-1)+B
   x_{2}\right)\right)\right\rbrace
\end{eqnarray}
\begin{eqnarray}
\nonumber \rho^{(2)\text{OPE}}_{1,5}(s)&=&\int_{0}^{1} dx  \frac{  m_{0}^2  \langle
\bar{s}s\rangle }{128 \pi ^4} \lbrace -m_{c} m_{s} (4
   m_{c}-m_{s})\rbrace\\
   \nn & +&\int_{0}^{1}\int_{0}^{1-x_{1}}dx_{1}~dx_{2} \frac{m_{0}^2 \langle
\bar{s}s\rangle  A}{384 \pi ^4
   \left(x_{1}^2+x_{1} B+B
   x_{2}\right)^5}  \left\lbrace 9
   m_{c}^3 \left(x_{1}^3+  x_{1}^2 (2 x_{2}-1)\right. \right. \\
   \nn &+&\left. \left.  2
   x_{1} B x_{2}+B
   x_{2}^2\right)^2-8 m_{c}^2 m_{s} x_{1} x_{2}
   \left(x_{1}^4+x_{1}^3 (3 x_{2}-2)+x_{1}^2 \left(4
   x_{2}^2-5 x_{2}+1\right) \right. \right. \\  
   \nn &+&\left. \left.  x_{1} x_{2} \left(3
   x_{2}^2-5 x_{2}+2\right)+B^2
   x_{2}^2\right)-15 m_{c} s x_{1} x_{2} 
   \left(x_{1}^4+x_{1}^3 (3 x_{2}-2)\right. \right.\\
   \nn &+& \left. \left. x_{1}^2 \left(4
   x_{2}^2-5 x_{2}+1\right)+x_{1} x_{2} \left(3
   x_{2}^2-5 x_{2}+2\right)+B^2
   x_{2}^2\right) \right. \\
    &+& \left. 24 m_{s} s x_{1}^2 x_{2}^2
   A^2\right\rbrace
\end{eqnarray}
\begin{eqnarray}
\nonumber \rho^{(2)\text{OPE}}_{1,6}(s)&=&\int_{0}^{1} dx \frac{\langle
\bar{s}s\rangle^2 \left(g_{s}^2 m_{c} m_{s}+18 \pi ^2
   \left(4 m_{c}^2-3 m_{c} m_{s}-3 m_{s}^2 (x-1)
   x\right)\right)}{864 \pi ^4}\\
   \nn & +& \int_{0}^{1}\int_{0}^{1-x_{1}}dx_{1}~dx_{2} \frac{-g_{s}^2 \langle
\bar{s}s\rangle^2 x_{1} x_{2}
   A^2}{324 \pi ^4
   \left(x_{1}^2+x_{1} B+B
   x_{2}\right)^5}  \left\lbrace m_{c}^2
   \left(x_{1}^3+x_{1}^2 (2 x_{2}-1)\right. \right. \\
   &+& \left. \left. 2 x_{1}
   B x_{2}+B x_{2}^2\right)-3 s
   x_{1} x_{2} A\right\rbrace
\end{eqnarray}
\begin{eqnarray}
\nonumber \rho^{(2)\text{OPE}}_{1,7}(s)&=&\int_{0}^{1} dx \frac{-\langle
\alpha_{s} \frac{G^{2}}{\pi}\rangle  \langle
\bar{s}s\rangle m_{c}}{384 \pi ^2} + \int_{0}^{1}\int_{0}^{1-x_{1}}dx_{1}~dx_{2} \frac{\langle
\alpha_{s} \frac{G^{2}}{\pi}\rangle  \langle
\bar{s}s\rangle}{1536
   \pi ^2 }\\
   \nn &\times &  \frac{1}{B \left(x_{1}^2+x_{1}
   B+B x_{2}\right)^4} \left\lbrace m_{c} \left(-16
   x_{1}^6-32 x_{1}^5 B+ x_{1}^4 \left(-11
   x_{2}^2+27 x_{2}-16\right)\right. \right. \\
   \nn &+& \left. \left. 2 x_{1}^3 B
   x_{2}^2 + x_{1}^2 x_{2} \left(2 x_{2}^3+9
   x_{2}^2-16 x_{2}+5\right)+5 x_{1} B^2
   x_{2}^2 (x_{2}+1)\right. \right. \\
    &-& \left. \left. 16 B^2 x_{2}^4\right)+ 4
   m_{s} x_{1} B x_{2} \left(x_{1}^3+2
   x_{1}^2 B +x_{1} \left(2 x_{2}^2-3
   x_{2}+1\right)+B^2 x_{2}\right)\right\rbrace
\end{eqnarray}
\begin{equation}
 \rho^{(2)\text{OPE}}_{1,8}(s)=0
\end{equation}

\subsection{Spectral densities of tetraquark tensor current $(J^{(2)}_{\mu\nu})$}
\begin{eqnarray}
\nonumber \rho^{(2)\text{OPE}}_{2,0}(s)&=&\int_{0}^{1}\int_{0}^{1-x_{1}}dx_{1}~dx_{2}\frac{\left( s x_{1} x_{2} A-m_{c}^2
   \left(x_{1}^3+x_{1}^2 (2 x_{2}-1)+2 x_{1} B
   x_{2}+B x_{2}^2\right)\right)^2}{384 \pi ^6
   A \left(x_{1}^2+x_{1}
   B+B x_{2}\right)^8}  \\
   \nn &\times &\left\lbrace m_{c}^4
   x_{1} x_{2} \left(x_{1}^3+x_{1}^2 (2 x_{2}-1)+2
   x_{1} B x_{2}+B x_{2}^2\right)^2+6
   m_{c}^3 m_{s} (x_{1}+x_{2})^2\right. \\
  \nn &\times & \left.  \left(x_{1}^2+x_{1} B+B
   x_{2}\right)^3+2 m_{c}^2 \left(x_{1}^2+x_{1}
   B+B x_{2}\right) \left(18 m_{s}^2
   \left(x_{1}^2+x_{1} B+B
   x_{2}\right)^3\right. \right. \\
   \nn &-& \left. \left. 5 s x_{1}^2 x_{2}^2
   \left(x_{1}^2+x_{1} (2 x_{2}-1)+B
   x_{2}\right)\right)-24 m_{c} m_{s} s x_{1} x_{2}
   \left(x_{1}^2+x_{1} B+B
   x_{2}\right)^2\right. \\
    &\times &\left.  \left(x_{1}^2+x_{1} (2
   x_{2}-1)+B x_{2}\right)+15 s^2 x_{1}^3
   x_{2}^3 A^2\right\rbrace
\end{eqnarray}
\begin{eqnarray}
\nonumber \rho^{(2)\text{OPE}}_{2,3}(s)&=&\int_{0}^{1}\int_{0}^{1-x_{1}}dx_{1}~dx_{2}\frac{\langle
\bar{s}s\rangle}{8 \pi ^4 \left(x_{1}^2+x_{1}
   B+B x_{2}\right)^6} \left\lbrace m_{c}^5 \left(-\left(x_{1}^3+x_{1}^2
   (2 x_{2}-1)\right. \right. \right.  \\
   \nn &\times & \left. \left. \left. +2 x_{1} B x_{2}+B
   x_{2}^2\right)^3\right) +2 m_{c}^4 m_{s} x_{1}
   x_{2} A \left(x_{1}^3+x_{1}^2 (2
   x_{2}-1)+2 x_{1} B x_{2}+B
   x_{2}^2\right)^2\right. \\
   \nn &+&\left.  4 m_{c}^3 s x_{1} x_{2}
   A \left(x_{1}^3+x_{1}^2 (2 x_{2}-1)+2
   x_{1} B x_{2}+B x_{2}^2\right)^2- 12
   m_{c}^2 m_{s} s x_{1}^2 x_{2}^2
   A^2 \right. \\
   \nn &\times & \left. \left(x_{1}^3+x_{1}^2 (2
   x_{2}-1)+2 x_{1} B x_{2}+B
   x_{2}^2\right)-3 m_{c} s^2 x_{1}^2 x_{2}^2
   A^2 \left(x_{1}^3+x_{1}^2 (2
   x_{2}-1)\right. \right. \\
 &+&\left. \left.  2 x_{1} B x_{2}+B
   x_{2}^2\right)+12 m_{s} s^2 x_{1}^3 x_{2}^3
   A^3\right\rbrace
\end{eqnarray}
\begin{eqnarray}
\nonumber \rho^{(2)\text{OPE}}_{2,4}(s)&=&\int_{0}^{1}\int_{0}^{1-x_{1}}dx_{1}~dx_{2} \frac{\langle
\alpha_{s} \frac{G^{2}}{\pi}\rangle}{4608 \pi ^4 B
   A \left(x_{1}^2+x_{1}
   B+B x_{2}\right)^6} \left\lbrace m_{c}^4 x_{1} B
   x_{2} \left(52 x_{1}^2\right. \right. \\
   \nn &+ & \left. \left. x_{1} (66-97 x_{2})+ 52
   x_{2}^2+66 x_{2}-54\right) \left(x_{1}^3+x_{1}^2 (2
   x_{2}-1)+2 x_{1} B x_{2}+B
   x_{2}^2\right)^2\right. \\
   \nn &+&\left.  12 m_{c}^3 m_{s}
   \left(x_{1}^2+x_{1} B+B
   x_{2}\right)^2 \left(8 x_{1}^7+2 x_{1}^6 (13
   x_{2}-9)+x_{1}^5 \left(37 x_{2}^2-49
   x_{2}+12\right)\right. \right. \\
   \nn &+&\left. \left.  x_{1}^4 \left(35 x_{2}^3-68 x_{2}^2+35
   x_{2}-2\right)+x_{1}^3 x_{2} \left(40 x_{2}^3-91
   x_{2}^2+63 x_{2}-12\right)\right. \right. \\
   \nn &+&\left. \left.  x_{1}^2 x_{2}^2 \left(27
   x_{2}^3-78 x_{2}^2+71 x_{2}-20\right)+3 x_{1}
   B^2 x_{2}^3 (7 x_{2}-4)+2 B^2
   x_{2}^4 (5 x_{2}-1)\right)\right. \\
   \nn &+&\left.  4 m_{c}^2
   \left(x_{1}^2+x_{1} B+B
   x_{2}\right) \left(3 m_{s}^2 \left(x_{1}^2+x_{1}
   B+B x_{2}\right)^2 \left(8 x_{1}^5+16
   x_{1}^4 B\right. \right. \right. \\
   \nn &+& \left. \left. \left. x_{1}^3 \left(7 x_{2}^2-15
   x_{2}+8\right)-x_{1}^2 B^2 x_{2}-x_{1}
   B^2 x_{2}^2+8 B^2 x_{2}^3\right)-4 s
   x_{1}^2 B x_{2}^2 \left(7 x_{1}^4 \right. \right. \right. \\
   \nn &+&\left. \left. \left.  5 x_{1}^3
   (x_{2}+1)+x_{1}^2 \left(-4 x_{2}^2+38
   x_{2}-24\right)+x_{1} \left(5 x_{2}^3+38 x_{2}^2-48
   x_{2}+12\right)\right. \right. \right. \\
   \nn &+& \left. \left. \left. x_{2} \left(7 x_{2}^3+5 x_{2}^2-24
   x_{2}+12\right)\right)\right)-18 m_{c} m_{s} s x_{1}
   x_{2} \left(8 x_{1}^9+34 x_{1}^8 B\right. \right. \\
   \nn &+&\left. \left.  7
   x_{1}^7 \left(9 x_{2}^2-17 x_{2}+8\right)+x_{1}^6
   \left(78 x_{2}^3-199 x_{2}^2+165
   x_{2}-44\right)+x_{1}^5 \left(76 x_{2}^4\right. \right. \right. \\
   \nn &-& \left. \left. \left. 249
   x_{2}^3+278 x_{2}^2-121 x_{2}+16\right)+x_{1}^4
   B^2 \left(62 x_{2}^3-131 x_{2}^2+49
   x_{2}-2\right)\right. \right. \\
   \nn &+& \left. \left. x_{1}^3 B^3 x_{2} \left(52
   x_{2}^2-73 x_{2}+12\right)+x_{1}^2 B^3
   x_{2}^2 \left(38 x_{2}^2-57 x_{2}+20\right)\right. \right. \\
   \nn &+& \left.\left. x_{1}
   B^3 x_{2}^3 \left(23 x_{2}^2-27
   x_{2}+12\right)+2 B^4 x_{2}^4 (5
   x_{2}-1)\right)+s^2 x_{1}^3 B x_{2}^3
   A^2 \left(18 x_{1}^2\right. \right.\\
    &+&\left. \left. 11 x_{1}
   (x_{2}+12)+6 \left(3 x_{2}^2+22
   x_{2}-25\right)\right)\right\rbrace
\end{eqnarray}
\begin{eqnarray}
\nonumber \rho^{(2)\text{OPE}}_{2,5}(s)&=&\int_{0}^{1} dx  \frac{ - m_{0}^2  \langle
\bar{s}s\rangle m_{c} m_{s} (6
   m_{c}-m_{s})}{48 \pi ^4}  + \int_{0}^{1}\int_{0}^{1-x_{1}}dx_{1}~dx_{2} \frac{m_{0}^2 \langle
\bar{s}s\rangle  A}{48 \pi ^4
   \left(x_{1}^2+x_{1} B+B
   x_{2}\right)^5} \\
   \nn &\times & \left\lbrace 3 m_{c}^3
   \left(x_{1}^3+x_{1}^2 (2 x_{2}-1)+2 x_{1} B
   x_{2}+B x_{2}^2\right)^2-2 m_{c}^2 m_{s}
   x_{1} x_{2} \left(x_{1}^4+x_{1}^3 (3
   x_{2}-2)\right. \right. \\
   \nn &+& \left. \left. x_{1}^2 \left(4 x_{2}^2-5
   x_{2}+1\right)+x_{1} x_{2} \left(3 x_{2}^2-5
   x_{2}+2\right)+B^2 x_{2}^2\right)-6 m_{c} s
   x_{1} x_{2} \left(x_{1}^4+x_{1}^3 (3
   x_{2}-2)\right. \right. \\
    &+&\left. \left.  x_{1}^2 \left(4 x_{2}^2-5
   x_{2}+1\right)+x_{1} x_{2} \left(3 x_{2}^2-5
   x_{2}+2\right)+B^2 x_{2}^2\right)+8 m_{s} s
   x_{1}^2 x_{2}^2 A^2\right\rbrace
\end{eqnarray}
\begin{eqnarray}
\nonumber \rho^{(2)\text{OPE}}_{2,6}(s)&=&\int_{0}^{1} dx \frac{-\langle
\bar{s}s\rangle^2  \left(2 g_{s}^2 m_{c} m_{s}+27 \pi ^2
   \left(8 m_{c}^2-4 m_{c} m_{s}-9 m_{s}^2 (x-1)
   x\right)\right)}{648 \pi ^4}\\
   \nn & +& \int_{0}^{1}\int_{0}^{1-x_{1}}dx_{1}~dx_{2} \frac{g_{s}^2 \langle
\bar{s}s\rangle^2 x_{1} x_{2}
   A^2}{162 \pi ^4 \left(x_{1}^2+x_{1}
   B+B x_{2}\right)^5}\left\lbrace m_{c}^2
   \left(x_{1}^3+x_{1}^2 (2 x_{2}-1)\right. \right.\\
   &+& \left. \left. 2 x_{1} B
   x_{2}+B x_{2}^2\right)-4 s x_{1} x_{2}
   A\right\rbrace
\end{eqnarray}
\begin{eqnarray}
\nonumber \rho^{(2)\text{OPE}}_{2,7}(s)&=&\int_{0}^{1} dx \frac{\langle
\alpha_{s} \frac{G^{2}}{\pi}\rangle  \langle
\bar{s}s\rangle m_{c}}{144 \pi ^2} + \int_{0}^{1}\int_{0}^{1-x_{1}}dx_{1}~dx_{2} \frac{\langle
\alpha_{s} \frac{G^{2}}{\pi}\rangle  \langle
\bar{s}s\rangle}{9216 \pi ^2 B
   \left(x_{1}^2+x_{1} B+B
   x_{2}\right)^4 }\\
   \nn &\times &  \left\lbrace 2 m_{c} \left(128
   x_{1}^6+240 x_{1}^5 B+x_{1}^4 \left(-153
   x_{2}^2+57 x_{2}+96\right)-x_{1}^3 B^2 (473
   x_{2}-16)\right. \right. \\
   \nn &-&\left. \left.  16 x_{1}^2 x_{2} \left(29 x_{2}^3-67
   x_{2}^2+51 x_{2}-13\right)-x_{1} B^2
   x_{2}^2 (263 x_{2}-199)\right. \right. \\
   \nn &+& \left. \left. B^2 x_{2}^3 (121
   x_{2}+7)\right)+m_{s} x_{1} B x_{2}
   \left(80 x_{1}^3+5 x_{1}^2 (49 x_{2}-32)\right. \right. \\
   &+& \left. \left. 4 x_{1}
   \left(59 x_{2}^2-79 x_{2}+20\right)+71 B^2
   x_{2}\right)\right\rbrace
\end{eqnarray}
\begin{equation}
 \rho^{(2)\text{OPE}}_{2,8}(s)=0
\end{equation}
%
%
\twocolumngrid


\end{document}